%% file: main.tex
\documentclass[12pt]{iopart}
\usepackage{graphicx}
\usepackage{subfigure}
\usepackage{amsthm}
\usepackage{amsmath}
\usepackage{amssymb}
\usepackage{amsfonts}
\usepackage{dsfont}
\usepackage{bm}
\usepackage{tikz}
\usepackage{tikz-3dplot}
\usepackage{tkz-euclide}
\usepackage{epstopdf}
\usepackage{hyperref}
\begin{document}

\title[Brownian Molecules Formed by Delayed Harmonic Interactions]{Brownian Molecules Formed by Delayed Harmonic Interactions}

\author{Daniel Geiss$^{1,2,\star}$, Klaus Kroy$^{2, \ddagger}$, Viktor Holubec$^{2,3, \dagger}$}

\address{$^1$ Max Planck Institute for Mathematics in the Sciences, Inselstr. 22, D-04103 Leipzig, Germany}
\address{$^2$ Institut f\"ur Theoretische Physik, Universit\"at Leipzig,  Postfach 100 920, D-04009 Leipzig, Germany}
\address{$^3$ Charles University,  Faculty of Mathematics and Physics, 
 Department of Macromolecular Physics, V Hole{\v s}ovi{\v c}k{\' a}ch 2, 
 CZ-180~00~Praha, Czech Republic}
\ead{$\star$ daniel.geiss@mis.mpg.de}
\ead{$\ddagger$ klaus.kroy@uni-leipzig.de}
\ead{$\dagger$ viktor.holubec@mff.cuni.cz}

\vspace{10pt}
\begin{indented}
\item[]August 2017
\end{indented}

\begin{abstract}
A time-delayed response of individual living organisms to information exchanged within flocks or swarms leads to the emergence of complex collective behaviors. A recent experimental setup by Khadka~et~al.,~Nat.~Commun.~{\bf 9},~3864~(2018), employing synthetic microswimmers, allows to emulate and study such behavior in a controlled way, in the lab. Motivated by these experiments, we study a system of $N$ Brownian particles interacting via a retarded harmonic interaction. For $N \le 3$, we characterize its collective behavior analytically, by solving the pertinent stochastic delay-differential equations, and for $N > 3$ by Brownian dynamics simulations. The particles form molecule-like non-equilibrium structures which become unstable with increasing number of particles, delay time, and interaction strength. We evaluate the entropy and information fluxes maintaining these structures and, to quantitatively characterize their stability, develop an approximate time-dependent transition-state theory to characterize transitions between different isomers of the molecules.  
For completeness, we include a comprehensive discussion of the analytical solution procedure for systems of linear stochastic delay differential equations in finite dimension, and new results for covariance and time-correlation matrices.
\end{abstract}

%
\vspace{2pc}

%
\submitto{\NJP}
%
%
%

\input{Content/Introduction.tex}
\input{Content/GeneralModel.tex}
\input{Content/Dimer.tex}
\input{Content/Trimer.tex}
\input{Content/StructureAndEntropy.tex}
\input{Content/IsomerTransition.tex}
\input{Content/Extension.tex}
\input{Content/Conclusion.tex}

\clearpage

\input{Content/Appendix.tex}

\section*{References}

\bibliographystyle{iopart-num}
\bibliography{bibfile}

\end{document}

%% file: Content/Introduction.tex
\section{Introduction}

\subsection{Feedback systems}

From the synchronized response of a flock of starlings \cite{cavagna2010starling} avoiding an attack of a predator to the formation of colonies of living bacteria \cite{ben1994bacteria,zhang2010bacteria}, the surging field of active matter provides a wide range of fascinating phenomena. Its ultimate aim is to develop a microscopic understanding of the behavior of large numbers of interacting, active and energy consuming ``agents''  \cite{elgeti2015swimmers,ramaswamy2010mechanics}, with a focus on emergent collective behavior \cite{vicsek2012collective}. Most of the quantitative models, such as the Vicsek model \cite{vicsek1995model}, neglect the finite speed of information transmission between the individual particles. However, recent studies \cite{attanasi2014information,piwowarczyk2018collective,mijalkov2016engineering,khadka2018active} have shown that a time delay in the interaction may significantly affect the system dynamics. Moreover, experimental realizations mimicking natural interacting systems require implementing the non-physical interactions, such as a reaction of a bird to its environment, via a feedback loop \cite{mijalkov2016engineering,khadka2018active,Brambilla2013,bauerle2018quorum}. Finite processing of the information in the feedback loop then inevitably introduces time delay into the system dynamics. 

Current (mainly optical) micromanipulation techniques allow to realize such feedback systems on microscale \cite{khadka2018active,bauerle2018quorum,gibbs1981optics,arecchi1992optics,masoller2002feedback,kanter2010optics,Baraban2013,Qian2013} . Many \cite{bauerle2018quorum,Baraban2013,Qian2013}  of these techniques are based on spherical Janus particles \cite{walther2008janus,jiang2010janus} with hemispheres coated with different materials in order to excite surface flows to propel them actively upon illumination or in presence of other energy sources (e.g. chemical fuel added to the solvent). In order to steer these particles, one usually has to wait until the rotational diffusion reorients them towards the desired location. This issue was resolved by the setup introduced by Khadka et al.~\cite{khadka2018active} based on Brownian particles, symmetrically decorated by gold nanoparticles, that thermophoretically self-propel in the direction determined by the position of the laser focus on their circumference. In the feedback experiment, the particles are tracked with a camera with finite exposure time and the position of the heating laser is determined by positions of the particles in the previous frame. The setup allows to create arbitrary time-delayed interactions in the many-body system. In Ref.~\cite{khadka2018active}, an interaction leading to constant absolute values of velocities of the individual particles was considered.

In the present paper, we theoretically analyze a system similar to that considered by Khadka et al.~\cite{khadka2018active}, but with harmonic interactions between the individual particles. Similarly to the case of Ref.~\cite{khadka2018active}, the two-dimensional $N$-particle system is described by a set of $2N$ coupled non-linear stochastic delay differential equations. For small enough values of the delay, highly symmetric non-equilibrium molecular-like structures form after a transient period, which fluctuate due to thermal noise. The resulting structures strongly differ from the molecules created with the constant-velocity protocol studied in Ref.~\cite{khadka2018active}, which oscillated, even for vanishing noise amplitude, due to the nonzero delay. Another difference between the two realizations is that our setup leads, for large delays, to oscillations with amplitudes exponentially increasing in time, while, in the setup of Khadka et al., they are always bounded. The specific form of the interaction considered in our setup moreover allows us to linearize the underlying set of stochastic delay differential equations and to study many aspects of the model behavior analytically. For dimer ($N=2$) and trimer ($N=3$), we use the linearized model to calculate properties of the resulting Gaussian probability distributions for the bond length, namely its mean values, covariance matrix, and time-correlation matrix. We verify the validity of these results by Brownian dynamics (BD) simulations of the complete model. Moreover, we use the BD simulations to show that the behavior of larger molecules ($N>3$) is qualitatively the same as that of the dimer and the trimer, with the difference that the critical value of the delay, beyond which the molecules become unstable, decreases inversely in the particle number $N$. If we would scale the interaction strength by the particle number, as it is common in toy models of many-body systems, the critical value of the delay would thus be constant. If we label the individual particles, we can distinguish between several isomers of the respective molecules according to their ordering.  In the course of time, the noise induces jumps of a given molecule between the individual isomers. We utilize our analytical results for the dimer and for the trimer to evaluate the corresponding transition rates using Kramers' theory \cite{kramers1940original,hanggi1990review} and a more recent theory by Bullerjahn et al.~\cite{bullerjahn201spectroscopy}. We compare the results with the rates calculated from our BD simulations and identify a useful formula for the transition rate that provides good predictions for small and moderate values of the delay.

\subsection{Stochastic delay differential equations}

In general, delay differential equations (DDE's) \cite{bellman1963delay,atay2010complex} may generate rich dynamics \cite{michiels2007stability}. Their solutions may converge to fixed points or limit cycles, behave chaotically, and exhibit multistability \cite{foss1996multistability}. For systems affected by noise, the DDEs are generalized to stochastic delay differential equations (SDDE) \cite{longtin2009stochastic}, which exhibit non-Markovian dynamics. Due to delay-induced temporal correlations, the corresponding Fokker-Planck equation (FPE) cannot be written in a closed form \cite{guillouzic1999small,frank2003fokker,loos2017force}. Instead, one obtains an infinite hierarchy of coupled FPEs for the $n$-time joint probability densities for which generally no finite closure is known. 

For non-linear systems, there are three established approximate approaches how to tackle the infinite hierarchy: (i) the so called small delay approximation \cite{guillouzic1999small}, which employs a Taylor expansion in the delay to make the equations time local; (ii) also, if the delayed terms in the SDDE are small so that the system dynamics is almost Markovian, a perturbation theory can be applied, leading to closed FPEs for the individual joint probability densities \cite{frank2005delay}; (iii) a closed equation for the 1-time probability density valid in the steady state can be obtained by linearization of all equations of the FPE hierarchy except for the first one \cite{loos2017force}.

So far, the only exactly solved problem is a one-dimensional linear stochastic delay equation with Gaussian white noise, whose $n$-time probability densities are given by multivariate Gaussians. Its stationary solution and the conditions for its existence were discussed in Refs.~\cite{kuchler1992delay,guillouzic1999small,frank2003fokker}. Recently, a full time-dependent solution for 1- and 2-time probability densities was found by Giuggioli et al. \cite{giuggioli2016fokker}. Employing the so-called time-convolutionless transform introduced in the 1970s \cite{adelman1976fokker,fox1977generalized, hanggi1978correlation,sancho1982analytical,hernandez1983joint},
these authors transformed the non-Markovian linear delayed Langevin equation (LE) into a time-local form. Afterwards, they utilized this result in a derivation of analytically solvable time-local FPEs for 1- and 2-time probability densities.

Even though an analytical treatment is thus rather complicated, there is a great interest in understanding DDE's and SDDE's, due to their broad range of applications. Prominent examples are found in population dynamics \cite{gopalsamy2013population,mao2005population}, where the delay results from maturation times, economics \cite{voss200economic,stoica2005finance,mackey1989price,gao2009finance} when the limited reaction times of the market participants matters, or engineering  \cite{kyrychko2010engineering}. In biology, finite transition times can play a significant role in physiological systems \cite{beuter1993feedback,chen1997coordination,Novak2008} and neural networks  \cite{haken2007brain,marcus1989stability,sompolinsky1991visual, foss1996multistability}. Recently~\cite{rosinberg2015thermodyn,van2018thermodynamic,loos2019heat}, first efforts were also made to incorporate a time delay into the language of stochastic thermodynamics \cite{seifert2012thermodyn,sekimoto2010stochastic} in order to evaluate energy and entropy fluxes in time-delayed stochastic system.

\subsection{Outline}

The rest of the paper is structured as follows. In Sec.~\ref{Sec.:Dynamics} we first introduce the general model and formulate the underlying equations of motion in terms of SDDEs. After appropriate linearization, we study them analytically, considering both transient and stationary properties of the probability distributions for ``bond'' lengths in ``molecules'' self-assembling from two and three particles. Larger systems are studied in Sec.~\ref{sec:structure}. In Sec.~\ref{sec:entropy_fluxes}, we apply the obtained results for evaluation of the entropy outflux (or information influx) from the system due to the feedback maintaining the non-equilibrium stationary structures. In order to obtain a more quantitative characterization of the stability of the non-equilibrium molecules, we utilize our analytical description of the dimer and trimer for analytical and numerical investigation of the isomer transitions and back up the results by BD simulations in Sec.~\ref{Sec.: TST}. In order to assess the robustness of our findings, Sec.~\ref{Sec.: Extension} addresses the role of the functional form of the memory kernel considering negative delays and exponential memories. We summarize our findings and conclude in Sec.~\ref{Sec.:Conclusion}. Most of the technical details are given in \ref{App:A1} -- \ref{app:time_corr_matr}. In \ref{App:A1}, we review the known results concerning the solution of systems of LDDEs. In \ref{App:B}, we show how to generalize these results for linear SDDEs. Finally, we apply the obtained results in \ref{app:time_corr_matr} for the calculation of the time-correlation matrix and the covariance matrix for systems of linear SDDEs. 

\begin{figure}[t!]
\centering
	\begin{tikzpicture}
	\node (img1)  {\includegraphics[width=0.49\columnwidth]{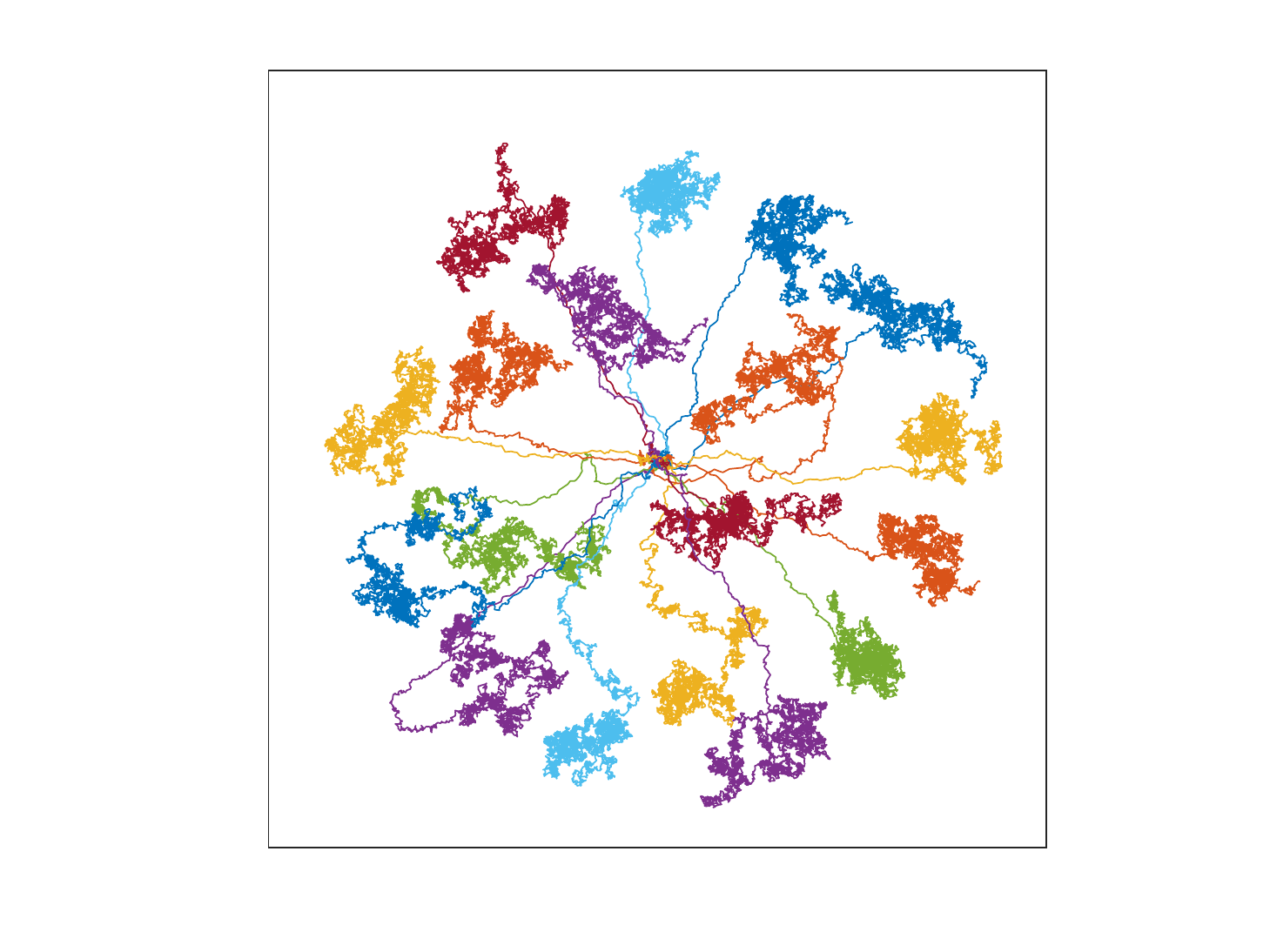}};
	\node[right=of img1, node distance=0.0cm, yshift=0cm,xshift=0cm] (img2)
	{\includegraphics[width=0.3\columnwidth]{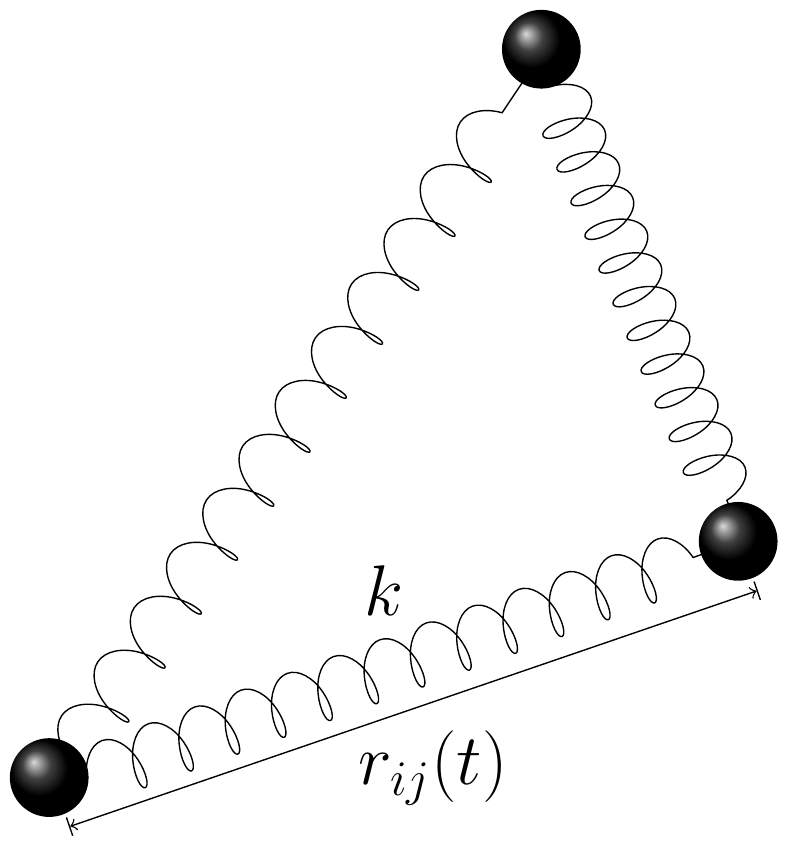}};
	\node[above=of img1, node distance=0.0cm, yshift=-1.8cm,xshift=-2.9cm] {a)};
	\node[above=of img2, node distance=0.0cm, yshift=-1.8cm,xshift=-2.9cm] {b)};	
	\end{tikzpicture}	
	\caption{Panel a) displays stochastic trajectories of $N=18$ Brownian particles bound by delayed harmonic forces. At long times and short enough delays, the particles form molecular-like vibrating structures, while for long delays, they exhibit exponentially diverging oscillations. In b), we depict the model as a system of $N=3$ Brownian particles interconnected by ideal springs with stiffness $k$ and equilibrium lengths $R$, whose response to deformation is time-delayed by evaluating $r_{ij}$ at an earlier time $t-\tau$.}	
	\label{fig:modelsketch}	
\end{figure}

%% file: Content/GeneralModel.tex
\section{Stochastic Dynamics}\label{Sec.:Dynamics}
We consider a two-dimensional system of $N$ overdamped Brownian particles coupled via time-delayed harmonic pair interactions given by the potential
\begin{equation}
V_2[r_{ij}(t-\tau)] = \frac{k}{2} \left[ r_{ij}(t-\tau) - R \right]^2,
\label{eq:two_particle_V}
\end{equation}
depicted in Fig.\ref{fig:modelsketch} by springs connecting the individual particles. In Eq.~\eqref{eq:two_particle_V}, $R>0$ denotes the equilibrium spring length, $k$ their stiffness, and $r_{ij}(t-\tau) = \left| \mathbf{r}_i(t-\tau) - \mathbf{r}_j(t-\tau) \right|$ is the distance between the particles $i$ and $j$ located at positions $\mathbf{r}_i$ and $\mathbf{r}_j$ at an earlier time $t-\tau$. Clearly, the picture of linear springs can properly represent the time-delayed interactions only for a vanishing time delay $\tau$.

Altogether, the particles diffuse in the composity potential 
\begin{equation}
V = \frac{1}{2}\sum_{(i,j)} V_2(r_{ij}),
\label{eq:total_energy}
\end{equation}
where the summation runs over all pairs $(i,j)$, so that the $i$th particle is driven by the force $\mathbf{F}_i = - \nabla_i V = (\partial_{x_i} V, \partial_{y_i} V)$. Because at time $t$ the particle feels the value of the potential corresponding to its position at time $t-\tau$, here $x_i$ and $y_i$ denote the Cartesian coordinates of the position vector $\mathbf{r}_i$ \emph{in the past}. In effect, the $N$-particle system therefore obeys the set of \emph{nonlinear} delayed Langevin equations
\begin{equation}
	\dot{\mathbf{r}}_i(t) = -\frac{k}{\gamma} \sum_{j\neq i} \left[r_{ij}(t-\tau) -R\right]\mathbf{e}_{ij}(t-\tau) + \sqrt{2D_0} \bm{\eta}_i(t)\,,\quad i = 1,\dots, N.
\label{eq:Langevin_n}
\end{equation}
The unit vector $\mathbf{e}_{ij} = \mathbf{r}_{ij}/r_{ij}$ points from particle $j$ to particle $i$ and the diffusion coefficient $D_0=(\beta\gamma)^{-1}$ is related to the inverse temperature $\beta=1/k_{\rm B}T$ and the friction coefficient $\gamma$ via the Einstein relation ($k_{\rm B}$ denotes the Boltzmann constant). The vectors $\bm{\eta}_i$ comprise independent Gaussian white noises satisfying the relations
\begin{equation}
	\langle \bm{\eta}_i(t)\rangle=0,\qquad \left< \eta^\alpha_i(t_1)\eta^\beta_j(t_2)\right> = \delta_{ij} \delta_{\alpha\beta} \delta_\mathrm{D}(t_1-t_2).  \label{noise}
\end{equation}
The numbers $\alpha$ and $\beta$ label the components of the vector $\bm{\eta}_i$, while $i$ refers to the specific particle. Note that the noise and the friction in Eq.~\eqref{eq:Langevin_n} are related by the fluctuation-dissipation theorem~\cite{Kubo1966} for a vanishing delay $\tau=0$ only, and that the system is always out of thermodynamic  equilibrium for $\tau>0$~\cite{loos2019heat}. In order to obtain a model that would obey the fluctuation-dissipation theorem, one should consider a different noise correlation function \eqref{noise}.

For small enough values of the time delay, the particles form, after an initial transient period, highly symmetric molecular-like structures, some of which are displayed in Fig.~\ref{fig:CompareBondNumber} a) in Sec.~\ref{sec:structure}. For $N=2$ (dimer) and $N=3$ (trimer) the steady-state structures occupy the global minimum of the potential $V$. For $N > 3$ the global minimum becomes inaccessible due the chosen two-dimensional geometry and the resulting structures are thus frustrated in the sense that some of the springs do not reach their equilibrium length in the steady-state. The structures are dynamical, due to the Brownian motion of the particles, which persistently kicks the system out of the minimum of the potential energy~\eqref{eq:total_energy}. The effect of the delay is that the system may exhibit exponentially decaying oscillations on its return to the energy minimum. The decay rate of these oscillations decreases with increasing delay, and, for delays larger than a certain threshold, their amplitude exponentially increases. This is because large delays induce in the system a ``swing effect'', when the repulsive force from one side of the potential propels the particle to a ``higher'' position at its opposite side, and so on.

Within the equilibrium model that obeys the fluctuation-dissipation theorem, the stationary probability density function (PDF) for positions of the individual particles would simply be the Boltzmann distribution $\exp(-\beta V)/Z$ with potential $V$, inverse temperature $\beta$, and normalization $Z$. However, the physical situation at hand, where the delay is interpreted as a result of a feedback control mechanism and thus is independent of the noise, requires the more involved description with Eq.~\eqref{noise} that leads to non-trivial non-equilibrium steady states. Consequently, the Boltzmann distribution as stated above has to be modified. For the simplest case of a dimer one finds an (approximate) Gaussian distribution as discussed later while for larger molecules the description is more complicated.

A similar system with a quasi-constant force between the particles (constant upto a change of sign at distance $R$), i.e. obeying the set of Langevion equations
\begin{equation}
	\dot{\mathbf{r}}_i(t) = -\frac{k}{\gamma} \frac{\sum_{j\neq i} {\rm sign}\!\left[r_{ij}(t-\tau) -R\right]\mathbf{e}_{ij}(t-\tau)}{\left|\sum_{j\neq i} {\rm sign}\!\left[r_{ij}(t-\tau) -R\right]\mathbf{e}_{ij}(t-\tau)\right|} + \sqrt{2D_0} \bm{\eta}_i(t),
\label{eq:LangevinCV_n}
\end{equation}
$i = 1,\dots, N$, with ${\rm sign}(x)$ denoting the signum function, was discussed earlier in \cite{khadka2018active}. The main difference from our setting~\eqref{eq:Langevin_n} is that, in Eq.~\eqref{eq:LangevinCV_n}, the absolute value of the force does not depend on the interparticle distances and  the particle number $N$. The main benefit of assuming the harmonic potential in Eq.~\eqref{eq:Langevin_n} is that it allows much more complete analytical treatment. To allow for an easy comparison of the two models, we illustrate our results using parameters inspired by Ref.~\cite{khadka2018active}. In the following, we first review some analytical results for stochastic dynamics of dimers and trimers. On this basis, we will return to the discussion of the emerging structures in Sec.~\ref{sec:structure}.

%% file: Content/Dimer.tex
\subsection{Center of Mass}
\label{sec:CMS}

Similarly as for the dynamics considered in Ref.~\cite{khadka2018active}, the center of mass coordinate $\mathbf{r}_\mathrm{c} \equiv \left(\sum_{i=1}^n\mathbf{r}_i\right)/N$ of the system obeys the Langevin equation
\begin{equation}
\dot{\mathbf{r}}_\mathrm{c}(t) = \sqrt{2D_c} \bm{\eta}_\mathrm{c}(t)\,,
\label{COM_Langevin}
\end{equation}
where $\bm{\eta}_\mathrm{c}\equiv \sum_{i=1}^N \bm{\eta}_i /\sqrt{2}$ denotes Gaussian white noise satisfying Eq.~\eqref{noise} (with the labels $i,j$ replaced by $ {\rm c}$) and the diffusion coefficient $D_c = D/N$. Regardless of the interactions, the center of mass performs ordinary Brownian motion and, assuming the center of mass is at time $t=0$ located at the point $\mathbf{r}_{0}$, the probability density function (PDF) for $\mathbf{r}_{\rm c}$ reads 
\begin{equation}
	P^\mathrm{c}_N(\mathbf{r},t) = \sqrt{\frac{N}{4\pi D t}} \exp \left[- N \frac{(\mathbf{r} - \mathbf{r}_{0})^2}{4D t} \right]\,.
\end{equation}

\subsection{Dimer}\label{Section: Dimer}

Let us now consider the simplest case of two interacting particles. For $N=2$, Eq.~\eqref{eq:Langevin_n} yields the system of four coupled equations of motion:
\begin{align}
	\dot{\mathbf{r}}_1(t) &= -\frac{k}{\gamma} \left[r(t-\tau) -R\right]\mathbf{e}(t-\tau) + \sqrt{2D} \bm{\eta}_1(t)\,, \\
	\dot{\mathbf{r}}_2(t) &= +\frac{k}{\gamma} \left[r(t-\tau) -R\right]\mathbf{e}(t-\tau) + \sqrt{2D} \bm{\eta}_2(t)\,,
\end{align}
where we have used the abbreviations $\mathbf{e}(t)\equiv\mathbf{e}_{12}(t)$ and $r(t)\equiv r_{12}(t)$. In the previous section, we have already resolved the dynamics of the center of mass coordinate for arbitrary $N$. Now, we consider only the dynamics of the relative coordinate $\mathbf{r} = \mathbf{r}_{12}  = \mathbf{r}_1-\mathbf{r}_2$ which obeys the equation of motion
\begin{equation}
\dot{\mathbf{r}}(t) = - \omega \left[r(t-\tau) - R \right] \mathbf{e}(t-\tau) + \sqrt{4D} \bm{\eta}_\mathrm{r}(t)   \label{Rel Langevin}
\end{equation}
with frequency 
\begin{equation}
	\omega \equiv 2k/\gamma    \label{omega}
\end{equation}
and $\bm{\eta}_\mathrm{r}\equiv \left(\bm{\eta}_1-\bm{\eta}_2\right)/\sqrt{2}$ describing Gaussian white noise satisfying Eq.~\eqref{noise} with the vector components $i,j$ replaced by the label $\mathrm{r}$.

Projecting Eq.~\eqref{Rel Langevin} on the direction of the bond at time $t$ [by multiplication with $\mathbf{e}(t) = \left(\cos \varphi(t), \sin \varphi(t)\right)$] and on the direction perpendicular to the bond [by multiplying it with $\mathbf{e}_\varphi = \left(-\sin \varphi(t), \cos \varphi(t)\right)$], we obtain the equations
\begin{align}
\dot{r}(t) =& - \omega \left[ r(t-\tau) - R\right] \cos\left[ \varphi(t,t-\tau)\right] + \sqrt{4D}\eta_{\rm r}^r(t)\,,   
\label{eq:eqr2}
\\
\dot{\varphi}(t) =& \omega \frac{ r(t-\tau) -R}{r(t)}  \sin \left[\varphi(t,t-\tau)\right]  + \sqrt{\frac{4D}{r^2(t)}}\eta_{\rm r}^\varphi(t)\,,
\label{eq:eqphi2}
\end{align}
where $\varphi(t,t-\tau) = \varphi(t) - \varphi(t-\tau)$ denotes the change of orientation of the vector $\mathbf{e}(t-\tau)$ during time $\tau$. Above, we used the formulas $\mathbf{r} \equiv r \mathbf{e}$, $\dot{\mathbf{r}} \equiv \dot{r} \mathbf{e} + r\dot{\varphi} \mathbf{e}_\varphi$ and $\bm{\eta}_{\rm r} \equiv \eta_{\rm r}^r \mathbf{e} + \eta_{\rm r}^\varphi \mathbf{e}_\varphi$.

From symmetry considerations, it follows that the stationary PDF for the orientation must be constant in $\varphi$. To gain analytical insight into the dynamics and PDF of the bond-length $r$, we linearize the coupled Langevin equations \eqref{eq:eqr2} and \eqref{eq:eqphi2}. If the angle dependent stiffness $2k/\gamma \cos\left[ \varphi(t,t-\tau)\right]$ in Eq.~\eqref{eq:eqr2} is strong enough such that the terms proportional to $[r(t-\tau) - R]/R$ can safely be neglected independently of the noise strength, the formula~\eqref{eq:eqphi2} assumes the form~\footnote{In Eq.~\eqref{eq:eqr2}, we set $r = [(r-R)/R + 1]R$, expand it in $(r-R)/R$, and neglect all terms proportional to $(r-R)/R$.}
\begin{equation}
\dot{\varphi}(t) = \sqrt{\frac{4D}{R^2}}\eta_{\rm r}^\varphi(t). 	
\end{equation}
The corresponding transition PDF (Green's function) for change of the orientation by $\varphi(t,t-\tau) = \varphi(t) - \varphi(t-\tau)$ during the time interval $\tau$ reads \cite{Stephens1963random,Peruani2007,Selmke2018theory} $p\left[\varphi(t,t-\tau),\tau\right] = (2\pi)^{-1} + \pi^{-1} \sum_{m=1}^\infty \cos\left[m \varphi(t,t-\tau)\right]\exp\left[-2m^2\tau D/R^2\right]$. Using this function, we average Eq.~\eqref{eq:eqr2} over $\varphi(t,t-\tau)$ obtaining
\begin{equation}
\dot{r}(t) = -\omega_\tau  \left[ r(t-\tau)-R \right] +\sqrt{4D}\eta_{\rm r}^r(t),
\label{eq:eqr3}
\end{equation}
where $\omega_\tau = \omega \exp(-2D\tau/R^2)$ is the natural relaxation rate. Note that the same formula with $\omega_\tau$ substituted by $\omega$ is obtained by simply assuming that the change of the orientation $\varphi(t,t-\tau)$ of the bond per delay time $\tau$ is small, i.e. for $2\tau D/R^2 \ll 1$. The main difference between the two approximations is that $D/R^2$ does not necessarily have to be small in the first case. We will discuss the regime of validity of the Eq.~\eqref{eq:eqr3} in more detail around Eq.~\eqref{eq:dimer_coditions} below. 

Equation~\eqref{eq:eqr3} is a linear stochastic delay differential equation which can be solved analytically for $r \in (-\infty,\infty)$. In our setting, $r \ge 0$ and thus we should solve Eq.~\eqref{eq:eqr3} with a reflecting boundary at the origin. However, since we have assumed that $|r(t-\tau)-R|/r(t) \ll 1$, we already work in the regime where $r$ only seldom significantly deviates from $R$ and thus the solution of Eq.~\eqref{eq:eqr3} on the full real axis should approximate well the desired solution on the positive half-line. The solution of Eq.~\eqref{eq:eqr3} for $r \in (-\infty,\infty)$ and $t\ge 0$ in terms of deviations of the bond length from its equilibrium length (which we call as \emph{shifted bond length}), 
\begin{equation}
x(t) = r(t) - R,
\label{eq:shifted_r}
\end{equation}
can be derived by several methods. We review two of them (time-convolutionless transform and Gaussian ansatz) for a general multidimensional linear SDDE
in \ref{App:A1}--\ref{app:noise_sol3}. Here, we present just the main formulas. Assuming that the system was initially in state $x(0) = x_0$ and that $x(t)=0$ for $t<0$, the formal solution of Eq.~\eqref{eq:eqr3} for $r \in (-\infty,\infty)$ and $t\ge 0$ reads
\begin{equation}
	x(t) = x_0\lambda(t) + \sqrt{4D} \int_0^t \mathrm{d}s\ \lambda(t-s)\eta_{\rm r}^r(s),  \label{eq:solution_SDDE_MT}
\end{equation}
where the dimensionless Green's function
\begin{equation}
	\lambda(t) = \sum_{k=0}^{\infty} \frac{\left(-\omega_\tau \right)^k}{k!}\left(t-k\tau\right)^k \theta(t-k\tau)  \label{eq:fundamental_sol_MT}
\end{equation}
is obtained by solving Eq.~\eqref{eq:eqr3} with $D=0$ for the specfic initial conditions $\lambda(t)=0$ for $t<0$ and $\lambda(0)=1$. The symbol $ \theta(x)$ in Eq.~\eqref{eq:fundamental_sol_MT} stands for the Heaviside step function. For an arbitrary initial condition $x(t)$, $t<0$ and $x(0) = x_0$, the expression  $x_0 \lambda(t)$ in Eq.~\eqref{eq:solution_SDDE_MT} must be substituted by $x_0 \lambda(t) - \omega_\tau \int_{-\tau}^0 \mathrm{d}s\ \lambda (t-\tau-s)x(s)$. Based on the value of the \textit{reduced delay} $\omega_\tau \tau$ which is a dimensionless measure for the relevance of the delay relative to the natural relaxation time, the Green's function $\lambda(t)$ in Eq.~\eqref{eq:solution_SDDE_MT} exhibits three different dynamical regimes discussed in detail in \ref{App:A1}, in Fig.~\ref{fig:properties_of_lambda} and also below: (i) monotonic exponential decay to zero for short delays $0\leq\omega_\tau \tau\leq1/e$, (ii) oscillatory exponential decay to zero for intermediate (`resonant') delays $1/e\leq\omega_\tau \tau\leq \pi/2$, and (iii) oscillatory exponential divergence for long delays $\omega_\tau \tau>\pi/2$.

The stochastic process $x(t)$ in Eq.~\eqref{eq:solution_SDDE_MT} arises as a linear combination of white noises and thus the corresponding PDFs must be Gaussian. Indeed, we find that one- and two- time conditional PDFs for $x(t)$ with the initial condition $\delta(x)$ for $t<0$ and $\delta(x-x_0)$ at $t=0$ read
\begin{align}
	&P_1(x,t|x_0,0) = \frac{1}{\sqrt{2\pi \nu(t)}} \exp\bigg\{ -\frac{1}{2} \bigg(\frac{x- \mu(t)}{\sqrt{\nu(t)}} \bigg)^2 \bigg\}, \label{eq:W_1_prob_distr}  \\
	&P_2(x,t;x',t'|x_0,0) = \frac{1}{\sqrt{ 4\pi^2\nu(t)\nu(t')[1-w^2(t',t)]}}  \label{eq:W_2_prob_distr} \\
	&\qquad \times \exp\bigg\{ \frac{1}{2[1-w^2(t',t)]} \bigg[\bigg(\frac{x- \mu(t)}{\sqrt{\nu(t)}} \bigg)^2 + \bigg(\frac{x'- \mu(t')}{\sqrt{\nu(t')}} \bigg)^2   \nonumber\\
	&\qquad - 2w(t,t') \bigg(\frac{x- \mu(t)}{\sqrt{\nu(t)}} \bigg) \bigg(\frac{x'- \mu(t')}{\sqrt{\nu(t')}} \bigg)   \bigg]  \bigg\},
	\nonumber
\end{align}
where $t\ge t' \ge 0$ and 
\begin{align}
    \mu(t) &\equiv \left<x(t)\right>  =  x_0\lambda(t),
    \label{eq:mean_bond_length}\\
	\nu(t) &\equiv \left<x^2(t)\right> - \mu^2(t)  = 4D \int_0^t \mathrm{d}s\ \lambda^2(s), 
	\label{eq:variance_dimer}\\
	w(t,t') &\equiv \frac{\left<x(t)x(t')\right>-\mu(t)\mu(t')}{\sqrt{\nu(t)\nu(t')}} = \frac{4D}{\sqrt{\nu(t) \nu(t')}}\int_0^{t'} \mathrm{d}s\ \lambda(t-s)\lambda(t'-s)
\end{align}
denote the mean of the shifted bond-length \eqref{eq:shifted_r}, its variance, and normalized time correlation, respectively.

The function $P_1(x,t|x_0,0){\rm d}x$ stands for the probability that the system which departs with certainty from state $x_0$ at time $0$ is found at time $t$ somewhere in the interval $(x,x+{\rm d}x)$. Similarly, $P_2(x,t;x',t'|x_0,0){\rm d}x{\rm d}x'$ denotes the probability that the (shifted) bond length is in the interval $(x',x'+{\rm d}x')$ at time $t'$ and at a later time $t$ in $(x,x+{\rm d}x)$ under the condition that it has started at time $t=0$ at $x_0$. The one-time PDF $P_1$ possesses the same structure as the corresponding PDF for $\tau = 0$ \cite{Risken1996}. The non-Markov character of the process \eqref{eq:eqr3} with nonzero delay manifests itself in the fact that the two-time PDF $P_2$ cannot be constructed from the one-time PDF $P_1$, while this is always possible for a Markov process. 

The FPEs corresponding to the PDFs \eqref{eq:W_1_prob_distr} and \eqref{eq:W_2_prob_distr} are given by Eqs.~\eqref{FPE W1} and \eqref{FPE W2} in \ref{App:B}, respectively. Interestingly enough, the diffusion and drift terms in the FPEs are given by the natural scales $2D$ and $\omega_\tau$ only in the limit $\tau\to 0$. Moreover, both coefficients acquire a time dependence, determined by the function $\lambda(t)$. Specifically, the diffusion and drift coefficients in Eq.~\eqref{FPE W1} for $P_1$ read $D_\tau(t) = D \lambda^2(t) {\rm d}\left[ \int_0^t {\rm d}s \lambda^2(s)/\lambda^2(t)\right]/{\rm d}t$ and  $\omega_\tau(t) = - \dot{\lambda}(t)/\lambda(t)$, respectively~\footnote{The case of $\lambda(t)=0$, where these coefficients diverge, is discussed in more detail in Sec.~\ref{sec:dimer_rate}.}. While the drift coefficient in Eq.~\eqref{FPE W2} for $P_2$ is also given by $\omega_\tau(t)$, the diffusion coefficient reads $2D_\tau(t) + 4D \lambda(t)\int_0^{t'}{\rm d}s\, {\rm d}\left[\lambda(t-s)\lambda(t'-s)/\lambda(t)\right]/{\rm d}t$. This difference in diffusion coefficients is the reason why the PDF $P_2$ can be constructed from $P_1$ in the standard way for Markov processes only for $\tau=0$, where both diffusion coefficients coincide.

\begin{figure}[t!]
\centering
	\begin{tikzpicture}
	\node (img1)  {\includegraphics[width=0.4\columnwidth]{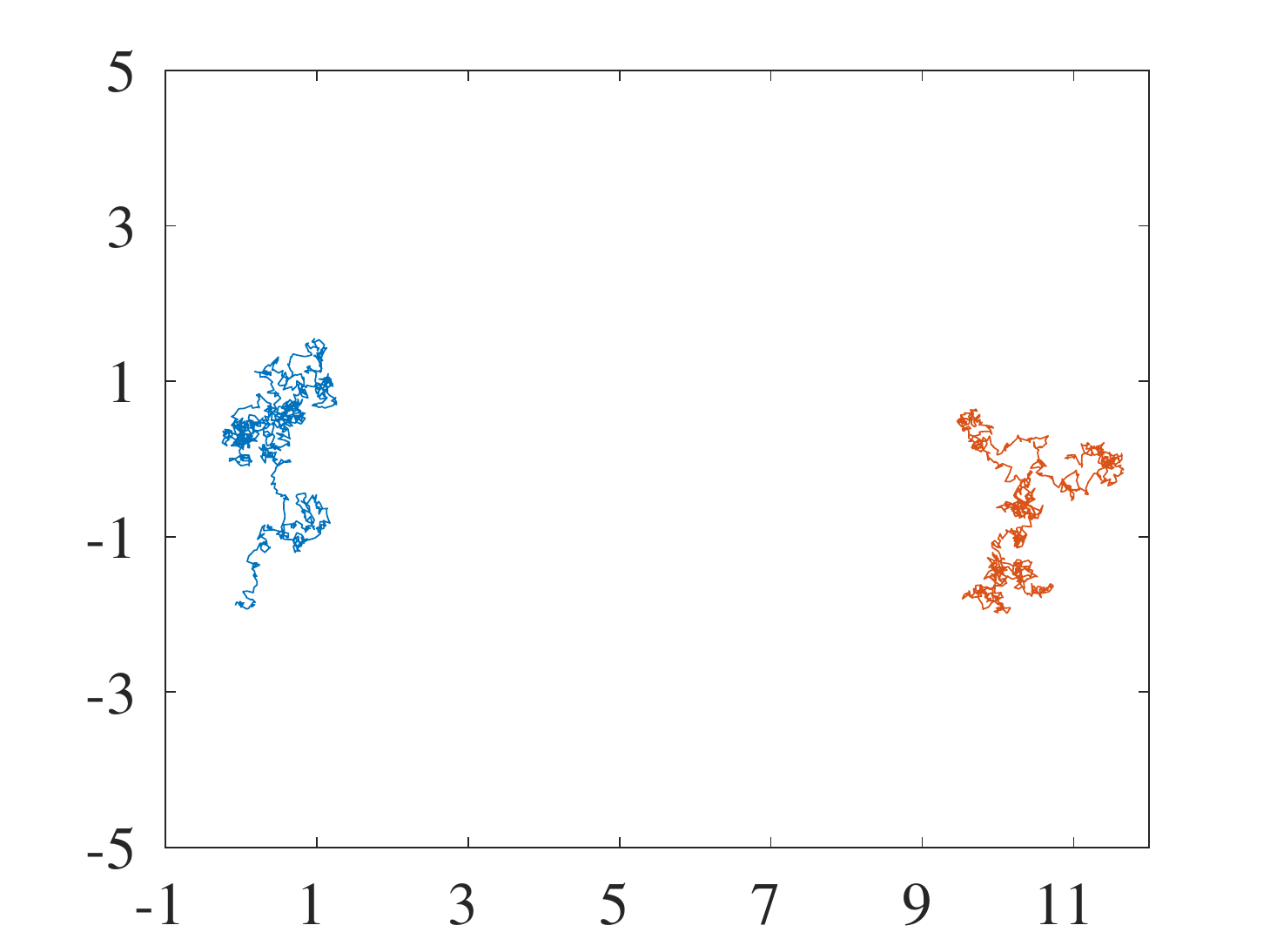}};
	\node[right=of img1, node distance=0.0cm, yshift=0cm,xshift=-1.0cm] (img2)
	{\includegraphics[width=0.4\columnwidth]{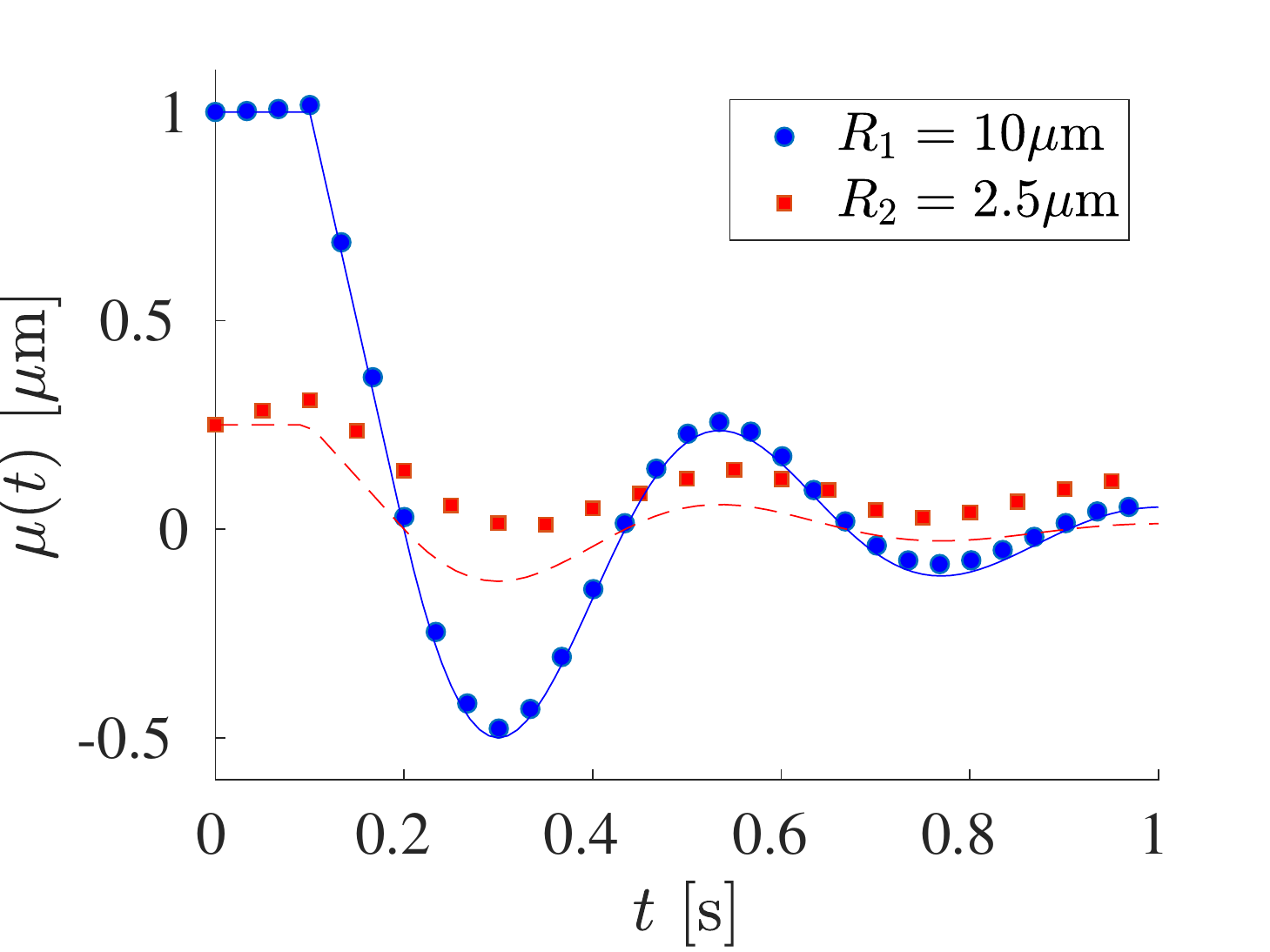}};
	\node[below=of img1, node distance=0.0cm, yshift=1cm,xshift=0.0cm] (img3)
	{\includegraphics[width=0.4\columnwidth]{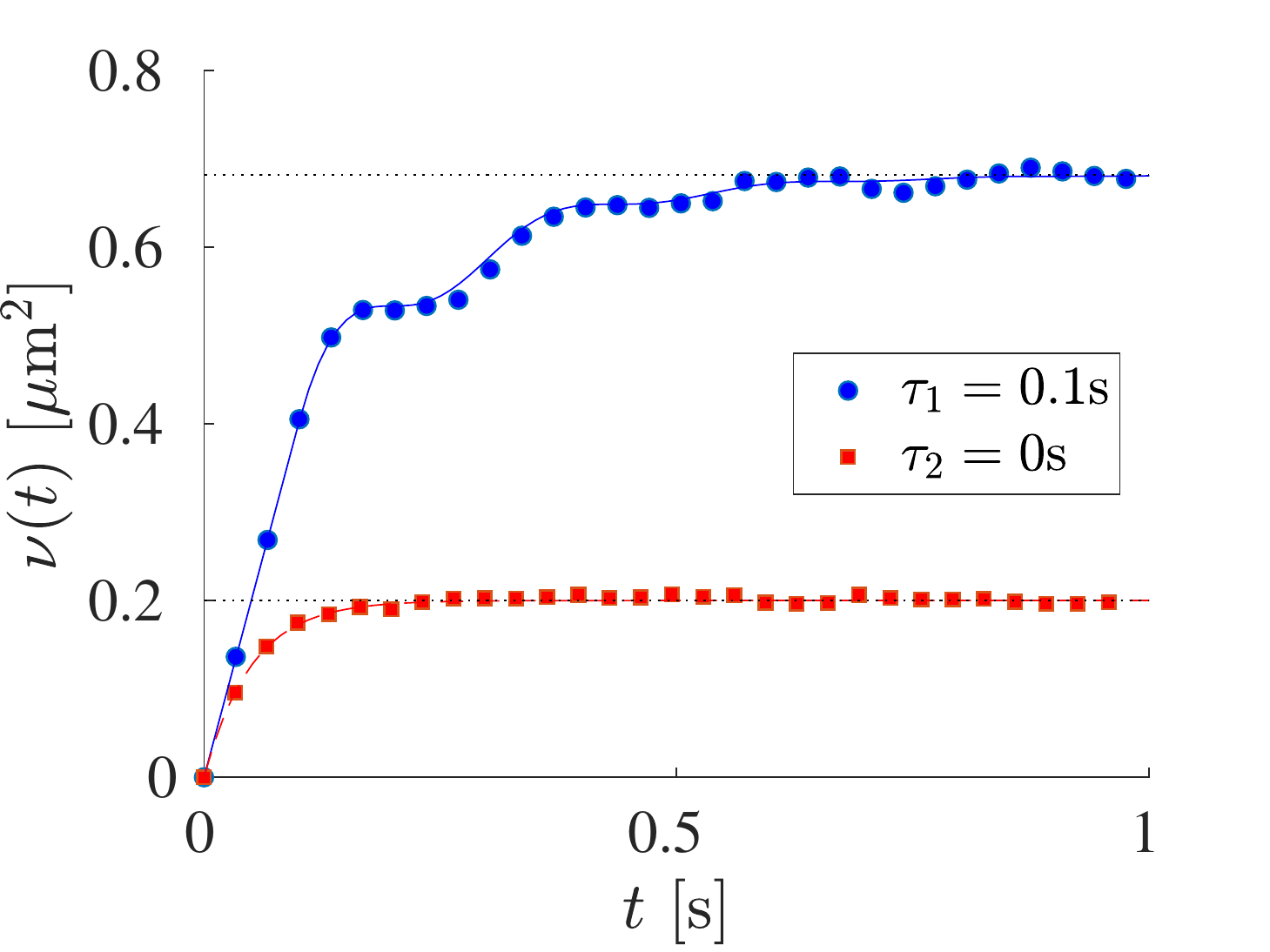}};
	\node[below=of img2, node distance=0.0cm, yshift=1cm,xshift=0.0cm] (img4)
	{\includegraphics[width=0.4\columnwidth]{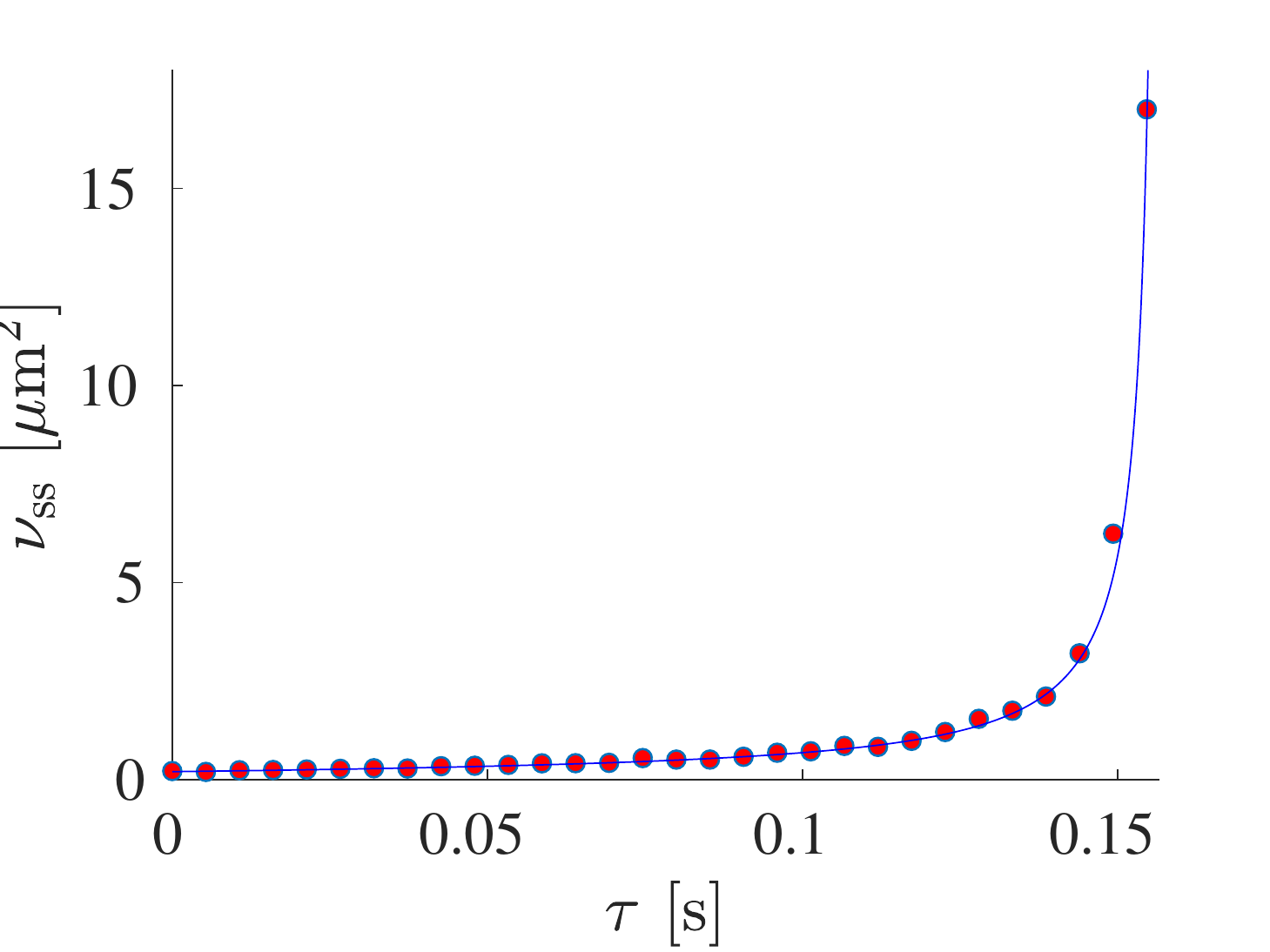}};
	\node[above=of img1, node distance=0.0cm, yshift=-1.8cm,xshift=-3.0cm] {a)};
	\node[above=of img2, node distance=0.0cm, yshift=-1.8cm,xshift=-3.0cm] {b)};	
	\node[above=of img3, node distance=0.0cm, yshift=-1.8cm,xshift=-3.0cm] {c)};
	\node[above=of img4, node distance=0.0cm, yshift=-1.8cm,xshift=-3.0cm] {d)};	
	\end{tikzpicture}	
	\caption{\textbf{Dimer dynamics in Brownian dynamics simulations and theory}  Comparison of the approximate analytical description of the bond dynamics [full and dashed lines in panels b) -- d)] and the behavior of the complete model obtained using Brownian dynamics simulations of Eqs.~\eqref{eq:eqr2}-\eqref{eq:eqphi2} (symbols). Panel a: Typical trajectory obtained from the simulation of a dimer with equilibrium bond length $R=R_1=10$ $\mu$m. Panel b), we show the average shifted bond length \eqref{eq:mean_bond_length} for a large value of $R=R_1$ and the initial value $x(0)= 1$ $\mu$m (solid line), where the analytical approximation works very well, and also for a moderate value of $R=R_2=2.5$ $\mu$m and $x(0)= 0.25$ $\mu$m (dashed line). The convergence of the variance \eqref{eq:variance_dimer} of the bond length to its stationary value $\nu_{\rm ss}$ \eqref{eq:SS_variance} (dotted lines) for $\tau_1=0.1$ s, where the system operates in the oscillatory regime (ii) (solid line), and for $\tau_2=0$ s, where the bond length decays exponentially to $R$ (dashed line), is shown in panel c). The divergence of the stationary value of the variance with increasing $\tau$ is depicted in panel d). If not specified otherwise in the description of the individual panels, we used the experimentally reasonable parameters: $\omega = 2k/\gamma =10$ $1/$s, $\tau_1=0.1$ s, $D=1$ $\mu$m$^2$/s, and $R=10$ $\mu$m. In the BD simulation, we averaged over $10^4$ trajectories with time step $dt=10^{-3}$ s.}	
	\label{fig:bond}	
\end{figure}

For $\tau=0$ the PDF $P_1$ always eventually relaxes to a time independent stationary state which does not depend on the initial condition and which is described by the equilibrium Gibbs formula $P_1(x,\infty;x_0,0) \propto \exp\left(-\omega x^2/2D\right)$. For a nonzero delay in the regimes (i) and (ii), i.e., when the noiseless solution \eqref{eq:fundamental_sol_MT} and thus the mean value $\mu(t) =  x_0\lambda(t)$ of $x$ converges to 0 for $t \to \infty$, the system relaxes into a time-independent non-equilibrium steady state $P_1(x,\infty;x_0,0) \propto \exp\left[-\omega_\tau(\infty) x^2/D_\tau(\infty)\right]$ with $\omega_\tau(\infty) \neq \omega$ and $D_\tau(\infty) \neq 2 D$, see Fig.~\ref{fig:transition_rate_time} in Sec.~\ref{sec:dimer_rate}. This means that, due to the delay, the diffusion coefficient and frequency in the Gaussian Boltzmann distribution for the approximate model have to be replaced by their effective counterparts. The corresponding state is characterized by a nonzero entropy production rate \cite{loos2019heat}. For long time delays, no stationary state exists. In comparison to the analogous setting with a piece-wise constant force discussed previously~\cite{khadka2018active}, this destabilization for long delay times $\tau$ is a new feature, due to increasingly high systematic forces that may occur for long delays.

In the regimes (i) and (ii), the variance $\nu(t) = \sqrt{\left<x^2(t) \right>  - \left<x(t) \right>^2}$
converges to the stationary value \cite{kuchler1992delay,guillouzic1999small,frank2003fokker}
\begin{equation}
	\nu_\mathrm{ss} = \lim_{t\to \infty} \nu(t) = \frac{2 D}{\omega_\tau} \frac{1+\sin(\omega_\tau\tau)}{\cos(\omega_\tau\tau)} . 
\label{eq:SS_variance}
\end{equation}
The derivation of this formula is given in \ref{app:time_corr_matr}, where we also derive an analytical expression for the stationary time correlation function $C(t) = \lim_{s\to \infty}\left<x(s)x(s+t)\right>$. Note that the variance~\eqref{eq:SS_variance} diverges upon entering the unstable regime (iii) for $\omega_\tau\tau\rightarrow \pi/2$.  

The formula~\eqref{eq:SS_variance} finally allows us to specify the regime $\nu_\mathrm{ss} \ll R^2$ where the approximation $\left[r(t-\tau)-R\right]/r(t) \approx 0$ used in the derivation of Eq.~\eqref{eq:eqr3} from Eqs.~\eqref{eq:eqr2} and \eqref{eq:eqphi2} is reasonable because the PDF for $r$ is relatively sharply peaked around the mean bond length $R$. As we already noted, Eq.~\eqref{eq:eqr3} is also valid in the case $2\tau D/R^2 \ll 1$ when the bond rotates only slightly in each delay interval and thus the angle $\varphi(t,t-\tau)$ in Eqs.~\eqref{eq:eqr2} and \eqref{eq:eqphi2} is small. However, also in this case we need to additionally assume that $\nu_\mathrm{ss} \ll R^2$ in order to 
ensure that the error caused by considering the wrong boundary condition at $r = -R$ is negligible. Altogether, the used approximation is expected to give good results if the condition
\begin{equation}
\nu_\mathrm{ss} \ll R^2
\label{eq:dimer_coditions}
\end{equation}
is fulfilled.

An example of the stochastic evolution of the dimer obtained from BD simulations of the exact system~\eqref{eq:eqr2} and \eqref{eq:eqphi2} is depicted in Fig.~\ref{fig:bond} a). In Fig.~\ref{fig:bond} b), we compare the results obtained from BD simulations with the time evolution of the average shifted bond length  \eqref{eq:mean_bond_length} for different values of the equilibrium length $R$. As expected, the approximate analytical formula~\eqref{eq:mean_bond_length} describes well the exact result for large enough $R$ satisfying the inequality~\eqref{eq:dimer_coditions}. For larger values of $\nu_\mathrm{ss}/R^2$, the analytical result underestimates the correct value. This is because the bond length in the BD simulation is obtained from Eq.~\eqref{eq:eqr2} with the reflecting boundary at the origin, while we allow negative values of $r(t)$ in the approximate analytical description. Similarly as for the mean value, the analytical formula~\eqref{eq:variance_dimer} for the bond length variance $\nu(t)$ approximates very well the value obtained from BD simulations for large enough $R$, as shown in Fig.~\ref{fig:bond} c). In Fig.~\ref{fig:bond} d), we depict the monotonous rapid divergence of the stationary variance~\eqref{eq:SS_variance} as the time delay $\tau$ reaches the unstable regime (iii). This means that the delay tends to delocalize structures. On the other hand, the variance can be optimized as a function of the frequency $\omega_\tau$. The best localized structure is obtained for the frequency fulfilling the formula $\cos(\omega_\tau \tau) = \omega_\tau \tau$, i.e. $\omega_\tau \approx 0.74/\tau$ and thus $2k\tau/\gamma \approx 0.74 \exp(2D\tau/R^2)$. For the corresponding value 0.74 of the reduced delay $\omega_\tau \tau$ the system is in the dynamical regime (ii) with exponentially decaying oscillations.

%% file: Content/Trimer.tex
\subsection{Trimer}\label{Section Trimer}

Let us now consider the system composed of three particles. Then, Eq.~\eqref{eq:Langevin_n} gives the system of six coupled equations of motion:
\begin{align}
	\dot{\mathbf{r}}_1(t) &= \frac{k}{\gamma} \left\{ \left[r_{12}(t-\tau) -R\right]\mathbf{e}_{21}(t-\tau) +  \left[r_{13}(t-\tau) -R\right]\mathbf{e}_{31}(t-\tau) \right\}+ \sqrt{2D} \bm{\eta}_1(t), \nonumber\\
	\dot{\mathbf{r}}_2(t) &= \frac{k}{\gamma} \left\{ \left[r_{12}(t-\tau) -R\right]\mathbf{e}_{12}(t-\tau) +  \left[r_{32}(t-\tau) -R\right]\mathbf{e}_{32}(t-\tau) \right\}+ \sqrt{2D} \bm{\eta}_2(t), \nonumber\\
	\dot{\mathbf{r}}_3(t) &= \frac{k}{\gamma} \left\{ \left[r_{32}(t-\tau) -R\right]\mathbf{e}_{23}(t-\tau) +  \left[r_{31}(t-\tau) -R\right]\mathbf{e}_{13}(t-\tau) \right\}+ \sqrt{2D} \bm{\eta}_3(t).   \nonumber
\end{align}
For the relative coordinates $\mathbf{r}_{12}(t) = \mathbf{r}_1(t) - \mathbf{r}_2(t)$ we obtain 
\begin{multline}
	\dot{\mathbf{r}}_{12}(t) = -\omega \left\{ \left[r_{12}(t-\tau) - R \right]\mathbf{e}_{12}(t-\tau) + \frac{1}{2}\left[r_{13}(t-\tau) - R \right]\mathbf{e}_{13}(t-\tau)\right. \\\left.+ \frac{1}{2}\left[r_{32}(t-\tau) - R \right]\mathbf{e}_{32}(t-\tau)  \right\} + \sqrt{2D} \left[ \bm{\eta}_1(t) - \bm{\eta}_2(t) \right]  ,
	\label{eq:r12_EQ1}
\end{multline}
where $\omega = 2k/\gamma$ and similarly for $\dot{\mathbf{r}}_{13}(t)$ and $\dot{\mathbf{r}}_{32}(t)$. To get analytical results for bond lengths $r_{ij}(t) = |\mathbf{r}_{ij}(t)|$, we multiply the formulas for $\dot{\mathbf{r}}_{ij}(t)$ by the corresponding unit vectors $\mathbf{e}_{ij}(t)=\mathbf{r}_{ij}(t)/|\mathbf{r}_{ij}(t)|$ and linearize the resulting equations. To this end, we need to deal with the expressions $\mathbf{e}_{\alpha}(t-\tau)\cdot \mathbf{e}_{\beta}(t)$, $\alpha,\beta = \mathrm{I},\dots,\mathrm{III}$, where we introduced roman numbers as a shorthand indexing $\mathrm{I}\equiv 12,$ $\mathrm{II}\equiv 13,$ $\mathrm{III} \equiv 32$. For a vanishing delay $\tau=0$, $\mathbf{e}_{\alpha}\cdot \mathbf{e}_{\alpha} = 1$ and the scalar products $\mathbf{e}_{\alpha}\cdot \mathbf{e}_{\beta}$ describe the angles of the triangle formed by the three particles (see Fig.~\ref{fig:modelsketch} in Sec.~\ref{Sec.:Dynamics}). We find that up to the leading order in the equilibrium bond length $R$ the triangle is equilateral and thus the internal angles are $\pi/3$, leading to the relations $\mathbf{e}_{\mathrm{I}}\cdot \mathbf{e}_{\mathrm{II}} = \mathbf{e}_{\mathrm{I}}\cdot \mathbf{e}_{\mathrm{III}} = - \mathbf{e}_{\mathrm{II}}\cdot \mathbf{e}_{\mathrm{III}} \approx 1/2  + O[(r_\alpha-r_\beta)/R]$ with a correction that is on the order of $(r_\mathrm{I}-r_\mathrm{II})/R$ for $\mathbf{e}_{\mathrm{I}}\cdot \mathbf{e}_{\mathrm{II}}$ and similarly for the other scalar products. The linearized equation for the relative coordinate $x_\alpha \equiv r_\alpha - R$ thus reads
\begin{align}
	\dot{x}_{\alpha}(t) = -\omega \bigg[ x_{\alpha}(t-\tau) + \frac{1}{4} x_{\alpha+\mathrm{I}}(t- \tau) + \frac{1}{4} x_{\alpha+\mathrm{II}}(t-\tau) \bigg]&   +\sqrt{2D} \left( \sum_{i=1}^3 \mathcal{A}_{\alpha i} \bm{\eta}_i(t) \right)   \cdot {\mathbf e}_\alpha(t),  \nonumber \\
	\mathcal{A}=\begin{pmatrix}
	1 & -1 & 0\\
	1 & 0 & -1\\
	0 & -1 & 1\\
	\end{pmatrix},&  \label{coupled LE}
\end{align}
where the lower index $\alpha \equiv \alpha \mod \mathrm{III}$ is considered as periodic with the period 3, i.e., $x_\mathrm{IV} \equiv x_\mathrm{I}$ and $x_\mathrm{V} \equiv x_\mathrm{II}$. Similarly as in the case of the dimer, Eq.~\eqref{coupled LE} describes the dynamics of $x_\alpha(t)$ well for large equilibrium bond lengths $R$ and for time delays small compared to reorientation times of the unit vectors $\mathbf{e}_{\alpha}$.

For an analytical treatment, it is advantageous to rewrite the system \eqref{coupled LE} in the matrix form
\begin{equation}
	\dot{\mathbf{x}}(t) = -\omega M\mathbf{x}(t-\tau) + \sqrt{2D} \bm{\xi}(t)  ,  \label{Trimer linear equ}
\end{equation}
for the three-dimensional column vector $\mathbf{x}(t) = \left[x_\mathrm{I}(t),x_\mathrm{II}(t),x_\mathrm{III}(t)\right]^\intercal$. In Eq.~\eqref{Trimer linear equ}, the noise vector $\bm{\xi}(t)$ is given by $\bm{\xi}(t) \equiv A^1 \bm{\eta}^1(t) + A^2\bm{\eta}^2(t)$ with the auxiliary noise vectors $ \bm{\eta}^j(t) \equiv [\eta_1^j,\eta_2^j,\eta_3^j]^\intercal$, $j=1,2$, containing the $j$th components of the original noises $\bm{\eta}_i(t)$, $i=1,\dots,3$. From the system~\eqref{coupled LE} follows that the matrices $M$, $A^1$ and $A^2$ read
\begin{equation}
M=\frac{1}{4}\begin{pmatrix}
	4 & 1 & 1\\
	1 & 4 & 1\\
	1 & 1 & 4\\
	\end{pmatrix},\,\,
	A^1=\begin{pmatrix}
	e_{\mathrm{I}}^1 & - e_{\mathrm{I}}^1 & 0\\
	e_{\mathrm{II}}^1 & 0 & -e_{\mathrm{II}}^1\\
	0 & -e_{\mathrm{III}}^1 & e_{\mathrm{III}}^1\\
	\end{pmatrix},\,\,
	A^2=\begin{pmatrix}
	e_{\mathrm{I}}^2 & - e_{\mathrm{I}}^2 & 0\\
	e_{\mathrm{II}}^2 & 0 & -e_{\mathrm{II}}^2\\
	0 & -e_{\mathrm{III}}^2 & e_{\mathrm{III}}^2\\
	\end{pmatrix},
\label{eq:matrices3D}
\end{equation}
where $e_{\alpha}^j$ denote $j$th component of the two-dimensional unit vector $\mathbf{e}_\alpha$. The time-correlations between the three components of the noise vector $\bm{\xi}(t)$ are not mutually independent and read
\begin{align}
	\left\langle \xi_\alpha(t)\xi_\beta(t') \right\rangle = \left( A^{1}A^{1\intercal} + A^{2}A^{2\intercal} \right)_{\alpha\beta} \delta_\mathrm{D}(t-t') = 2M_{\alpha\beta} \delta_\mathrm{D}(t-t').  \label{trimer noise corr} 
\end{align}

Due to the linearity of Eq.~\eqref{Trimer linear equ} and Gaussianity of the noise, the Green's function for the one-time PDF $P_1(\mathbf{r},t|\mathbf{r}_0,0)$ is Gaussian \cite{fox1977generalized}, and determined by the mean value $\bm{\mu}(t) = \left<\mathbf{x}(t)\right>$ and the covariance matrix $\mathcal{K}(t)=\left<\mathbf{x}(t) \mathbf{x}^\intercal(t)\right> - \bm{\mu}(t)\left[\bm{\mu}(t)\right]^\intercal$. We review in detail the derivation of these functions in \ref{App:A1} and \ref{app:noise_sol3}.

For the initial condition $\mathbf{x}(t) = 0$, $t<0$ and $\mathbf{x}(0) = \mathbf{x}_0$ we get
\begin{align}
\bm{\mu}(t) &= \lambda(t) \mathbf{x}_0,
\label{eq:mean_trimer}\\
\mathcal{K}(t) &= 4D M \int_0^t ds \lambda^2(s),
\label{eq:Ctrimer}
\end{align}
where $\lambda(t)$ denotes the Green's function for Eq.~\eqref{Trimer linear equ} given by
\begin{equation}
	\lambda(t) = \sum_{k=0}^{\infty} \frac{(-\omega M)^k}{k!} (t-k\tau)^k \theta(t-k\tau).  \label{Trimer Fundamental}
\end{equation}

Multiplying $\lambda(t)$ by the vector $[1,1,1]^\intercal$, using the formula $M [1,1,1]^\intercal = 3/2 [1,1,1]^\intercal$
and comparing the result to the one-dimensional Green's function~\eqref{eq:fundamental_sol_MT}, we find that $\lambda(t)$ undergoes with increasing $t$ (i) monotonous exponential decay to $0$ for $0 < 3\omega\tau/2 \le 1/e$, (ii) oscillatory exponential decay to $0$ for $1/e < 3\omega\tau/2 < \pi/2$ and (iii) oscillatory exponential divergence for $\pi/2 < 3\omega\tau/2$. In the regimes (i) and (ii) the stationary value of the covariance matrix reads
\begin{equation}
\mathcal{K}_{ss} = \lim_{t\to\infty} \mathcal{K}(t) = \frac{2D}{\omega}\frac{\mathcal{I}+\sin\left(\omega \tau M\right)}{\cos\left(\omega \tau M\right)},
\label{eq:CtrimerSS}
\end{equation}
where $\mathcal{I}$ denotes the identity matrix. This formula follows from the results of \ref{app:time_corr_matr} after substituting the matrices $\omega$ and $\sigma\sigma^\intercal$ from the formula~\eqref{eq:C_0_K_inf_scalar} in the appendix by $\omega M$ and $4D M$. In the appendix, we also derive an analytical expression for the stationary time correlation matrix $C(t) = \lim_{s\to \infty}\left<\mathbf{x}(s)\mathbf{x}^\intercal(s+t)\right>$. The regime of stability $3\omega\tau/2 < \pi/2$ of the trimer can also be determined from the condition that the matrix $\cos\left(\omega\tau M\right)$ is not singular, i.e., its determinant $\cos^2\left(3 \omega\tau/2\right)\cos\left(3 \omega\tau\right)$ is nonzero.

\begin{figure}[t!]
\centering
	\begin{tikzpicture}
	\node (img1)  {\includegraphics[width=0.4\columnwidth]{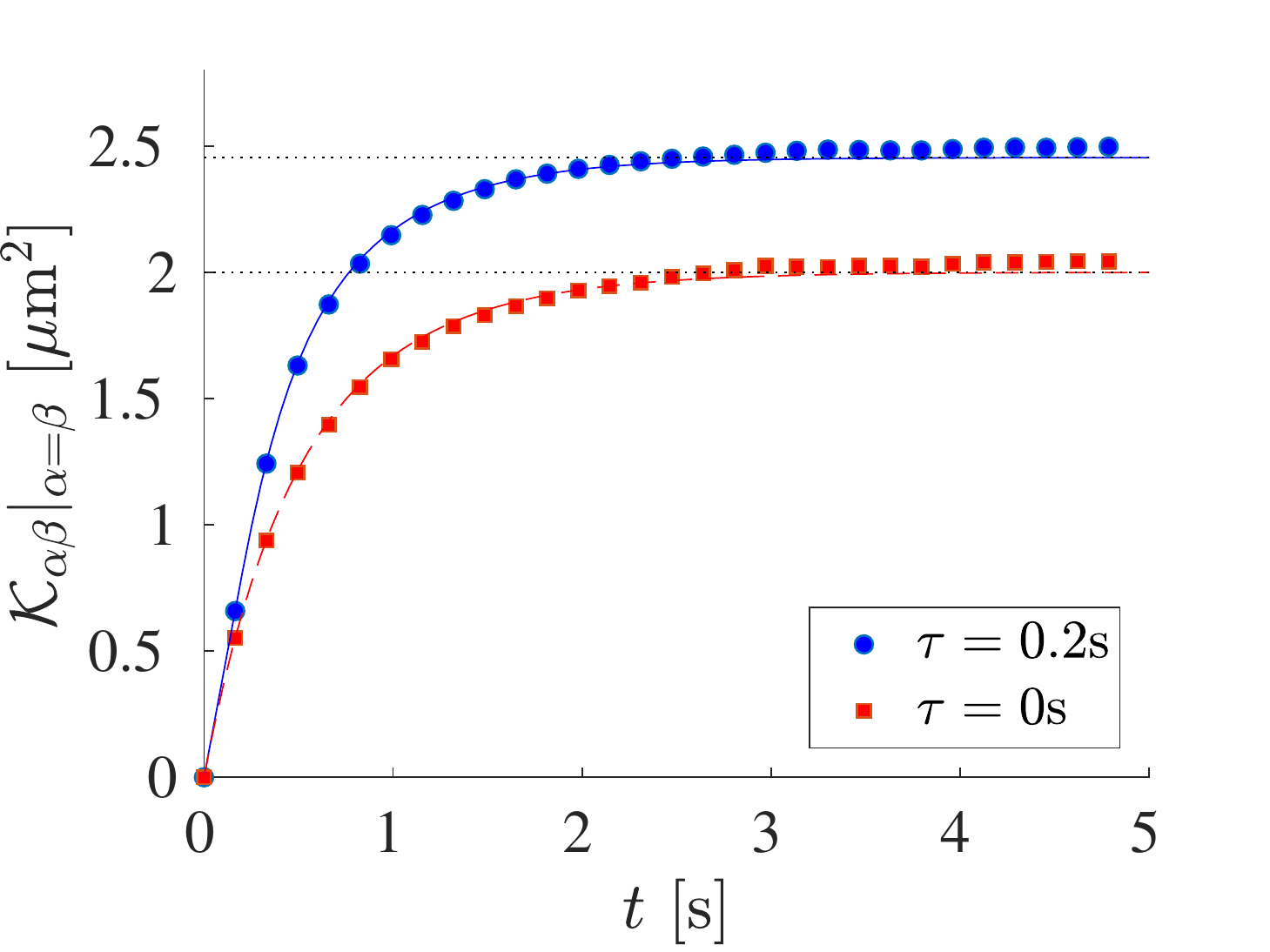}};
	\node[right=of img1, node distance=0.0cm, yshift=0cm,xshift=-1.0cm] (img2)
	{\includegraphics[width=0.4\columnwidth]{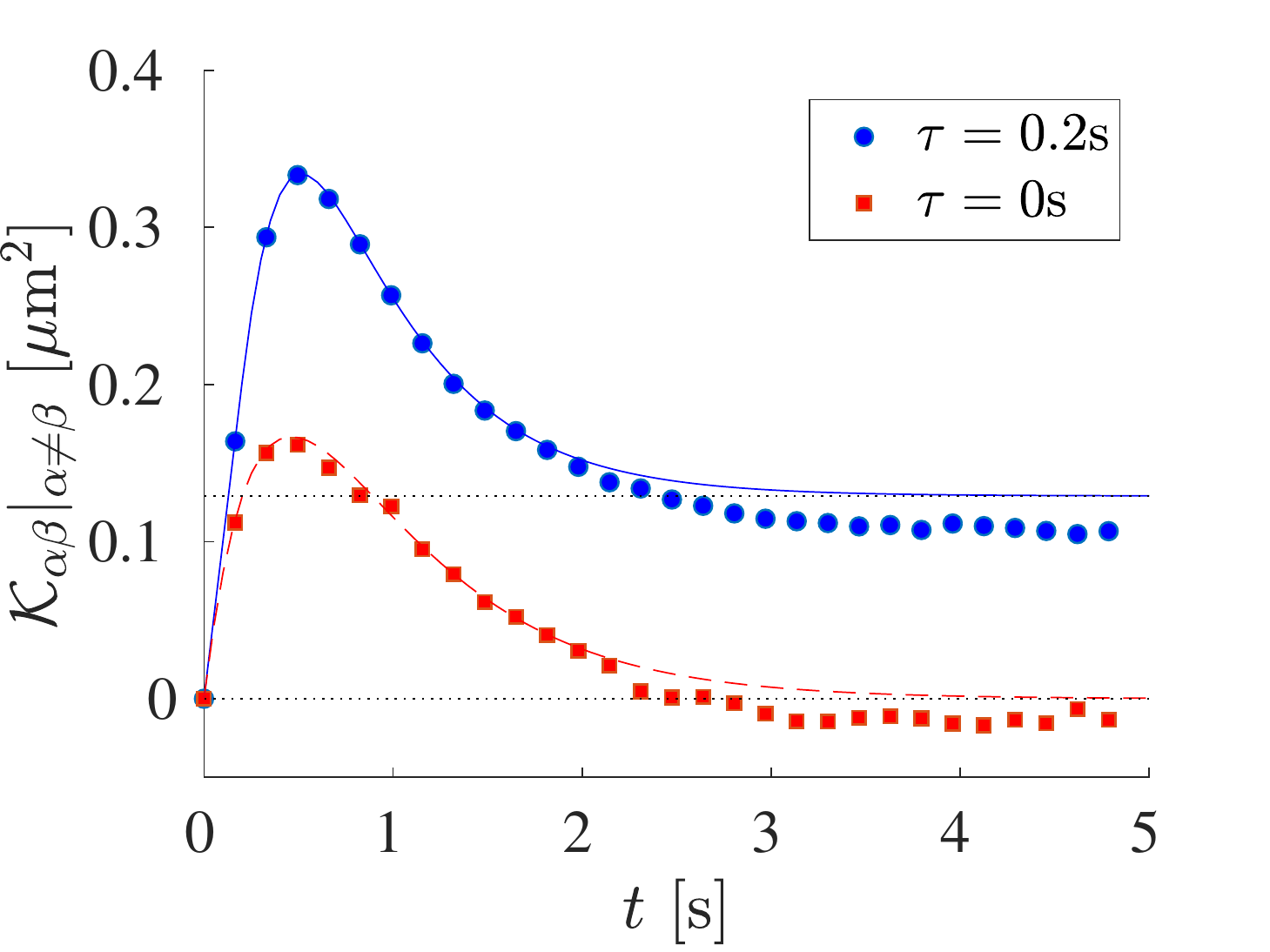}};
	\node[above=of img1, node distance=0.0cm, yshift=-1.8cm,xshift=-3.0cm] {a)};
	\node[above=of img2, node distance=0.0cm, yshift=-1.8cm,xshift=-3.0cm] {b)};	
	\end{tikzpicture}	
	\caption{a) Diagonal and b) off-diagonal elements of the covariance matrix $\mathcal{K}(t)$ as functions of time for two values $0$ s (dashed red lines and squares) and $0.2$ s (full blue lines and circles) of the delay $\tau$. The full and dashed lines are calculated from the approximate analytical formula~\eqref{eq:Ctrimer} and the symbols come from a BD simulation of the complete model. The horizontal dotted lines depict the elements of the stationary covariance matrix $\mathcal{K}_{ss}$ given by Eq.~\eqref{eq:CtrimerSS}. Parameters used: $\omega=$ 1/s, $D=1$ $\mu$m$^2$/s, $r_{ij}(0)=12$ $\mu$m, and $R=10$ $\mu$m. In the BD simulation, we averaged over $10^5$ trajectories with time step $dt=10^{-3}$ s.}	
	\label{fig:correlation}	
\end{figure}

Due to the symmetry of the problem, all diagonal elements of the matrix $\mathcal{K}_{ss}$ are identical and the same holds also for all its off-diagonal elements. The approximate analytical, time-dependent solution \eqref{eq:Ctrimer} for the covariance matrix is compared to the exact covariance matrix obtained by BD simulations of the complete model in Fig. \ref{fig:correlation}. Given the approximations made, we find very good agreement. The analytical results only slightly underestimate the diagonal elements (probably for the same reason as for the dimer) and overestimate the off-diagonal elements. The behavior of the covariance matrix as a function of the frequency $\omega$ and delay $\tau$ is similar to the behavior of the variance \eqref{eq:SS_variance} for the dimer. Specifically, the diagonal elements of $\mathcal{K}_{ss}$ monotonicly increase (the PDF for the bond lengths become broader) with $\tau$ and exhibit a minimum as functions of $\omega$, opening a possibility to optimize the width of the bond length PDF. The off-diagonal elements of $\mathcal{K}_{ss}$ monotonously increase
 (the individual bonds of the trimer become more correlated) both with the delay and with the natural relaxation frequency. 

%% file: Content/StructureAndEntropy.tex
\section{Structure Formation}\label{sec:structure}
The approximate analytical study of the dimer and trimer revealed that both systems obey three dynamical regimes: (i) and (ii) a monotonous and an oscillatory exponential relaxation towards a steady state with the average bond length $\mu(\infty) = R$, respectively, and (iii) an exponential divergence. The performed BD simulations confirmed that for large $R$, when the model is well described by the approximate analytical formulas, these regimes can indeed be observed also in the complete model~\eqref{eq:Langevin_n}. Furthermore, the analytical study predicted that the dimer is in the unstable regime (iii) for $\omega\tau>\pi/2$ and the trimer for $\omega\tau>\pi/3$. Let us now discuss how general the presented findings are.

\begin{figure}[t!]
\centering
	\begin{tikzpicture}
	\node (img1)  {\includegraphics[width=0.8\linewidth]{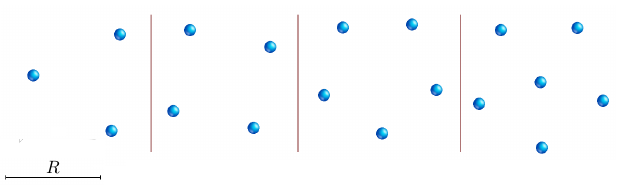}};
	\node[below=of img1, node distance=0.0cm, yshift=1cm,xshift=-3.0cm] (img2)  {\includegraphics[width=0.4\columnwidth]{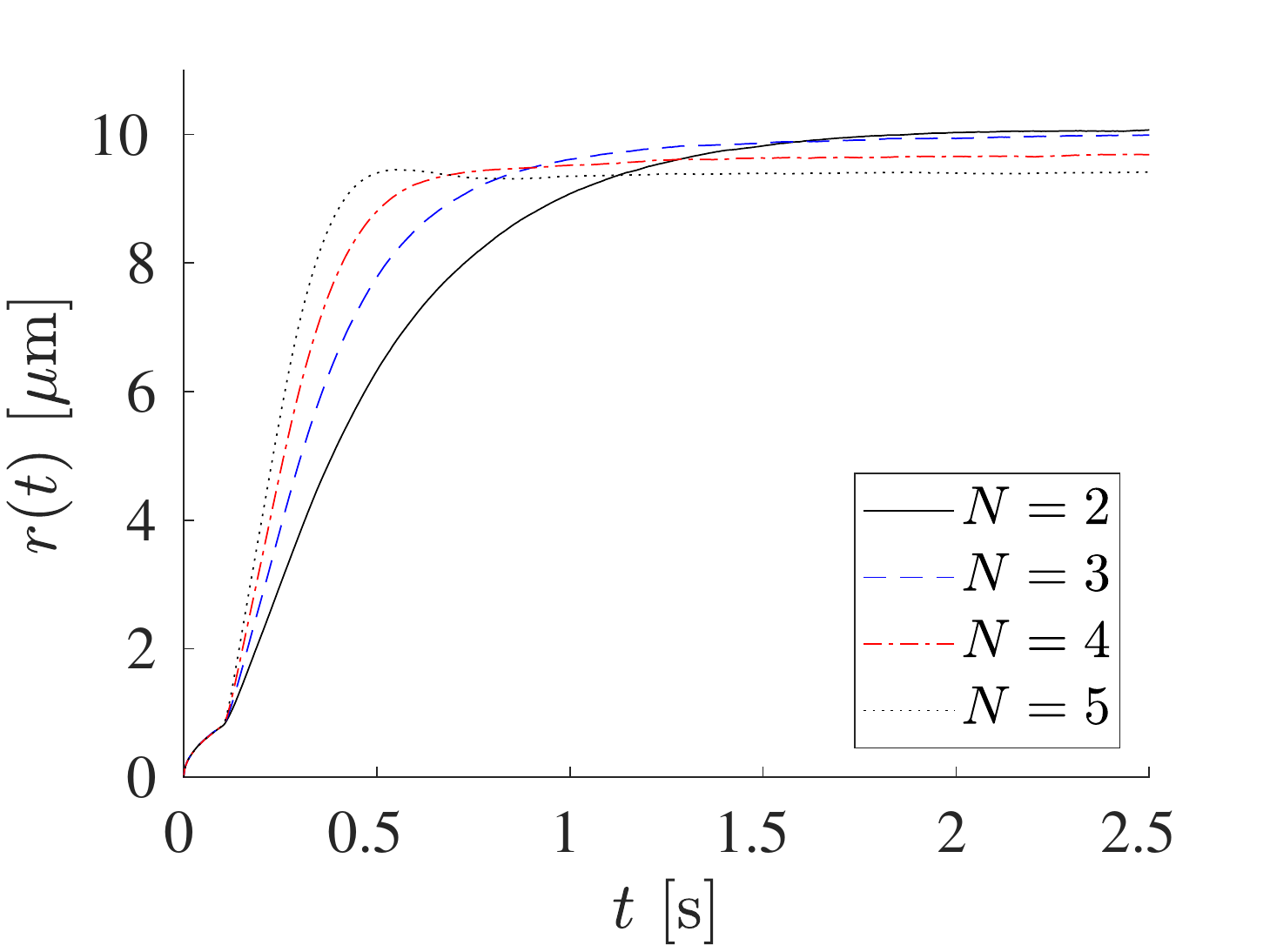}};
	\node[right=of img2, node distance=0.0cm, yshift=0cm,xshift=-1.0cm] (img3)
	{\includegraphics[width=0.4\columnwidth]{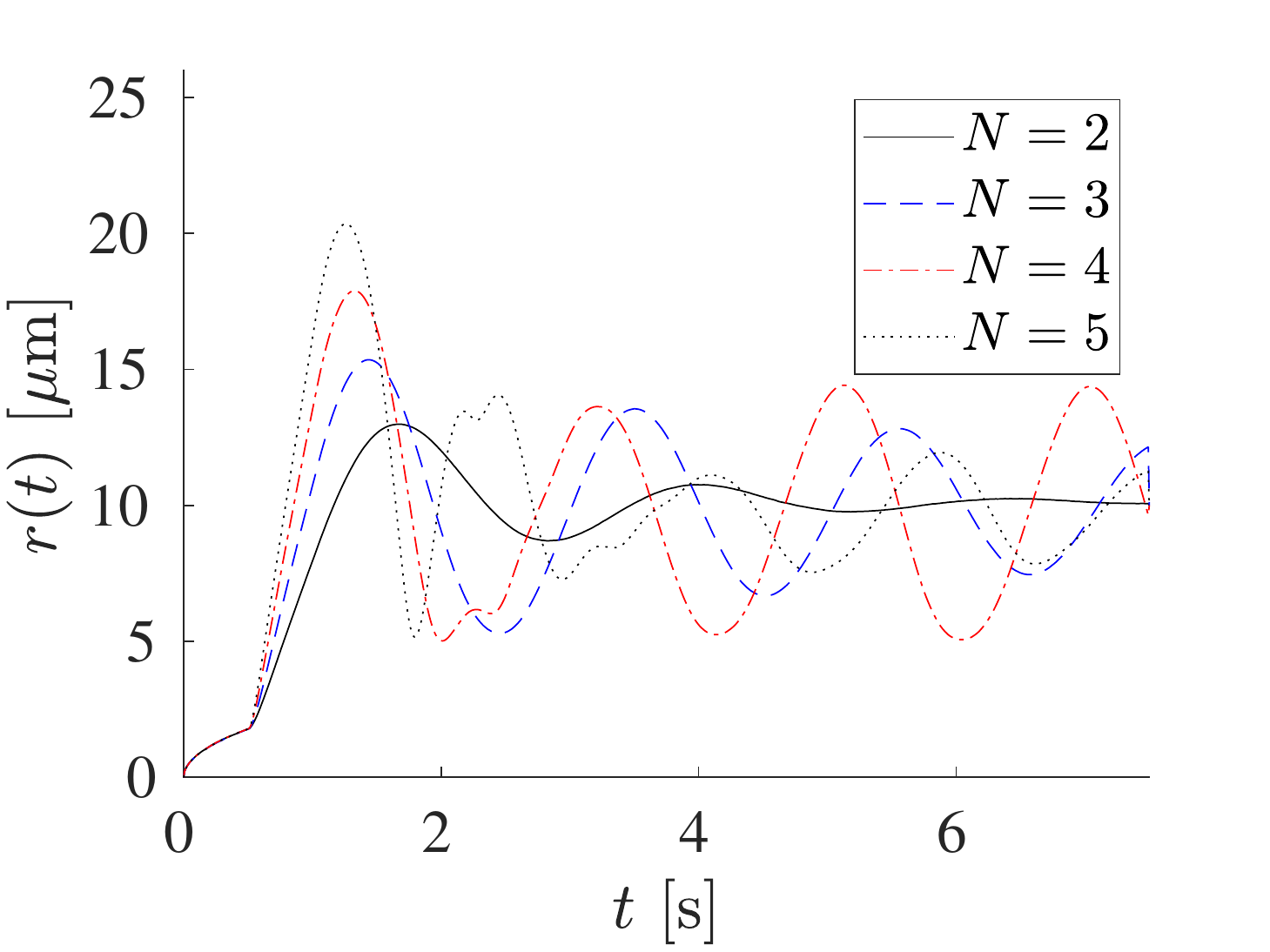}};
	\node[above=of img1, node distance=0.0cm, yshift=-1.8cm,xshift=-6.0cm] {a)};
	\node[above=of img2, node distance=0.0cm, yshift=-1.8cm,xshift=-3.0cm] {b)};
	\node[above=of img3, node distance=0.0cm, yshift=-1.8cm,xshift=-3.0cm] {c)};	
	\end{tikzpicture}	
	\caption{\textbf{Bound 'molecules' -- ground state and relaxation} In panel a) we show highly symmetric molecule-like structures formed in our model for a vanishing noise ($D=0$) and a small product $\omega\tau$ of the spring stiffness $\omega$ and the delay $\tau$. Due to the infinite range of the harmonic two-particle interaction, the inter-particle distances in the molecules decrease with increasing number of particles. For $D\neq 0$, the structures vibrate erratically due to the noise and may oscillate due to the delay. In panel b) we plotted the BD results average bond length $\mu(t)$ as a function of time for several values of $N$ for $\tau = 0.1$, where the oscillations do not arise [regime (i)]. The oscillations observed in panel c) with $\tau=0.5$ are only transient for $N \le 3$ [regime (ii)], and grow indefinitely for $N > 3$ [regime (iii)]. Other parameters used: $\alpha=$ $1/$s, $R=10$ $\mu$m, and $D=1$ $\mu$m$^2/$s. In the simulation, we averaged over $10^4$ stochastic trajectories with time step $dt=10^{-3}$ s.}	
	\label{fig:CompareBondNumber}	
\end{figure}
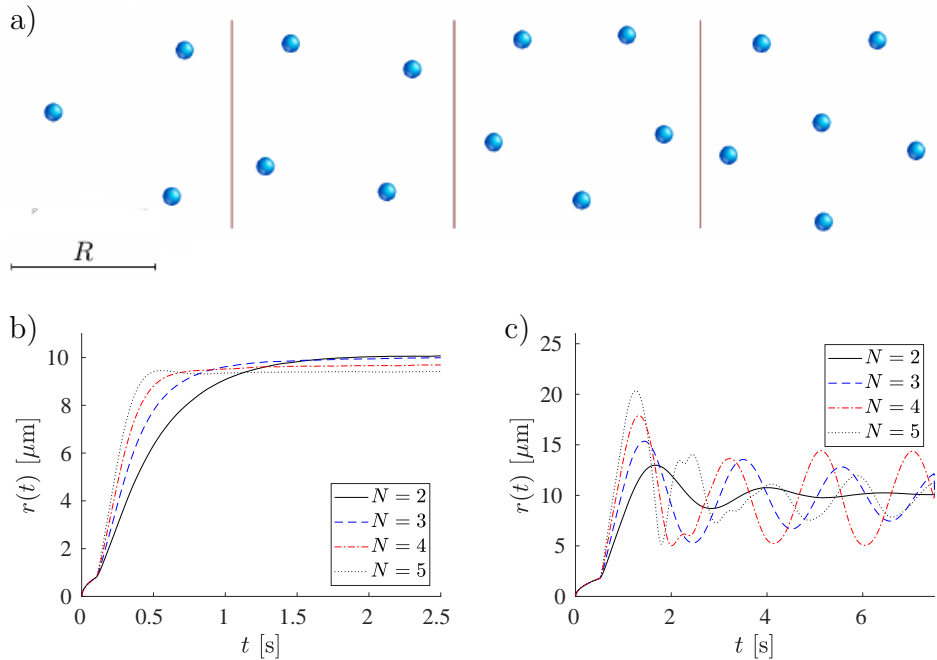

The stationary average bond length $\mu$ can be determined by minimizing the potential energy $V = \frac{1}{2}\sum_{(i,j)} V_2(r_{ij})$ with the two-particle potential $V_2$ given by Eq.~\eqref{eq:two_particle_V}. Minimizing the potential in our two-dimensional geometry yields the highly symmetric molecule-like structures shown in Fig.~\ref{fig:CompareBondNumber} a). Due to the confinement to 2d, the global minimum of the potential corresponding to $\mu = R$ is accessible only for the dimer ($N=2$) and the trimer ($N=3$). For larger molecules, the average bond length decreases as a result of the infinite range of the potential. The system asymptotically relaxes to the depicted structures if the noise $D$ vanishes and the reduced delay time $\omega \tau$ is small enough such that the system is in the dynamical regime (i) or (ii). Nonzero $D$ leads to fluctuations around the asymptotic structures and large $\omega \tau$ causes exponentially diverging oscillations.

We have solved the complete model using BD simulations and depict the behavior of the average bond length $r(t)$ for several values of $N$ in the dynamical regimes (i) and (ii)-(iii) in Fig.~\ref{fig:CompareBondNumber} b) and c), respectively. In the regime (i), we observe that larger systems relax faster than those with smaller $N$. Furthermore, in Fig.~\ref{fig:CompareBondNumber} c), we see that larger systems oscillate with larger amplitudes and that the threshold between the regimes (ii) and (iii) is reached at smaller values of $\omega\tau$. More precisely, all the curves in Fig.~\ref{fig:CompareBondNumber} c) are plotted using the same parameters except for $N$ and, while the curves for $N\le 3$ are in the regime (ii), the curves for $N>3$ correspond to the regime (iii). These observations are in accord with our analytical findings for the dimer and trimer.

\begin{figure}
	\centering
	\includegraphics[width=0.5\linewidth]{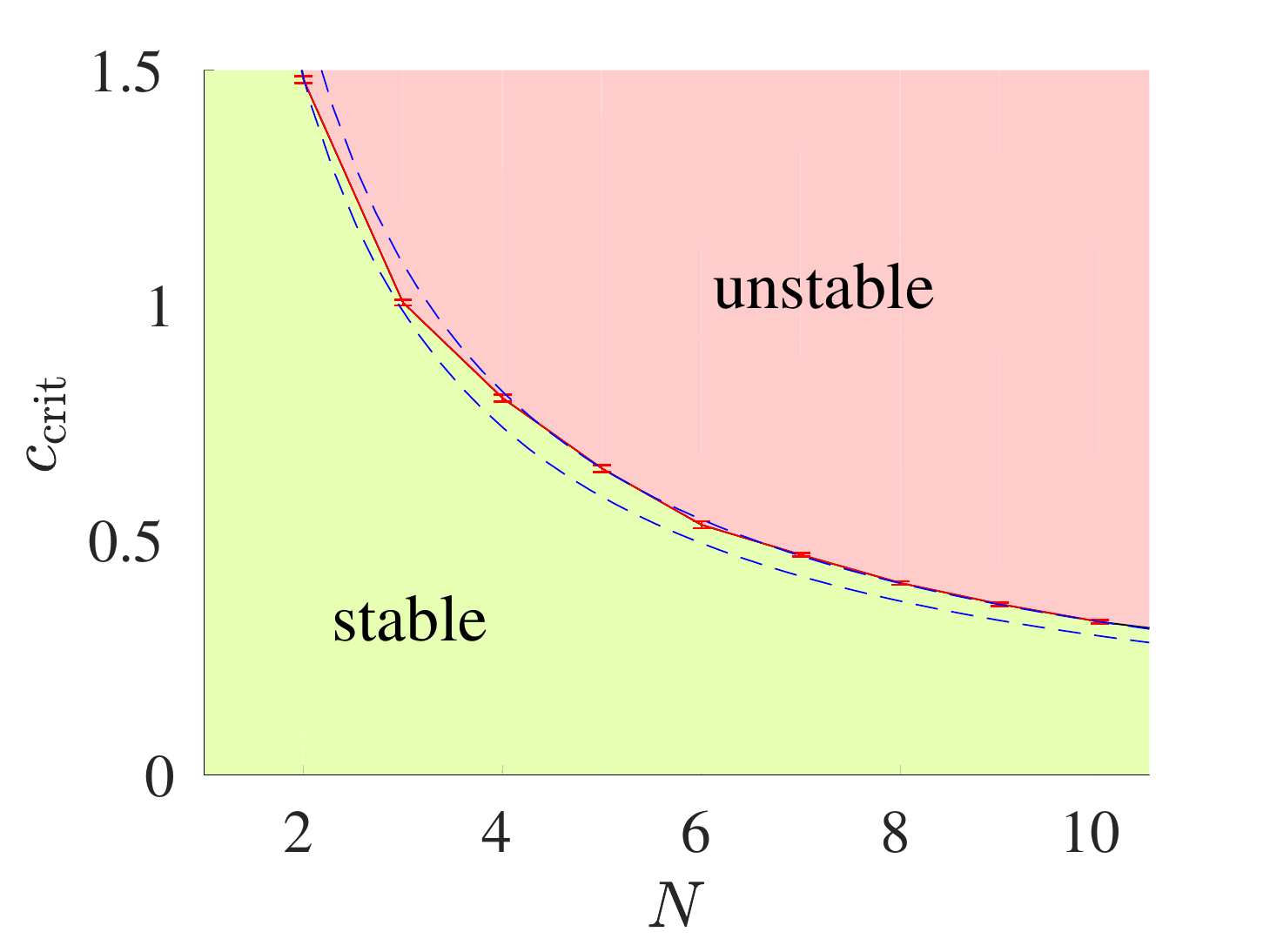}
	\caption{Stability factor~\eqref{eq:stability_factor} delimiting the threshold between the stable dynamical regimes (i-ii) and the unstable dynamical regime (iii) for systems consisting of $N=2,...,10$ particles obtained from BD simulation of the complete model~\eqref{eq:Langevin_n} (solid broken line). The stability factor is well approximated by the function $C/N$ (dashed lines). For the upper (lower) line we choose the constant $C$ such that the curve crosses the last (first) point obtained from the simulation.}
	\label{fig:crit delay}
\end{figure}

By analyzing the mean bond length at late times, we have evaluated the \textit{critical reduced delay} $(\omega \tau)^\star$ determining the threshold between the regimes (ii) and (iii) for $N=2,...,10$. In Fig. \ref{fig:crit delay}, we show its rescaled value 
\begin{equation}
c_\mathrm{crit}\equiv\frac{2}{\pi}(\omega \tau)^\star,
\label{eq:stability_factor}
\end{equation}
where the coefficient $2/\pi$ is introduced for the comparison to the approximate result for the dimer. We find that the stability factor is well described by $C/N$ as suggested by the approximate analytical results for the dimer [$(\omega \tau)^\star = \pi/2$] and trimer [$(\omega \tau)^\star = \pi/3$]. However, the analytical results would imply that the constant $C$ equals to 2, which is smaller than the value $C\approx 3$ obtained from the complete model. The actual dimer and trimer are thus more stable than their linearized versions considered in our analytical study. The extensive scaling $c_\mathrm{crit} \propto 1/N$ of the stability of the system could (almost) completely be removed by rescaling the potential stiffness according to $k \rightarrow k/N$ as it is often done in thermodynamical system with long-range interactions \cite{bouchet2010thermodynamics}.

To better understand this behavior, let us first consider the approximate analytical model described in Sec.~\ref{Section: Dimer}. Imagine a particle in a harmonic potential centered around $x=0$, which is initially located at $x(0)=x_0>0$. Assuming that $x(t)=0$ for $t<0$ and neglecting the noise, the particle does not feel any force in the time interval $t\in[0,\tau]$ such that $x(t)=x_0$ for all $t\in[0,\tau]$. In the subsequent time interval $t\in[\tau,2\tau]$, the particle experiences the force $F =-\omega_\tau x_0$ pushing it towards the opposite wall of the trap. For times $t>2\tau$ the force starts changing dynamically according to the earlier position at time $t-\tau$. The particle keeps its direction of motion until it reaches the position $x_1$ where the force changes its sign. For large delays, the particle may stop significantly later than crossing the minimum, so that $x_1<0$ and $|x_1|>|x_0|$. A similar process then repeats when the particle returns back, with the difference that now it reaches a maximum position $x_2>0$, $|x_2|>|x_1|>|x_0|$, etc. The amplitude thus increases after each half-period of oscillation causing a diverging behavior. 
  
In order to understand the difference between the approximate analytical and the complete (numerical) solutions of the model, it is helpful to project the latter to one dimension, where a particle moves in the double-well potential depicted Fig.~\ref{fig:PotentialPlot} in Sec.~\ref{sec:dimer_rate}. We assume that the particle starts in the right well and oscillates with increasing amplitude as discussed above. After some time the amplitude becomes large enough that the particle crosses the barrier to the left well. Due to the presence of the additional well, the potential 
now contains much wider low-energy region
compared to the purely harmonic case. The particle needs longer time to travel from one (unbounded) side of the potential to its other side, and hence also the (reduced) delay $\omega\tau$ required for inducing diverging oscillations is larger than in the harmonic case.
As a consequence, the complete model is seen to be more stable than foreseen by our analytical considerations. Moreover, our discussion reveals the existence of a fourth dynamical regime, preceding the unstable regime (iii), where the particle hops between the individual wells of the potential and the amplitude of the oscillations remains finite.

Compared to the quasi-constant velocity model investigated in Ref.~\cite{khadka2018active}, our analysis thus reveals two qualitative differences. First, the structures formed in the quasi-constant velocity model oscillate for arbitrary nonzero delay $\tau$, while in the harmonic model these oscillations appear only if $\omega \tau$ is large enough. Second, the amplitude of the oscillations in the quasi-constant velocity model is always constant in time, while the osculations in the harmonic model either vanish with time, if the system is in the regime (ii), or explode with time, if the system is in the regime (iii). The behavior observed in the harmonic model can be traced back to the increase of the force with the particle distance and thus we expect an analogous behavior also for other systems with time-delayed forces increasing with distance.

\section{Entropy Fluxes}
\label{sec:entropy_fluxes}

The investigated model, much as the model discussed in Ref.~\cite{khadka2018active}, is inspired by self-organized systems, where a feedback based on the information about the state of the system at a previous time leads to structure formation.
Interpreting the delayed interactions in our model as a result of such feedback control, we can investigate the entropy flow out of the system caused by the feedback. Due to the non-analyticity of the model with quasi-constant forces considered in Ref.~\cite{khadka2018active}, the analysis of entropy flows in the supplementary material therein was performed for vanishing delay only. Using the approximate Gaussian PDFs found in Secs.~\ref{Section: Dimer} and \ref{Section Trimer} for the dimer and the trimer, respectively, we can repeat this analysis with nonzero delay.

Without feedback, i.e., without the time-delayed harmonic interactions \eqref{eq:total_energy}, the particles would spread diffusively and the system entropy would increase accordingly. The feedback control utilizes the information about the particle positions to drive the system into a non-equilibrium steady state with a time independent PDF $P({\bf x})$ and thus with a time constant configurational entropy $\mathcal{\mathcal S} = - k_{\rm B} \int {\rm d}\mathbf{x} P(\mathbf{x}) \log P(\mathbf{x})$. The smaller the entropy of the non-equilibrium steady state, the more localized the steady-state structure and thus the better the result of the feedback control. Another measure of the performance of the feedback is the rate $-\dot{\mathcal S}_-$ of entropy taken \emph{from} the system per unit time that can also be interpreted as the amount of information pumped \emph{into} the system per unit time by the feedback device. This entropy flow balances the diffusive spreading in the steady state and is thus moreover a measure of the useful ``work'' (in units of J/K) performed by the feedback device against thermal dispersal. Evaluating the stationary entropy production $\dot{\mathcal S}_{\rm F}$ due to the feedback control mechanism and the stationary entropy production $\dot{\mathcal S}_{\rm D}$ due to the breaking of the fluctuation-dissipation theorem in Eqs.~\eqref{eq:Langevin_n} and \eqref{noise} for $\tau>0$, one can define the feedback efficiency as the ratio $\eta_{\rm F} = - \dot{\mathcal S}_-/(\dot{\mathcal S}_{\rm F} + \dot{\mathcal S}_{\rm D})$. The entropy production $\dot{\mathcal S}_{\rm D}$ can be calculated along the lines of Ref.~\cite{loos2019heat}. The entropy production $\dot{\mathcal S}_{\rm F} = q_{\rm H}/T$ is the housekeeping heat flux $q_{\rm H}$ flowing to the bath at temperature $T$, due to the overall operation of the feedback device, divided by the bath temperature. It clearly depends on the specific technical realization of the feedback. In all known relevant realizations of the feedback in microscopic systems \cite{khadka2018active,bauerle2018quorum,masoller2002feedback, Baraban2013,Qian2013}, the housekeeping heat flux is very large compared to the ``functional'' energy fluxes in the controlled system, resulting in a large $\dot{\mathcal S}_{\rm F}$ compared to $- \dot{\mathcal S}_-$, so that the efficiency $\eta_{\rm F}$ of such devices is usually negligibly small.

To evaluate the entropy flow due to the feedback (the time-delayed harmonic interaction) in the present setup, we proceed along the similar lines as in Refs.~\cite{khadka2018active,fuchs2016thermodynamics}. The center of mass coordinate of the system is not affected by the feedback and diffuses freely (see Sec.~\ref{sec:CMS}). The structure formation due to the feedback thus occurs only on the level of the bonds. Let us now consider the time-dependent PDF $P(\mathbf{x},t)$ for the bonds that converges to a time-independent non-equilibrium steady state due to the competition between feedback and diffusion. The rate of change of its Shannon entropy  $\mathcal{S}(t) = - k_{\rm B} \int {\rm d}\mathbf{x} P(\mathbf{x},t) \log P(\mathbf{x},t)$ can formally be written as
\begin{align}
\dot{\mathcal S}(t) = \dot{\mathcal S}_+(t) + \dot{\mathcal S}_-(t),
\label{eq:entropy_balance}
\end{align}
where $\dot{\mathcal S}_+(t)$ stands for the positive entropy flowing into the system due to the diffusive spreading of the particles and $\dot{\mathcal S}_-(t)$ corresponds to the outflow of entropy due to the feedback.

Assuming that the stochastic dynamics of the column vector $\mathbf{x}(t)$ describing the bonds obeys the generalized Langevin equation $\dot{\mathbf{x}}(t) = \mathbf{F}[\mathbf{x}(t), \mathbf{x}(t-\tau)] + \sigma \eta(t)$, where $\eta$ denotes a zero mean Gaussian white noise with the covariance matrix $\left<\eta_i(t)\eta_j(t')\right> = \delta_{ij}\delta(t-t')$, the dynamical equation for $P(\mathbf{x},t)$ can be written in the form \cite{guillouzic1999small,Frank2016} 
\begin{equation}
\frac{\partial}{\partial t}{P}(\mathbf{x},t) = \frac{1}{2}\sum_{ij}\left(\sigma \sigma^\intercal\right)_{ij}\frac{\partial^2}{\partial x_i \partial x_j} P(\mathbf{x},t) + \mathcal{L}[\mathbf{x},t].
\label{eq:dPEQ}
\end{equation}
In this equation, the fist term on the right stands for the diffusive spreading of the PDF. The term $\mathcal{L}[\mathbf{x},t]$ corresponds to the time-delayed force $\mathbf{F}[\mathbf{x}(t-\tau)]$ in the Langevin equation and thus it describes the effect of the feedback. Its concrete form is not relevant for the discussion below and thus we refer to the works \cite{guillouzic1999small,Frank2016} for more details about its structure.

Inserting Eq.~\eqref{eq:dPEQ} into the formal time derivative $\dot{\mathcal S}(t) = - k_{\rm B} \int {\rm d}\mathbf{x} [\partial_t P(\mathbf{x},t)] \log P(\mathbf{x},t)$ of the system entropy ${\mathcal S}(t)$, we find that 
\begin{align}
\dot{\mathcal S}_+(t) &= - \frac{k_{\rm B}}{2} \int \mathrm{d}\mathbf{x} \sum_{ij}\left(\sigma\sigma^\intercal\right)_{ij}\left(\frac{\partial^2}{\partial x_i \partial x_j} P\right) \log P =  - k_{\rm B} \int \mathrm{d}\mathbf{x}
(\nabla P)^\intercal \frac{\sigma\sigma^\intercal}{P}\nabla P,
\label{eq:Sp}\\
\dot{\mathcal S}_-(t) &=  - k_{\rm B} \int \mathrm{d}\mathbf{x} \mathcal{L}[\mathbf{x},t] \log P(\mathbf{x},t) = \dot{\mathcal S}(t) - \dot{\mathcal S}_+(t).
\label{eq:Sm}
\end{align}
The last equation allows us to calculate the amount of entropy taken out of the system due to the feedback per unit time, $\dot{\mathcal S}_-(t)$, from the PDF $P({\bf x},t)$ without knowing the explicit form of the operator $\mathcal{L}$. It is interesting to adopt the Seifert's idea of trajectory-dependent entropy~\cite{seifert2012thermodyn,Seifert2005} and use Eq.~\eqref{eq:Sm} to define the stochastic (position dependent) entropy flux 
\begin{equation}
\dot{s}_-(\mathbf{x},t) = k_{\rm B} \left[(\nabla P)^\intercal \frac{\sigma\sigma^\intercal}{P^2}\nabla P - \frac{1}{2}\partial_t (\log P)^2\right].
\label{eq:smflux}
\end{equation}
The average flux \eqref{eq:Sm} then follows as the average $\dot{\mathcal S}_-(t) = \left< \dot{s}_-(\mathbf{x},t)\right>$ either over the PDF $P(\mathbf{x},t)$ or over the individual stochastic trajectories generated in a BD simulation. To the best of our knowledge, the statistics of the entropy flux \eqref{eq:smflux} has not been investigated yet and thus it is not known whether its PDF fulfills some fluctuation symmetries. Such an investigation would clearly be beyond the scope of the present paper and we leave it for a future work.

Let us now evaluate the three entropy fluxes \eqref{eq:entropy_balance}, \eqref{eq:Sp} and \eqref{eq:Sm} for a general $d$-dimensional Gaussian PDF 
\begin{equation}
P(\mathbf{x},t) = \frac{1}{\sqrt{(2\pi)^d|{\mathcal K}(t)|}} \exp\left[- \frac{1}{2} \left(\mathbf{x}-\mu(t)\right)^\intercal {\mathcal K}^{-1}(t)\left(\mathbf{x}-\mu(t)\right)\right],
\label{eq:Gaussian_PDF}
\end{equation}
where $|{\mathcal K}(t)|$ denotes determinant of the covariance matrix ${\mathcal K}(t)$. The corresponding entropy ${\mathcal S}(t)$ reads 
\begin{equation}
\frac{{\mathcal S}(t)}{k_{\rm B}} = \frac{d}{2} + \frac{1}{2}\log \left[(2\pi)^d |{\mathcal K}(t)| \right]
\label{eq:system_entropy}
\end{equation}
leading to the rate of change
\begin{equation}
\frac{\dot{\mathcal S}(t)}{k_{\rm B}} = \frac{1}{2}\frac{\mathrm{d}}{\mathrm{d}t}\log |{\mathcal K}(t)|.
\label{eq:SdotGaussian}
\end{equation}
From Eq.~\eqref{eq:Sp} we then find that
\begin{equation}
\frac{\dot{\mathcal S}_+(t)}{k_{\rm B}} = \frac{1}{2}{\rm Tr}\left[\sigma\sigma^\intercal {\mathcal K}^{-1}(t)\right].
\label{eq:SpGauss}
\end{equation}
For finite times, all the entropy fluxes depend on the initial conditions and can be determined from Eqs.~\eqref{eq:Sm}, \eqref{eq:SdotGaussian} and \eqref{eq:SpGauss}. 

Let us now focus on the specific setups considered in Secs.~\ref{Section: Dimer} and \ref{Section Trimer}. For the dimer, we have investigated the PDF for the length of the single bond and thus $d=1$. Using our analytical findings with $\sigma\sigma^\intercal = 4D$ and ${\mathcal K}(t) = \nu(t)$, we get
\begin{equation}
\frac{{\mathcal S}(t)}{k_{\rm B}} = 1 + \frac{1}{2}\log[(2\pi)^2 \nu(t)]
\label{eq:dimer_entropy}
\end{equation}
and
\begin{equation}
\frac{\dot{\mathcal S}(t)}{2 D k_{\rm B}} = \frac{\lambda^2(t)}{\nu(t)},\quad
\frac{\dot{\mathcal S}_+(t)}{2 D k_{\rm B}} = \frac{1}{\nu(t)},\quad
\frac{\dot{\mathcal S}_-(t)}{2 D k_{\rm B}} = \frac{\lambda^2(t) - 1}{\nu(t)},
\label{eq:dimerSt}
\end{equation}
where $\lambda(t)$ is given by Eq.~\eqref{eq:fundamental_sol_MT} and the variance reads $\nu(t) = 4D \int_0^t \mathrm{d}s\ \lambda^2(s)$. Similarly, in our analytical investigation of the trimer, we have fixed the angles between the individual bonds and investigated the PDF for the three bond length only, implying that $d=3$. Using $\sigma\sigma^\intercal = 4D M$ and Jacobi's formula for the derivative of determinants, we obtain the expressions
\begin{equation}
\frac{{\mathcal S}(t)}{k_{\rm B}} = \frac{3}{2} + \frac{1}{2}\log[(2\pi)^3 |{\mathcal K}(t)|]
\label{eq:trimer_entropy}
\end{equation}
and
\begin{equation}
\frac{\dot{\mathcal S}(t)}{2Dk_{\rm B}} =  {\rm Tr}\left[\frac{M\lambda^2(t)}{{\mathcal K}(t)}\right],\quad
\frac{\dot{\mathcal S}_+(t)}{2Dk_{\rm B}} = {\rm Tr}\left[\frac{M}{{\mathcal K}(t)}\right],\quad
\frac{\dot{\mathcal S}_-(t)}{2Dk_{\rm B}} = {\rm Tr}\left[M\frac{\lambda^2(t) - {\mathcal I}}{{\mathcal K}(t)}\right],
\label{eq:trimerSt}
\end{equation}
where $\lambda(t)$ is given by Eq.~\eqref{Trimer Fundamental} and the covariance matrix ${\mathcal K}(t)$ reads $4D M \int_0^t {\rm d}s \lambda^2(s)$.

The formulas \eqref{eq:dimer_entropy} -- \eqref{eq:trimerSt}
are valid both in the stable regimes (i) and (ii), where the system at long times relaxes to a stationary time-independent structured state, and in the unstable regime (iii). In the unstable regime, the variance $\nu(t)$ and the covariance ${\mathcal K}(t)$ diverge in time. As a result, the system entropy ${\mathcal S}(t)$ diverges and the entropy flow $\dot{\mathcal S}_+(t)$ decays to zero, because the variance of the PDF is so large that the diffusion can hardly further increase it. On the other hand, the rate of entropy change $\dot{\mathcal S}(t)$ and thus also the entropy outflow $\dot{\mathcal S}_-(t)$ remain finite oscillating functions, as can be seen for the dimer by using the exponential long-time approximation \eqref{eq:xexp} for $\lambda(t)$, and similarly for the trimer.

For the purpose of structure formation, only the regimes (i) and (ii) are of interest, because only then the PDF reaches a time-independent non-equilibrium steady state at long times, i.e. $\lim_{t\to\infty}{\mathcal S}(t) = {\mathcal S}$, $\lim_{t\to\infty}\dot{\mathcal S}(t) = 0$ and $\dot{\mathcal S}_- \equiv \lim_{t\to\infty}\dot{\mathcal S}_-(t) = -\lim_{t\to\infty}\dot{\mathcal S}_+(t)$. Let us therefore now evaluate the long-time stationary system entropies $S$ and entropy fluxes $\dot{\mathcal S}_-(t)$ maintaining the molecular-like structures formed in our model for the dimer and the trimer in these two regimes. Using the asymptotic formulas \eqref{eq:SS_variance} and \eqref{eq:CtrimerSS} for the dimer bond-length stationary variance $\nu_{\rm ss}$ and trimer covariance ${\mathcal K}_{\rm ss}$, we find from Eqs.~\eqref{eq:dimer_entropy} -- \eqref{eq:trimerSt}
\begin{gather}
\frac{\mathcal S}{k_{\rm B}} = 1 + \log{(2\pi)}
+ \frac{1}{2}\log{\left[\frac{2 D}{\omega_\tau} \frac{1+\sin(\omega_\tau\tau)}{\cos(\omega_\tau\tau)}\right]},
\label{eq:SftyDimer}\\
\frac{\dot{\mathcal S}_-}{k_{\rm B}} = -\frac{\omega_\tau \cos\left(\omega_\tau\tau\right)}{1+\sin\left(\omega_\tau\tau\right)},
\label{eq:SminftyDimer}
\end{gather}
for the dimer and 
\begin{gather}
\frac{\mathcal S}{k_{\rm B}} = \frac{3}{2} + \frac{3}{2}\log{(2\pi)}
+ \frac{1}{2}\log{\left[(a - b)^2 (a + 2 b)\right]},
\label{eq:SftyTrimer}\\
\frac{\dot{\mathcal S}_-}{k_{\rm B}} = -2D{\rm Tr}\left[\frac{M}{{\mathcal K}_{\rm ss}}\right] = -6D \frac{(a+b)}{(a-b)(a+2b)},
\label{eq:SminftyTrimer}
\end{gather}
for the trimer. In the last two formulas, $a={\mathcal K}_{\rm ss}(1,1)$ denotes diagonal and $b={\mathcal K}_{\rm ss}(1,2)$ off-diagonal elements of the covariance matrix. The system entropies \eqref{eq:SftyDimer} and \eqref{eq:SftyTrimer} are determined by the width of the PDFs for the bonds. Therefore, they monotonously increase with temperature $T = \gamma D/k_{\rm B}$ and with the delay time $\tau$, and exhibit a minimum as functions of the frequencies $\omega_\tau$ (dimer) and $\omega$ (trimer), similarly as the variance $\nu_{\rm ss}$ and the diagonal matrix elements of the covariance matrix ${\mathcal K}_{\rm ss}$. The quality of the steady-state structures is thus in our model always unfavorably influenced by the delay and, for a given delay, one can tune the frequency in order to minimize this (usually unwanted) effect. 

\begin{figure}[t!]
\centering
	\begin{tikzpicture}
	\node (img1)  {\includegraphics[width=0.4\columnwidth]{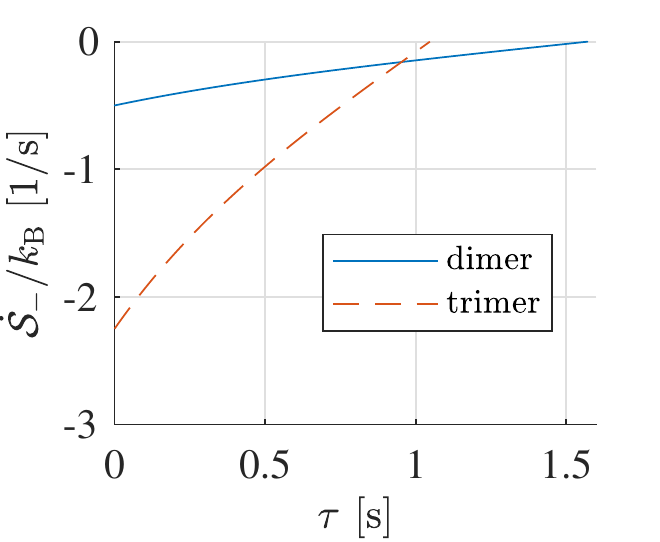}};
	\node[right=of img1, node distance=0.0cm, yshift=0cm,xshift=-1.0cm] (img2)
	{\includegraphics[width=0.4\columnwidth]{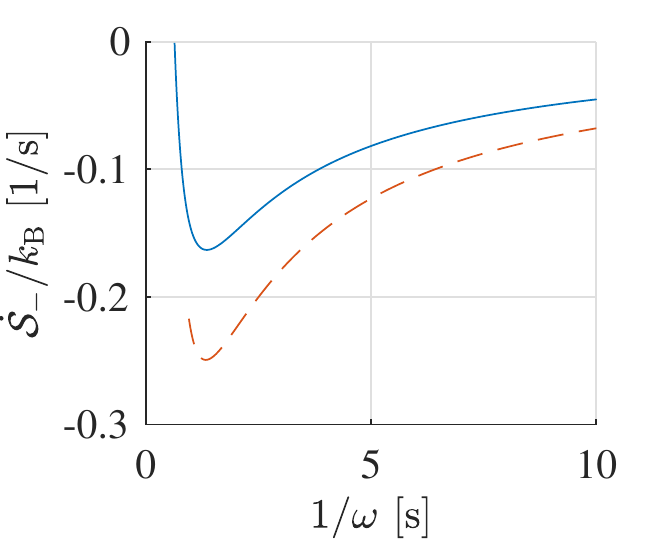}};
	\node[above=of img1, node distance=0.0cm, yshift=-1.6cm,xshift=-2.9cm] {a)};
	\node[above=of img2, node distance=0.0cm, yshift=-1.6cm,xshift=-3.0cm] {b)};	
	\end{tikzpicture}	
	\caption{The stationary structure maintaining entropy fluxes \eqref{eq:SminftyDimer} and \eqref{eq:SminftyTrimer} in the dimer (solid lines) and in the trimer (dashed lines) as functions of the delay [panel a)] and inverse frequency [panel b)], respectively. If not specified otherwise we used the parameters $D=1$ $\mu$m$^2/$s, $\tau = 1$ s and $\omega=$ $1/$s. For the dimer, we used the approximation $\omega_\tau = \omega$, which is accurate for $D\tau/R^2 \approx 0$.}	
	\label{fig:splus}	
\end{figure}

The two entropy fluxes \eqref{eq:SminftyDimer} and  \eqref{eq:SminftyTrimer} are negative, highlighting that they correspond to entropy outflows from (or information inflows into) the system. Interestingly enough, the entropy fluxes do not depend on the temperature $T$ (or noise strength $D$) as already predicted in Ref.~\cite{khadka2018active}. This means that the fluxes are discontinuous in the formal limit $T\to 0$ because they must inevitably vanish for zero noise, where the PDF for the system evolution is a $\delta$-function for all times. We plot the stationary entropy fluxes \eqref{eq:SminftyDimer} and \eqref{eq:SminftyTrimer} as functions of the delay $\tau$ in Fig.~\ref{fig:splus} a) and frequency $\omega$ in Fig.~\ref{fig:splus} b). Naturally, maintaining a stationary structure in a bigger system (trimer, dashed lines) requires a larger (more negative) entropy flux (or more information) than in the smaller one (dimer, solid lines). The maximum of the fluxes $|\dot{\mathcal S}_-|$, depicted in Fig.~\ref{fig:splus} b), arises as a result of a competition between stronger confinement, corresponding to larger frequencies $\omega$, and gradual destabilization with increasing $\omega\tau$, when the system enters the unstable regime (iii). For the dimer, it is located at the maximum of the variance \eqref{eq:SS_variance} defined by the equation $\cos (\omega_\tau \tau) = \omega_\tau \tau$ since $\dot{\mathcal S}_- = - \nu_{\rm SS}/2D$. This result is intuitive since more localized distributions spread due to the diffusion faster than broader ones. For the figure, we used for simplicity the approximation $\omega_\tau \approx \omega$ for the dimer.

%% file: Content/IsomerTransition.tex
\section{Transition Rates for Isomer Transformations} \label{Sec.: TST}

The particles within the molecular structures depicted in Fig.~\ref{fig:CompareBondNumber} a) may exchange their positions. Assuming the particles to be distinguishable, different arrangements of the same structure may arise which can be interpreted as different isomers of the same molecule. Their study can provide further insight into the stability properties of our non-equilibrium molecules. In fact, we find that the study for molecules made out of only a few particles is informative also for the phenomenology observed for large particle numbers. While for the purely deterministic motion then isomer transitions only appear for time delays $\tau$ in the unstable regime (iii), in a system affected by thermal fluctuations the transitions occur for arbitrary delays. The evaluation of the frequencies of such transitions, which measure the stability of the individual isomers, can thus provide insight into the overall energy landscape responsible for the non-equilibrium structure formation. It is the main topic of transition rate theory \cite{hanggi1990review}.

The transition rate $\kappa_{A\to B}(t)$ for switching from a conformation $A$ to a conformation $B$ at time $t$ can be (for arbitrary dynamics) found from the mean number of transitions $N_{A\to B}(t)$ from $A$ to $B$ during an infinitesimal time interval $\Delta t$ as $\kappa_{A\to B}(t) = N_{A\to B}(t)/\Delta t$. Alternatively, one can get it from the inverse mean first passage time for changing the two isomers, leading to the same results. In general, the deduced transition rates depend on the initial state of the system and on time and they can be calculated analytically only in few simple situations. While they in principle can straightforwardly be evaluated in simulations, this can take a (forbiddingly) long time if the transition rates are small.

For an analytical treatment, it is more convenient to define the transition rate via the so-called survival probability $S(t)$ that the system has not changed its initial isomer until time $t$. Our particular problem concerning the transitions between different isomers can be mapped to a particle moving in a high-dimensional energy landscape. We denote by $S_{A\rightarrow B}(t)$ the survival probability that the system, starting in the conformation $A$ with an absorbing boundary at the top of the barrier to conformation $B$ (and reflecting barriers elsewhere), will remain in the configuration $A$ until time $t$. The transition rate between the states $A$ and $B$ is then given by $\kappa_{A\rightarrow B}(t) = \dot{S}_{A\rightarrow B}(t)/S_{A\rightarrow B}(t)$. Hence, if the dynamical equation for the probability distribution for the state of the system with the correct boundary conditions is known, we can determine the transition rate numerically and, in some situations, even analytically. 

Considering standard Markovian Langevin dynamics, the asymptotic form $\lim_{t \to \infty} \kappa_{A\to B}(t)$ of the transition rate can (approximately) be calculated using Kramers' rate theory \cite{hanggi1990review,kramers1940original} which was originally developed to describe chemical reaction rates. The approximation works best for a high energy barrier compared to the thermal energy. Kramers' theory was extended to reaction rates for generalized Langevin equations (GLE's) describing non-Markovian systems. A crucial contribution in this direction came from Grote and Hynes \cite{grote1980ght} who derived a dynamical correction to Kramers' result. While their analysis was based on a parabolic barrier, Pollak \cite{pollak1990vtst} investigated the decay rate of an underdamped particle trapped in a symmetric cusp double well potential obeying the GLE with an arbitrary memory kernel satisfying the fluctuation-dissipation theorem. The time-dependent rate $\kappa_{A\to B}(t)$ for driven overdamped systems can be calculated using the recent theory of Bullerjahn et. al \cite{bullerjahn201spectroscopy} for forcible molecular bond breaking. 

To the best of our knowledge, the literature on the rate theory of time-delayed systems is scarce. The escape from a cubic metastable well under a time-delayed friction was investigated in Ref.~\cite{grabert1988delay}. Based on their small-delay approximation, Guillouzic et al. \cite{guillouzic2000rate} calculated the transition rate and the mean first passage time for an overdamped Brownian particle in a delayed quartic potential. From an experimental point of view, Curtin et al. \cite{curtin2004delay} studied transitions in a bistable system under time-delayed feedback. 

Our model does not belong to any of the previously investigated classes of systems. However, for a vanishing delay one can use Kramers' theory, since the system obeys a Markovian overdamped Langevin equation. Moreover, for nonvanishing delays in the stable regimes (i) and (ii), the one-time PDFs for dimer and trimer can be described by standard (time-local) FPEs with time-dependent coefficients, where Bullerjahn's theory applies and where one can evaluate the transition rate numerically. Furthermore, after long times, the coefficients in these FPEs become time independent suggesting that Kramers' theory may be applied also to obtain the long-time form of the transition rates for a nonzero delay.

Although looking promising, all the techniques above are based on the time-local FPE. For non-zero delay, they share one drawback, which may limit their applicability to small delays: the time-local FPE is derived from solutions to the delayed Langevin equations without the absorbing boundary condition. While this represents no problem for diffusion dynamics without delay, it can cause problems in our delayed system. In the following sections, we compare predictions of Kramers' theory, Bullerjahn's theory and direct numerical evaluation of the transition rates from the time-local FPE against BD simulations of the transition rates for dimer and trimer, demonstrating that the rates obtained from the time-local FPE are indeed accurate for small and moderate delays only.

\subsection{Dimer}
\label{sec:dimer_rate}

\begin{figure}
	\centering
	\includegraphics[width=0.5\linewidth]{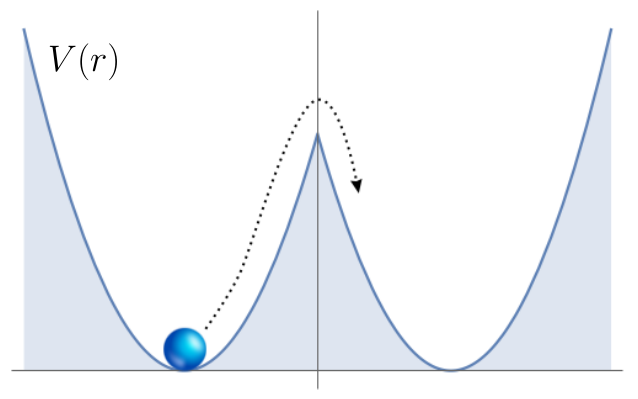}
	\caption{The exchange of positions of the two particles of a dimer in 1d can be mapped to the escape dynamics of a single particle in a cusped double-well potential.}
	\label{fig:PotentialPlot}
\end{figure}

To study transition rates, the simplest configuration of our model is the dimer with two distinguishable particles in one dimension. (Due to rotational symmetry, we cannot distinguish between dimer isomers in two dimensions). The setting is described by the approximate Langevin equation~\eqref{eq:eqr3} for the inter-particle distance $r = |\mathbf{r}_1-\mathbf{r}_2| > 0$. A transition occurs when the two particles exchange their positions, and can be assigned to the moment when the bond length $r$ vanishes. To illustrate the problem, it is useful to extend the domain of the distance variable $r$ such that it is positive for one isomer and negative for the other.
For vanishing delay, this redefined signed bond length  $\tilde{r}$ then diffuses in the cusped double-well potential $V(\tilde{r}) = \gamma\omega (|\tilde{r}|-R)^2/2$, depicted in Fig.~\ref{fig:PotentialPlot}, with the diffusion coefficient $2D$. 

For a nonzero delay, based on the approximate solution \eqref{eq:solution_SDDE_MT} to the Langevin equation~\eqref{eq:eqr3} assuming $r \in (-\infty,\infty)$ and $x(t) = 0$ for $t<0$, we have found that the one-time PDF $P_1 = P_1(x,t) = P_1(x,t|x_0,0)$ obeys the FPE~\eqref{FPE W1}, which reads
\begin{equation}
\frac{\partial}{\partial t}P_1 = \frac{\partial}{\partial x} \bigg[ \omega_\tau(t)x + 2D_\tau(t)\frac{\partial}{\partial x}\bigg] P_1 \equiv \mathcal{L}(x,t) P_1.  \label{Eq:FPE_W1_text}
\end{equation}
This equation describes diffusion in the harmonic potential 
\begin{equation}
	V(\tilde{r},t) = \gamma\omega_\tau(t) (|\tilde{r}|-R)^2/2 ,
\label{eq:cusped_pot}
\end{equation}
with the time-dependent stiffness $\gamma\omega_\tau(t)$ [given by \eqref{A(t)}] and the time-dependent diffusion coefficient $ 2D_\tau(t)$ [given by \eqref{D(t)}]. 

The validity of Eq.~\eqref{Eq:FPE_W1_text} for $P_1$ with natural boundary conditions suggests that one can further employ the analogy between the delayed dynamics and the (effective) Markovian model for calculating the transition rate $\kappa(t)$ for switching between the isomers. In the Markovian case, the transition rate for surpassing the (effective) energy barrier $\gamma\omega_\tau(t) R^2/2$ at $r=0$ to the other isomer can be calculated from Eq.~\eqref{Eq:FPE_W1_text} with an absorbing boundary at $x=-R$ \cite{Risken1996}. We now review several methods suitable for this task, and compare the results to BD simulations of the complete model with energy barrier $\gamma\omega R^2/2$ and delayed dynamics. In order to study the transition rate between the isomers of the dimer analytically, it is enough to consider the dynamics of the system in one of the wells of the potential, i.e. for $x = \tilde{r} - R > 0$.

\subsubsection*{Numerical Method}

We first consider the situation when the system dwells in the state $x(t) = 0$ for $t\le 0$ and then starts to diffuse in the time-dependent potential~\eqref{eq:cusped_pot}. Then, the time-dependent Markovian rate $\kappa_{\rm M}(t)$ can be determined from the equation
\begin{equation}
\partial_t \tilde{P}_{\rm a}(x,t) = \mathcal{L}(x,t) \tilde{P}_{\rm a}(x,t) + \kappa_{\rm M}(t)\tilde{P}_{\rm a}(x,t),
\label{eq:absorbing}
\end{equation}
for the normalized PDF $\tilde{P}_{\rm a}(x,t) = P_{\rm a}(x,t)/S_{\rm a}(t)$ for the position of the particle surviving in the right well of the cusped potential ~\cite{Ornigotti2018}. Here, $P_{\rm a}(x,t)$ is the solution to the FPE~\eqref{Eq:FPE_W1_text} with \emph{absorbing boundary} at $x=-R$ and $S_{\rm a}(t)= \int_{-R}^\infty \mathrm{d}x P_{\rm a}(x,t)$ is the probability that the particle survives in the right well until time $t$. Equation~\eqref{eq:absorbing} follows from Eq.~\eqref{Eq:FPE_W1_text} by inserting the definitions of the PDF $\tilde{P}_{\rm a}$ and of the transition rate
\begin{equation}
\kappa_{\rm M}(t) = -\dot{S_{\rm a}}(t)/S_{\rm a}(t).
\label{eq:rate_nt}
\end{equation} 
We solved it numerically using the method presented in Ref.~\cite{Holubec2018}. 

\subsubsection*{Bullerjahn's Method}

Alternatively, one can determine the rate approximately using the analytical theory developed by Bullerjahn et al. in Ref.~\cite{bullerjahn201spectroscopy}. Therein, the rate is constructed from the (Gaussian) solution $P_1$~\eqref{eq:W_2_prob_distr} of the FPE \eqref{Eq:FPE_W1_text} with natural boundary conditions. Specifically, one approximates the probability current 
\begin{equation}
j(-R,t) = \dot{S_{\rm a}}(t) = -\left[\omega_\tau(t)x + 2D_\tau(t)\partial_x\right] P_{\rm a}(x,t)|_{x=-R}
\end{equation}
across the absorbing boundary by~\footnote{Rescaling the diffusion coefficient by factor 2 corrects for the part of the diffusive flux that can not return to the system due to the absorbing boundary, see Ref.~\cite{bullerjahn201spectroscopy} for details.}
\begin{equation}
j^*(-R,t) \equiv - \left[ \omega_\tau(t)x +4D_\tau(t)\partial_x\right] P_1(x,t) |_{x=-R},
\end{equation}
and the survival probability $S_{\rm a}(t)$ by
\begin{equation}
S_1(t) = \int_{-R}^\infty \mathrm{d}x \ P_1(x,t) = \frac{1}{2} \left[ 1 + \mathrm{Erf}\left(R/\sqrt{2\nu(t)}\right)\right]. 
\end{equation}
In the last expression, the symbol $\mathrm{Erf}$ denotes the error function and the variance $\nu(t)$ is given by Eq.~\eqref{eq:variance_dimer}. The approximate Markovian transition rate is then given by
\begin{equation}
\kappa_{\rm B}(t) = -j^*(-R,t)/S_1(t).
\label{eq:rate_appox_t}
\end{equation}

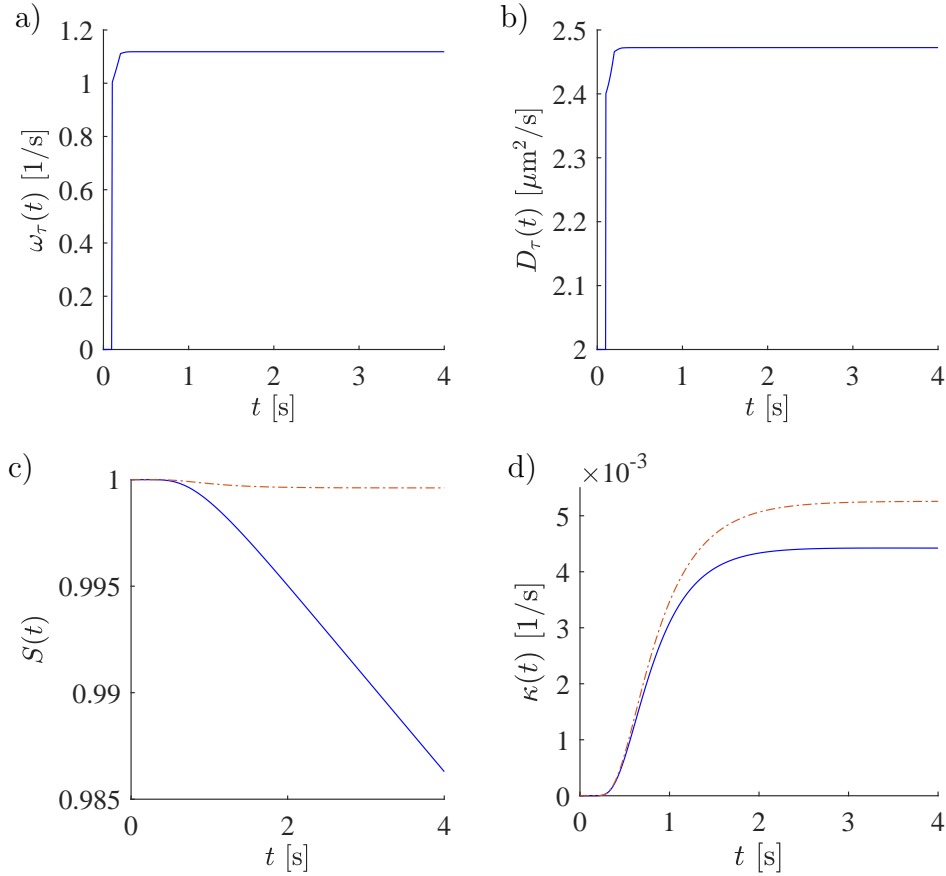
\begin{figure}[t!]
\centering
	\begin{tikzpicture}
	\node (img1)  {\includegraphics[width=0.4\columnwidth]{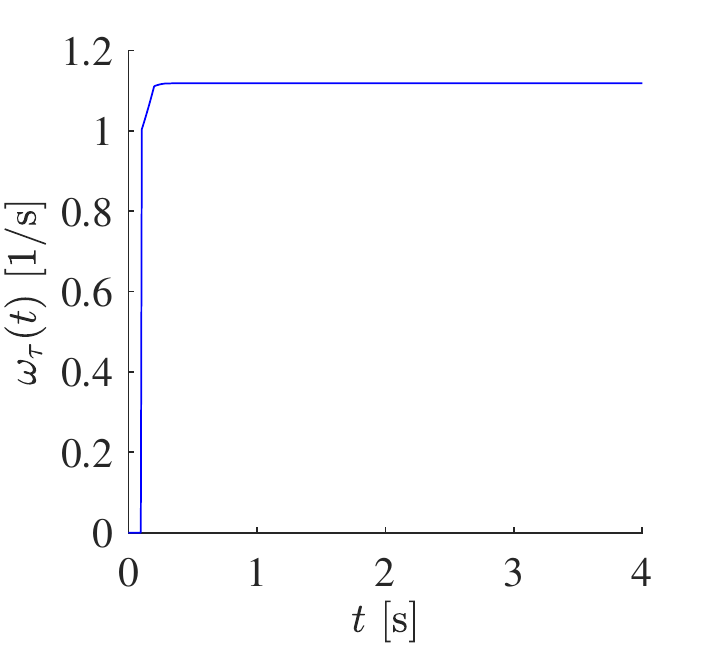}};
	\node[right=of img1, node distance=0.0cm, yshift=0cm,xshift=-1.0cm] (img2)
	{\includegraphics[width=0.4\columnwidth]{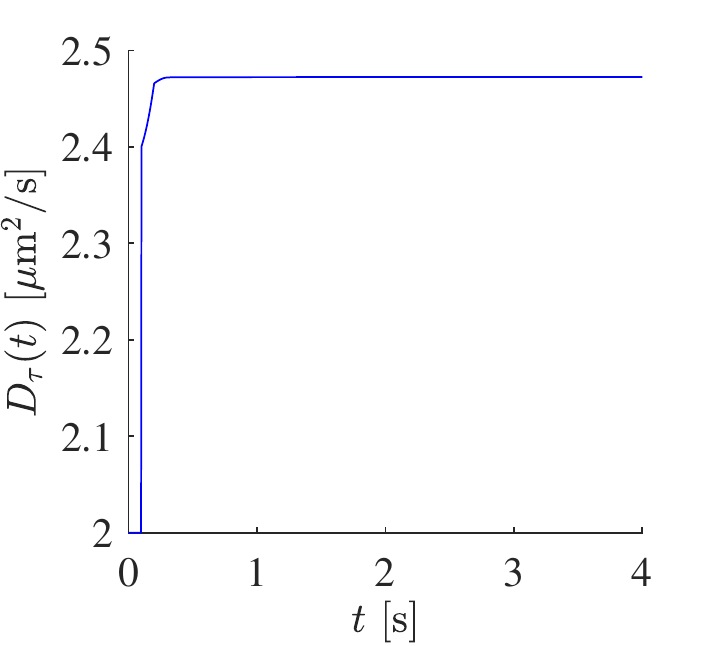}};
	\node[below=of img1, node distance=0.0cm, yshift=1cm,xshift=0.0cm] (img3)
	{\includegraphics[width=0.4\columnwidth]{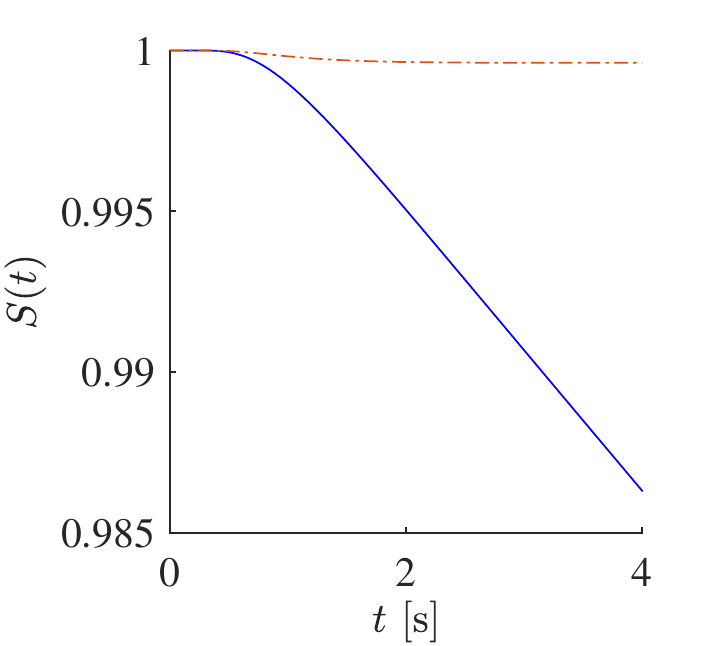}};
	\node[below=of img2, node distance=0.0cm, yshift=1cm,xshift=0.0cm] (img4)
	{\includegraphics[width=0.4\columnwidth]{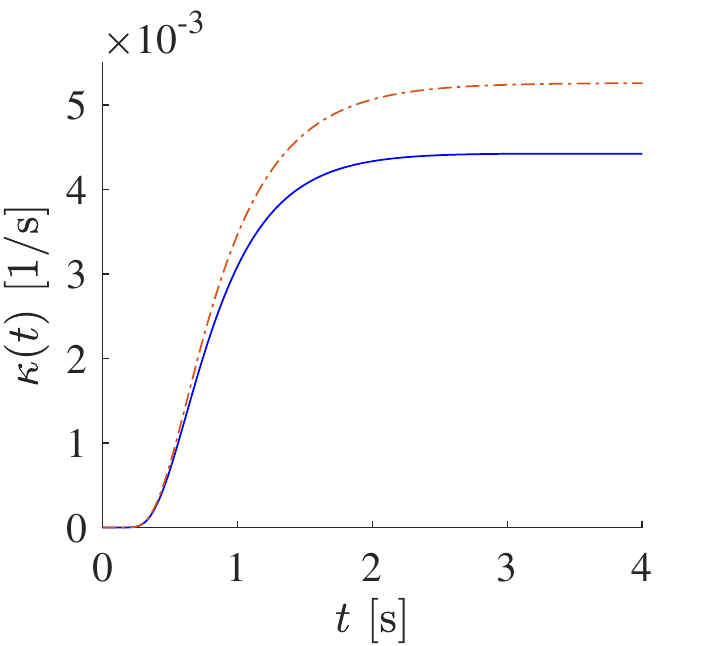}};
	\node[above=of img1, node distance=0.0cm, yshift=-1.8cm,xshift=-3.0cm] {a)};
	\node[above=of img2, node distance=0.0cm, yshift=-1.8cm,xshift=-3.1cm] {b)};	
	\node[above=of img3, node distance=0.0cm, yshift=-1.8cm,xshift=-3.1cm] {c)};
	\node[above=of img4, node distance=0.0cm, yshift=-1.8cm,xshift=-3.0cm] {d)};	
	\end{tikzpicture}	
	\caption{The frequency $\omega_\tau(t)$ [panel a)] and the diffusion coefficient $2D_\tau(t)$ [panel b)] from the Markovian FPE~\eqref{Eq:FPE_W1_text} as functions of time. The solid lines in c) and d) show the time dependence of the survival probability $S_{\rm a}(t)$ and transition rate $\kappa_{\rm M}(t)$ calculated numerically from Eqs.~\eqref{eq:absorbing} and \eqref{eq:rate_nt}, respectively. The dot-dashed lines in c) and d) depict the corresponding variables $S_1(t)$ and $\kappa_{\rm B}(t)$ obtained by approximate analytical solution of Eq.~\eqref{eq:absorbing}, see Eq.~\eqref{eq:rate_appox_t} and the text above. Parameters used: $\omega =$ 1/s, $D = 1$ $\mu$m$^2$/s, $R = 5$ $\mu$m, and $\tau = 0.1$ s. The system is in the dynamical regime (i).}	
	\label{fig:transition_rate_time}	
\end{figure}

In Fig.~\ref{fig:transition_rate_time}, we show the frequency $\omega_\tau(t)$, the (effective) diffusion coefficient $2D_\tau(t)$, survival probabilities $S_{\rm a}(t)$ and $S_1(t)$ and the transition rates $\kappa_{\rm M}(t)$ and $\kappa_{\rm B}(t)$ as functions of time $t$ for parameters in the dynamical regime (i). One can observe that both the parameters $\omega_\tau(t) = \dot{\lambda
}(t)/\lambda(t)$ and $D_\tau(t) = 2D\lambda(t)^2 + \omega_\tau(t) \nu(t)$ and the rates saturate with time. The relaxation time of $\omega_\tau(t)$ is determined by the time in which the Green's function $\lambda(t)$ approaches the long-time exponential representation~\eqref{eq:xexp}, and the corresponding stationary value, $\omega_\tau(\infty) = 1/t_{\rm R}$, is controlled by the relaxation time $t_{\rm R}$ for decay of $\lambda(t)$ to $0$, see also Fig.~\ref{fig:properties_of_lambda} in \ref{App:A1}. The effective diffusion coefficient converges to the value $D_\tau(\infty) = 2D [1 + \sin(\omega_\tau \tau)]/\cos(\omega_\tau \tau) \ge 2D$, determined by the stationary variance $\nu(\infty) = \nu_{\mathrm{ss}}$, see Eq.~\eqref{eq:SS_variance}. The transition rates relax with the relaxation time $t_{\rm R}$, similarly as the corresponding PDFs $P_{\rm a}$ and $P_1$. The analytical expressions for $S_1(t)$ and $\kappa_{\rm B}(t)$ approximate the numerical results for $S_{\rm a}(t)$ and $\kappa_{\rm M}(t)$ best for short times $t\ll t_{\rm R}$, where the PDFs $P_{\rm a}$ and $P_1$ are still hardly affected by the different boundary conditions at $x=-R$. For long times and up to moderate values of time delay, the approximate analytical transition rate $\kappa_B$ overestimates the corresponding exact rate $\kappa_M$, see also Figs.~\ref{fig:transition rate depending on delay} a) and b) below. For long delays, the (effective) barrier height over the (effective) thermal energy decreases so that the assumptions of the transition state theory are not valid any more, and $\kappa_B < \kappa_M$, see Fig.~\ref{fig:transition rate depending on delay} c). 

The situation of low barriers can be understood from the behavior at vanishing potential strength $\omega \to 0$, when $D_\tau(t)=2D$ and the finite time transition rates $\kappa_{\rm M}(t)$ and $\kappa_{\rm B}(t)$ can be calculated analytically. Namely, the PDFs $\tilde{P}_{\rm a}$ and $P_{1}$ in the definitions \eqref{eq:rate_nt} and \eqref{eq:rate_appox_t} of the rates read $\tilde{P}_{\rm a}(x,t) = \left\{\exp(-x^2/4Dt) - \exp[-(x+2R)^2/4Dt]\right\}/\sqrt{4\pi D t}$  and $P_1(x,t) = \exp(-x^2/4Dt)/\sqrt{4\pi D t}$ \cite{Risken1996} leading to the formulas 
\begin{align}
\kappa_{\rm M}(t) &= \frac{2R}{t} \frac{\exp(-R^2/4Dt)}{\int_{-R}^\infty {\rm d}x \left\{\exp(-x^2/4Dt) - \exp[-(x+2R)^2/4Dt]\right\}},\\
\kappa_{\rm B}(t) &= \frac{2R}{t} \frac{\exp(-R^2/4Dt)}{\int_{-R}^\infty {\rm d}x \exp(-x^2/4Dt)}.
\end{align}
Since the denominator in the expression for the rate $\kappa_{\rm M}(t)$ is smaller than that for $\kappa_{\rm B}(t)$, we conclude that, for low energy barriers, the inequality $\kappa_{\rm B}(t) \le \kappa_{\rm M}(t)$ holds.
On the other hand, for very high barriers, the PDFs $\tilde{P}_{\rm a}$ and $P_1$ are (almost) identical because the absorbing boundary at $x=-R$ is effectively inaccessible. In such a case, the probability current $j(-R,t)$ is (almost) zero, $S(t) \approx 1$, and $j^* \approx j - 2D \partial_x P_1|_{x=-R} < 0$ leads to the inequality $\kappa_{\rm B}(t) \ge \kappa_{\rm M}(t)$.

To gain a deeper insight into the behavior of the transition rates, let us consider the stationary regime, $t\to \infty$. In this regime, $\tilde{P}_{\rm a}(-R,\infty) = 0$, due to the absorbing boundary condition at $x=-R$, and $j_1(x,\infty) \equiv - \left[ \omega_\tau(\infty)x + 2D_\tau(\infty)\partial_x\right] \tilde{P}_1(x,\infty) |_{x=-R} = 0$, due to the conservation of probability $\partial_t \tilde{P}_1(x,\infty) = \partial_t [P_1(x,\infty)/S_1(\infty)] = -\partial_x j_1(x,\infty)$, where $\partial_t \equiv \partial/\partial t$ and $\partial_x \equiv \partial/\partial x$. Then the transition rates $\kappa_{\rm M}(\infty) = - j_{\rm Da}$ and $\kappa_{\rm B}(\infty)= -j_{\rm D1}$ are determined solely by the diffusive fluxes $j_{\rm Da}\equiv 2D_\tau(\infty) \partial_x \tilde{P}_{\rm a}(x,\infty)|_{x=-R}$ and $j_{\rm D1}\equiv 2D_\tau(\infty) \partial_x \tilde{P}_{\rm 1}(x,\infty)|_{x=-R}$. The smaller the frequency $\omega$, the wider the PDFs $\tilde{P}_1$ and $\tilde{P}_{\rm a}$. For small $\omega$, the boundary at $x=-R$ is in the region where the PDF $\tilde{P}_1$ has its maximum and it is also close to the maximum of $\tilde{P}_{\rm a}$. In such a case, the PDF $\tilde{P}_{\rm a}$, which must vanish at $x=-R$, changes near the boundary faster than $\tilde{P}_{\rm 1}$, leading to $|j_{\rm D1}| < |j_{\rm Da}|$ and $\kappa_{\rm B}(\infty) \le \kappa_{\rm M}(\infty)$, in accord with the argument put forward in the previous paragraph. With increasing $\omega$, the boundary shifts away from the maxima of $\tilde{P}_1$ and $\tilde{P}_{\rm a}$ towards their tails. Due to the trajectories trapped in the absorbing state~\cite{Ornigotti2018,Siiler2018}, the maximum of $\tilde{P}_{\rm a}$ is slightly farther away from the absorbing boundary than the maximum of $\tilde{P}_1$, and thus, with increasing $\omega$, the tail of $\tilde{P}_{\rm a}$, with small derivative (small $j_{\rm Da}$), hits the boundary at $x=-R$ before the corresponding tail of $\tilde{P}_1$. Hence, for large enough barrier height, the inequality between the rates crosses over to $\kappa_{\rm B}(\infty) \ge \kappa_{\rm M}(\infty)$. Finally, for very stiff traps ($\omega \to \infty$), both $j_{\rm Da}$ and $j_{\rm D1}$ vanish and $\kappa_{\rm M}(\infty)=\kappa_{\rm B}(\infty)=0$.

As shown in the figure~\ref{fig:transition_rate_time} d), the transition rates converge with time to constant values in regime (i), where the limits $\lim_{t \to \infty}\omega_\tau(t)$ and $\lim_{t \to \infty}D_\tau(t)$ exist. Also in regime (ii), the PDF $P_1$ assumes, after long times, the time-independent stationary form $P_1 = \exp\left(-x^2/2\nu_{\rm ss}\right)/\sqrt{2\pi \nu_{\rm ss}}$ with the variance $\nu_{\rm ss}$ given by Eq.~\eqref{eq:SS_variance}. This suggests that, also in this regime, the transition rate should saturate at long times. However, both the frequency $\omega_\tau(t)$ and the diffusion coefficient $2D_\tau(t)$ actually exhibit diverging oscillations caused by the oscillations in the Green's function $\lambda(t)$ \eqref{eq:fundamental_sol_MT}, in regime (ii). These divergences cause problems both in the FPE and in the approximate calculation of the rate using Bullerjahn's method. As a consequence, the (effective) Markov description can not be valid in the dynamical regime (ii). Nevertheless, let us now investigate to what extent the long-time transition rate $\kappa=\kappa(\infty)$ obtained from BD simulations of the Langevin equation~\eqref{eq:eqr3} is captured by the predictions \eqref{eq:rate_nt} and \eqref{eq:rate_appox_t} above.

\subsubsection*{Long-Time Behavior and Kramers' Method}

Assuming that at long times the PDF $\tilde{P}$ is time-independent and the limits $\lim_{t \to \infty}\omega_\tau(t)$ and $\lim_{t \to \infty}D_\tau(t)$ exist, we can rewrite the formula~\eqref{eq:absorbing} as the eigenvalue problem 
\begin{equation}
\mathcal{L}(x,\infty) \tilde{P}_{\rm a}(x,\infty) = \kappa_{\rm M}\tilde{P}_{\rm a}(x,\infty),
\label{eq:rateMtinfty}
\end{equation}
for the long time Markovian transition rate $\kappa_{\rm M} = \kappa_{\rm M}(\infty)$. We solve this formula numerically using the method described in Refs.~\cite{Ornigotti2018,Holubec2018}. The steady state value of the transition rate predicted with Bullerjahn's method reads 
\begin{equation}
\kappa_{\rm B} = -j^*(\infty)/S_1(\infty)
\label{eq:rateBtinfty}
\end{equation}
with the survival probability $S_1(\infty)=\left[ 1+ \mathrm{Erf}\left(R/\sqrt{2}\nu_\mathrm{ss}\right) \right]/2$ and the probability current $j^*(\infty) = \omega_{\tau}(\infty)R \exp\left(- R^2/2\nu_\mathrm{ss} \right)/\sqrt{2\pi\nu_\mathrm{ss}}$. 

For high barriers, where $S_1(\infty) \approx 1$, the long time form of Bullerjahn's transition rate coincides with the classical prediction by Kramers~\cite{hanggi1990review,kramers1940original} for the transition rate for leaving one of the wells of a cusped potential with barrier height $E_{\rm b}$, 
\begin{equation}
	\kappa_\mathrm{K} = 	
	\frac{2}{\sqrt{\pi}R^2} \frac{(\beta E_{\rm b})^{3/2}}	
	{\beta\gamma}\exp \left( -\beta E_{\rm b} \right)  = \frac{\omega_{\tau}(\infty)R}{\sqrt{2\pi\nu_\mathrm{ss}}}\exp\left(- \frac{R^2}{2\nu_\mathrm{ss}} \right) = j^*(\infty)
	.  \label{eq:kramers}
\end{equation}
In the penultimate equality, we used the appropriate inverse thermal energy $\beta=1/2 \gamma D_\tau(\infty)$ and barrier height $E_{\rm b} = \gamma \omega_\tau(\infty)R^2/2$, and also the asymptotic form of the diffusion coefficient $D_{\tau}(\infty) = \omega_{\tau}(\infty)\nu_\mathrm{ss}$, which follows from the condition $\lim_{t\to \infty}\lambda(t) = 0$, valid in the dynamical regimes (i) and (ii).

\subsubsection*{Renormalized Transition Rates}

Interestingly, the term $\omega_\tau(t)$, which causes divergences of the diffusion coefficient and the frequency of the potential in the FPE~\eqref{Eq:FPE_W1_text}, does not enter the argument of the exponential in the rates $\kappa_\mathrm{B}$ and $\kappa_\mathrm{K}$. This means that it just determines the kinetic prefactor, as can be also observed directly from the long-time form $\partial_tP_1 = \omega_\tau(t) \partial_x \left( x + 2\nu_\mathrm{ss} \partial_x\right) P_1$ of the FPE~\eqref{Eq:FPE_W1_text}. In the dynamical regime (ii), the kinetic prefactor in Eq.~\eqref{eq:kramers} cannot be correct, due to the diverging oscillations in the time-dependent frequency $\omega_\tau(t)$. Nevertheless, the exponential term seems to be reasonable, and thus it is tempting to use in the prefactor of the transition rates simply $\omega$, instead of the problematic $\omega_\tau(\infty)$. This substitution gives the correct pre-exponential factor of the rate for vanishing delay $\tau = 0$, where $\omega_\tau(\infty)=\omega$ and $2D_\tau(\infty) = 2D$. We denote the rates with the renormalized prefactor as 
\begin{equation}
\tilde{\kappa}_\mathrm{x} = \omega \kappa_\mathrm{x}/\omega_\tau(\infty)
\label{eq:scaled_rates}
\end{equation}
with $\mathrm{x}=$ M, B or K indicating Markov, Bullerjahn, or Kramers, respectively.

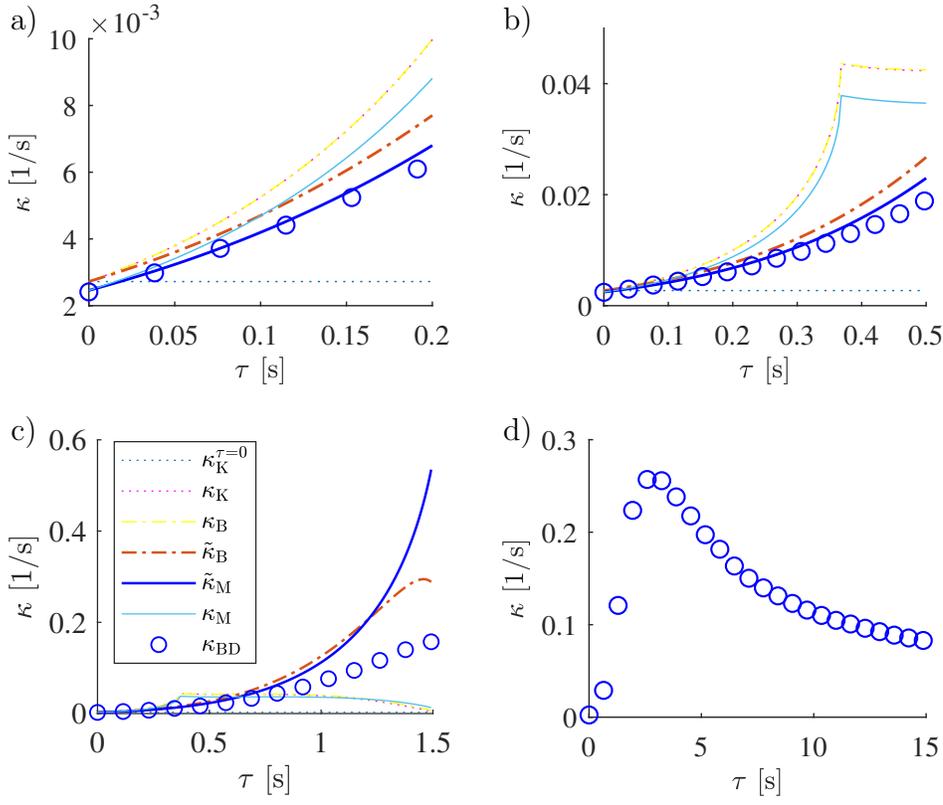
\begin{figure}[t!]
\centering
	\begin{tikzpicture}
	\node (img1)  {\includegraphics[width=0.4\columnwidth]{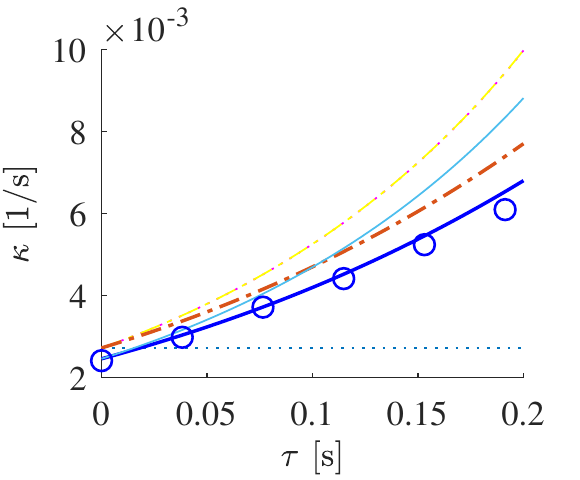}};
	\node[right=of img1, node distance=0.0cm, yshift=0cm,xshift=-1.0cm] (img2)
	{\includegraphics[width=0.4\columnwidth]{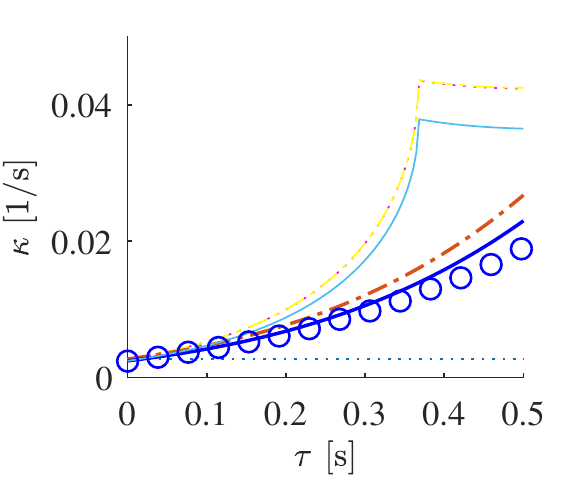}};
	\node[below=of img1, node distance=0.0cm, yshift=1cm,xshift=0.0cm] (img3)
	{\includegraphics[width=0.4\columnwidth]{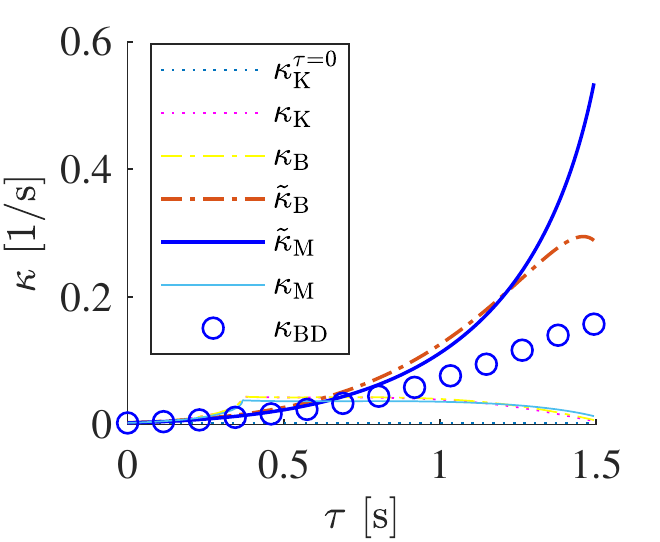}};
	\node[below=of img2, node distance=0.0cm, yshift=1cm,xshift=0.0cm] (img4)
	{\includegraphics[width=0.4\columnwidth]{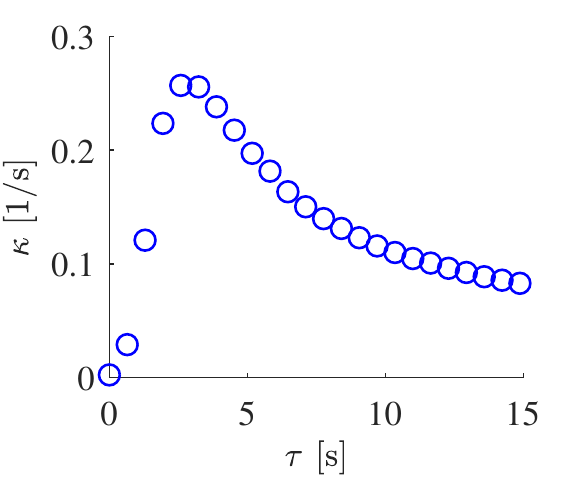}};
	\node[above=of img1, node distance=0.0cm, yshift=-1.8cm,xshift=-2.9cm] {a)};
	\node[above=of img2, node distance=0.0cm, yshift=-1.8cm,xshift=-2.9cm] {b)};	
	\node[above=of img3, node distance=0.0cm, yshift=-1.8cm,xshift=-2.9cm] {c)};
	\node[above=of img4, node distance=0.0cm, yshift=-1.8cm,xshift=-2.9cm] {d)};	
	\end{tikzpicture}	
	\caption{Panels a) - c) show typical $\tau$ behavior of the predictions $\kappa_{\rm M}$ (pale blue solid line), $\tilde{\kappa}_{\rm M}$ (dark blue solid line), $\kappa_{\rm B}$ (yellow dot-dashed line), $\tilde{\kappa}_{\rm B}$ (dark orange dot-dashed line) and $\kappa_{\rm K}$ (non-horizontal red dotted line) for transition rates obtained from Eqs.~\eqref{eq:absorbing}--\eqref{eq:scaled_rates}. The horizontal dotted lines correspond to Kramers' prediction $\kappa^{\tau=0}_{\rm M}$ for $\tau = 0$ s. The symbols ($\kappa_{\rm BD}$) in all the panels were obtained from $10^4$ simulated trajectories of the Langevin equation \eqref{eq:eqr3} with the time step $\mathrm{d}t=10^{-4}$ s. The individual panels differ only in the scale of the $\tau$-axis. We used the same parameters as in Fig.~\ref{fig:transition_rate_time}, where $\omega = 1$/s and thus the boundaries between the dynamical regimes (i), (ii) and (iii) approximately correspond to the values of $\tau$ $0.39$ s and $1.57$ s.}	
	\label{fig:transition rate depending on delay}	
\end{figure}

The necessity to change the kinetic prefactor in the rates stems from the fact that, although the absorbing boundary condition we used in Eq.~\eqref{eq:rateMtinfty} is correct for Markov dynamics ($\tau = 0$), it can not be precisely valid for the time-delayed dynamics ($\tau > 0$). To see this, it is enough to realize that the delayed system arriving at the boundary at a time $t$ does not feel the energy barrier $E_{\rm b} = \gamma \omega R^2/2$, but the barrier with energy $\gamma \omega [r(t-\tau)]^2/2$.

In Figs.~\ref{fig:transition rate depending on delay} a) -- c), we compare the various analytical predictions $\kappa_\mathrm{x}$ and $\tilde{\kappa}_\mathrm{x}$, $x=$ M, B or K, for the long-time transition rate (lines) with the asymptotic rate $\kappa$ (symbols) calculated from BD simulations of Eq.~\eqref{eq:eqr3} using the inverse first passage time for reaching the absorbing boundary at $r=0$. For a broad range of parameters fulfilling $k_B T \ll V(-R,0)$, we have found that the rate $\kappa$ can be predicted reasonably well only for values of $\tau$ in the dynamical regime (i) ($\omega\tau < 1/{\rm e} \approx 0.37$) and in the first part of the dynamical regime (ii) ($\omega\tau < \pi/2 \approx 1.57$). The rate $\kappa$ is best approximated by the expression $\tilde{\kappa}_\mathrm{M}$ obtained numerically from Eq.~\eqref{eq:rateMtinfty} with $\omega$ substituted for $\omega_\tau(\infty)$ in the operator $\mathcal{L}$. From the analytical expressions the re-scaled Bullerjahn expression $\tilde{\kappa}_\mathrm{B}$ [see Eqs.~\eqref{eq:rateBtinfty} and \eqref{eq:scaled_rates}] works best. However, in the figure we have used parameters leading to a high barrier $E_{\rm b}$, and thus Kramer's and Bullerjahn's predictions, $\kappa_\mathrm{K}$ and $\kappa_\mathrm{B}$, almost coincide. As a consequence, the line for $\tilde{\kappa}_\mathrm{K}$ in the figure overlaps with $\tilde{\kappa}_\mathrm{B}$, similarly as the line $\kappa_\mathrm{K}$ (suppressed in the figure for better readability) overlaps with $\kappa_\mathrm{B}$.

\subsubsection*{Delay-Dependence of $\kappa$}

In Fig.~\ref{fig:transition rate depending on delay} d), we also show the behavior of the rate $\kappa$ in the parameter regime (iii) inaccessible to the analytical and numerical formulas due to the diverging oscillation. The simulated transition rate in Figs.~~\ref{fig:transition rate depending on delay} a) -- d) is first approximately exponential and thus convex in $\tau$, then its curvature changes to concave and it runs through a maximum and, finally, the rate starts to decrease. The value of $\tau$ where the curvature changes sign coincides with the boundary $\pi/2\omega\tau \approx 1.57$ between the dynamical regimes (ii) and (iii). Interestingly, no qualitative change of $\kappa(\tau)$ is observed at the boundary $\omega
\tau=1/{\rm e} \approx 0.37$ between the regimes (i) and (ii). It is tempting to attribute, the (approximate) exponential increase of the rate with $\tau$ in regimes (i) and (ii) to the increase of the steady-state variance $\nu_{\rm ss}$, which is given as the ratio of the effective energy barrier $V(-R,\infty)$ and the effective thermal energy $\gamma/2D_\tau(\infty)$. Although this explanation may work well for small delays, it breaks down for values of $\tau$ in the second half of the regime (ii), where the actual rate $\kappa$ is no longer well approximated by our analytical and numerical predictions. This means that the identification of the parameters $\gamma \omega_\tau(t)$ and $2D_\tau(t)$ with the effective potential stiffness and the effective diffusion coefficient, respectively, suggested by the effective Markov equation~\eqref{Eq:FPE_W1_text}, is reasonable for relatively small values of $\tau$ only. In the dynamical regime (iii), the particle undergoes oscillations with an amplitude that increases both with $\tau$ and $t$. The corresponding transition rate, obtained from the BD simulations, thus decreases with the delay $\tau$ as a result of oscillations leading away from the transition boundary at $r = -R$. Note that, in this regime, the stationary transition rate actually does not exist, since the amount of time spent distant from the boundary increases with $t$ (so that the transition rate decreases with $t$), where $t$ is the duration of the simulation.

\subsection{Trimer}
\begin{figure}
	\centering
	\includegraphics[trim={0 6cm 0 0},width=0.7\linewidth]{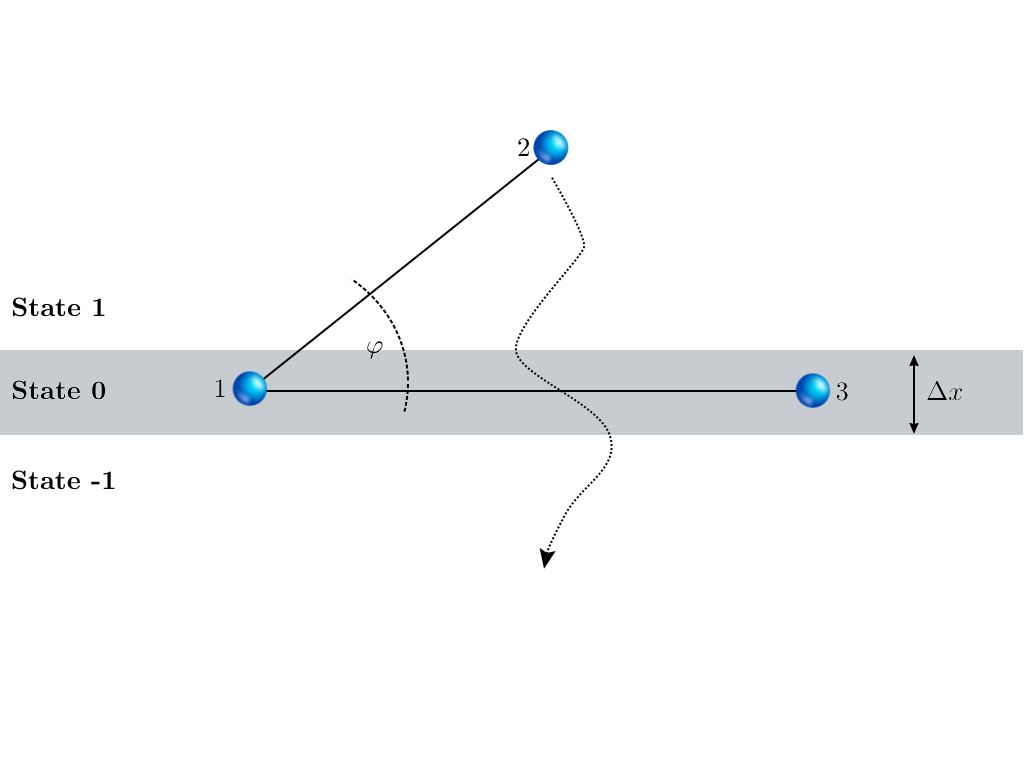}
	\caption{Transition from the clockwise to the counterclockwise isomer of a trimer molecule. To calculate the transition rate from the clockwise to the counterclockwise isomer, we choose the coordinate frame such that $x$-axis points from particle 1 to particle 3. By this choice of frame, all transitions are mapped to the crossing of the $x$-axis by particle 2. The grey region of width $\Delta x$ depicts a transition zone or ``neutral region'' that helps to avoid that small fluctuations around the axis are counted as isomer transitions.}
	\label{fig:isomer}
\end{figure}

Consider the trimer depicted in Fig.~\ref{fig:isomer} with the three distinguishable particles labeled by the numbers 1 to 3. We can count the particles either clockwise or anticlockwise and thus two different isomers can form in two dimensions. As in the case of the dimer discussed above, transitions between the two isomers occur with a transition rate depending on the diffusion coefficient, the coupling strength, and the equilibrium spring length. 

There are several ways how the clockwise isomer may turn into the anticlockwise one and vice versa. For example, the particles 1 and 2 can switch their positions, or the particle 2 can migrate from above the line connecting particles 1 and 3 to below that line, to name a few. The transition rate for hopping between the two isomers is then given as a sum of the transition rates for all the possible realizations of the transition. In order to make an analytical prediction for the transition rate, we choose the coordinate frame in such a way that the $x$-axis always points from particle 1 to particle 3 (see Fig.~\ref{fig:isomer}). Then all possible transitions between the two isomers boil down to a single event when particle 2 crosses the $x$-axis.  In particular, this also includes the transitions due to exchanging particles 1 and 3. In this case, the direction of the $x$-axes changes and thus the particle 2 effectively moves to its other side. In the following, we estimate the long-time transition rate $\kappa = \kappa(\infty)$ for the isomer transition in the steady state by means of Kramers' theory and compare it to BD simulations.

In the BD simulations, we have evaluated the rate $\kappa$ using the angle $\varphi$ between the abscissas $|12|$ and $|13|$ and a neutral region of width $\Delta x = \sqrt{3/16} R$ as exemplarily shown in Fig. \ref{fig:isomer}. A neutral state $\left|0\right>$ is introduced to avoid over counting due to fluctuations of $\varphi$ around $0$ and $\pi$ . It is  occupied if the smallest height of the triangle formed by the three particles, is smaller $\Delta x/2$. i.e. either if $|\varphi| < \phi_> \equiv \max\left[\arcsin(\Delta x/2r_{12}),\arcsin(\Delta x/2r_{13})\right]$ or $|\varphi| > \phi_<\equiv\arccos(\Delta x/2r_{12}) + \arccos(\Delta x/2r_{13})$. For $\varphi\in[\phi_>,\phi_<]$ the system is said to be in the clockwise state $\left|1\right>$, while for $\varphi\in[-\phi_>,-\phi_<]$ it is in the counter-clockwise state $\left|-1\right>$. To calculate the transition rate, we have counted the number of transitions between the states $\left|\pm1\right>$ during a specific simulation time window, where the transition occurred if the system underwent the sequence of states $\left|1\right> \to \left|0\right> \to \left|-1\right>$ or $\left|-1\right> \to \left|0\right> \to \left|1\right>$.

\begin{figure}[t!]
	\centering
	\begin{tikzpicture}
	\node (img1)  {\includegraphics[width=0.44\columnwidth]{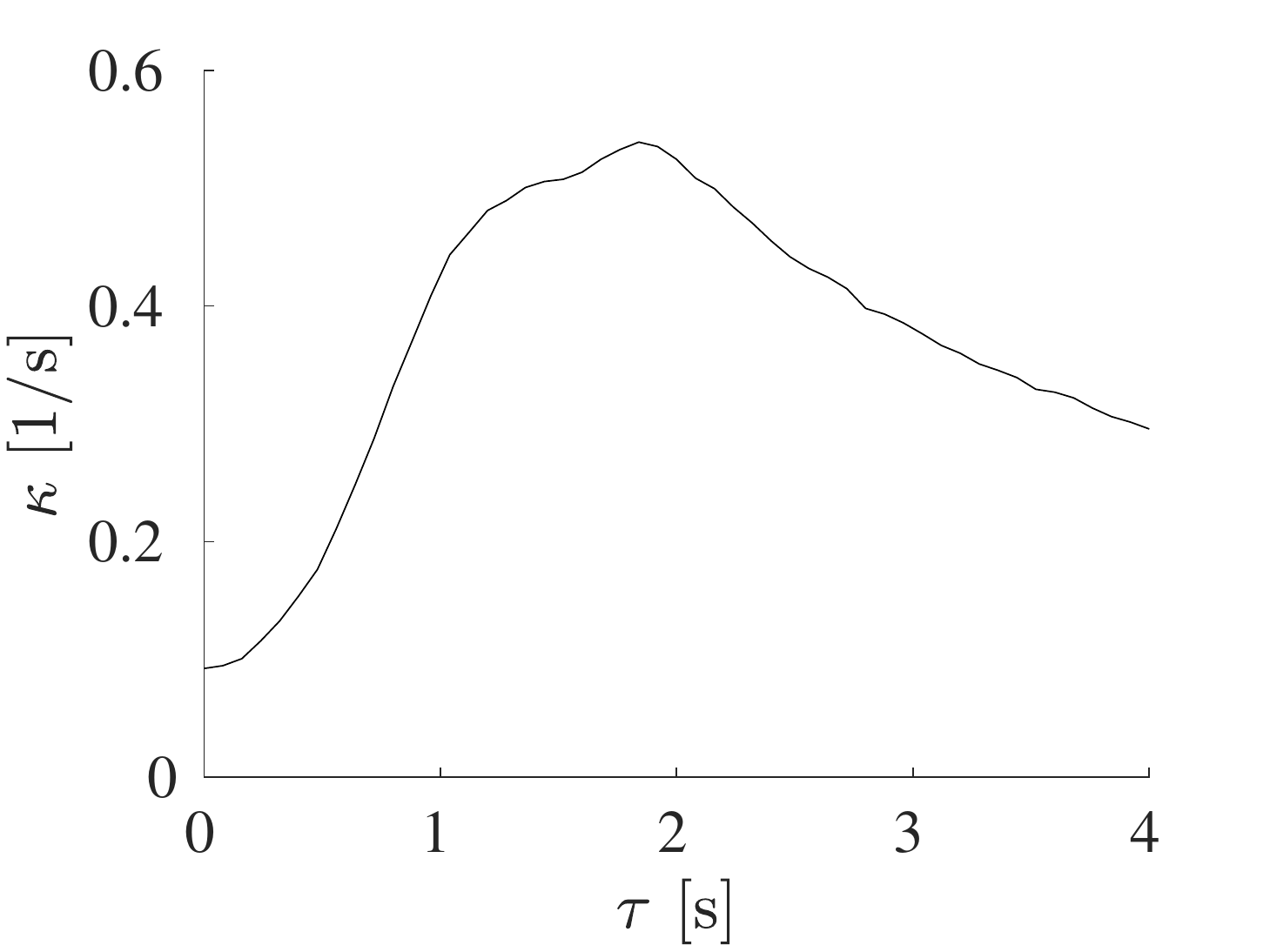}};
	\node[right=of img1, node distance=0.0cm, yshift=0cm,xshift=-1.0cm] (img2)
	{\includegraphics[width=0.44\columnwidth]{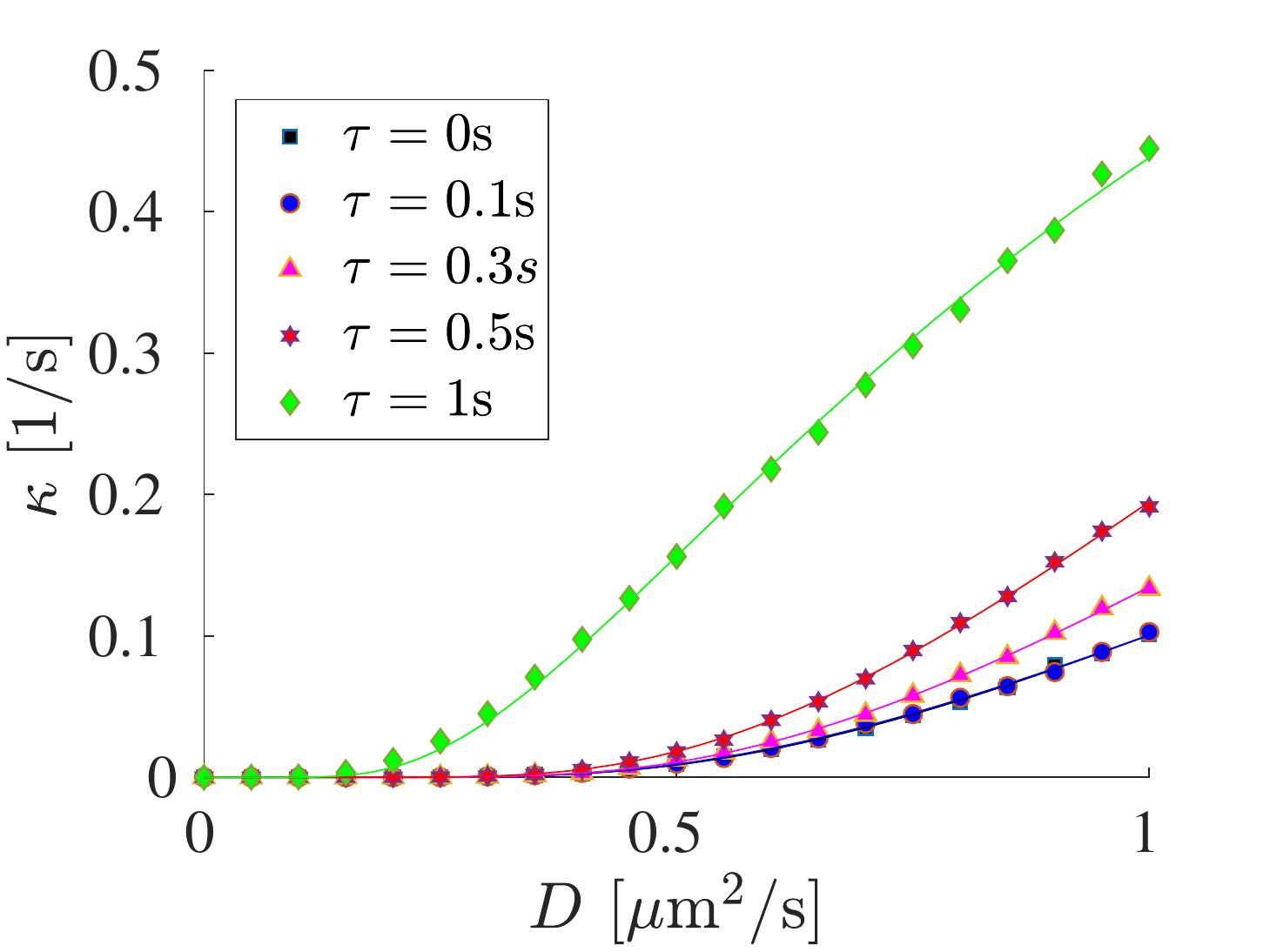}};
	\node[above=of img1, node distance=0.0cm, yshift=-1.8cm,xshift=-3.5cm] {a)};
	\node[above=of img2, node distance=0.0cm, yshift=-1.8cm,xshift=-3.5cm] {b)};
	\end{tikzpicture}	
	\caption{a) Steady-state transition rate for the trimer as a function of time delay obtained by BD simulations.   b) Steady state transition rate for the trimer as a function of the diffusion coefficient for several values of time delay. The curves were obtained by fitting the formula \eqref{eq: trimer tst} to the simulation data (symbols). The curves of $\tau=0\ $s and $\tau=0.1\ $s are lying on top of each other. Parameters used: $\omega= 1/$s, $R=5\ \mu$m, $D=1\ \mu$m$^2/$s. In BD simulations, we have averaged over $10^3$ trajectories with time step $dt=10^{-3}$ s. }	
	\label{fig:transition rate trimer}
\end{figure}

The resulting transition rate $\kappa$ as a function of the time delay $\tau$ is depicted in Fig.~\ref{fig:transition rate trimer} a). Therein, one can observe the four dynamical regimes described in Sec.~\ref{sec:structure}. For small delays the rate is approximately exponential, then the curvature of $\kappa(\tau)$ changes from positive to negative, after which the derivative of the rate starts to increase due to the appearance of the fourth dynamical regime where the particle hops between the individual minima of the potential, and for large delays in the unstable regime (iii) the rate drops, while the particle exhibits diverging fluctuations. The transition rate is qualitatively similar to that obtained for the dimer in Fig.~\ref{fig:transition rate depending on delay}. The only difference is that for the dimer we have not observed the fourth dynamical regime, because we have calculated the rate using the first-passage time method, which is insensitive to the potential shape beyond the boundary. 

For an approximate analytical treatment, we map the situation to a one-dimensional transition problem, to which we apply Kramers' theory. Specifically, we focus on the distance $r_\mathrm{b}$ of the particle 2 from the $x$-axis and construct an appropriate effective energy barrier $E_\mathrm{b,eff}$ and diffusion coefficient $D_\mathrm{eff}$. In the steady state, the three particles are most likely found at the vertices of an equilateral triangle. Using the notation of Sec.~\ref{Section Trimer}, we express the vector $\mathbf{r_b}$ from the particle 2 to the center of the abscissa $|13|$ as $\mathbf{r}_b = \mathbf{r}_{12} + \mathbf{r}_{31}/2$. Hence the Gaussian white noise $\bm{\eta}_\mathrm{b}(t)$ corresponding to the coordinate $\mathbf{r}_b$ is given by 
\begin{align}
	& \sqrt{4D} \left( \bm{\eta}_1(t) - \bm{\eta}_2(t) \right) + \frac{1}{2} \sqrt{4D} \left( \bm{\eta}_3(t) - \bm{\eta}_1(t) \right) 
	\equiv \sqrt{2 \left( 3D \right) } \bm{\eta}_\mathrm{b}(t) ,
\end{align}
where we have used the relations~\eqref{noise}. Based on these considerations, we approximate the effective diffusion coefficient by $D_\mathrm{eff}=3D$.

For the corresponding effective energy barrier, we make the ansatz $E_\mathrm{b,eff}=C(\tau) \gamma\omega R^2/12$. Namely, we first consider the minimum value $E_\mathrm{b}= k R^2/6 = \gamma\omega R^2/12$ of the energy difference between a configuration with all the three particles aligned ($V = kR^2/6$) and the configuration where the particles form an equilateral triangle with side length $R$ ($V = 0$), where $V$ is given by Eq.~\eqref{eq:total_energy}. The (possibly delay-dependent) unknown dimensionless prefactor $C(\tau)$ accounts for the time delay and all additional ways the particle 2 may take in the multidimensional energy profile in order to pass from $\left|1\right>$ to $\left|-1\right>$. The resulting Kramers' rate for the transition from $\left|1\right>$ to $\left|-1\right>$ thus reads
\begin{align}
	\kappa_{\mathrm{TST}} = a(\tau)
	\exp\left\{- \frac{C(\tau)\omega R^2}{36D} \right\},
	\label{eq: trimer tst}
\end{align}
where $a(\tau)$ is a further unknown prefactor that may depend on all parameters of the model [see, for example, Eq.~\eqref{eq:kramers} for Kramers' rate in a cusped potential].

\begin{table}
	\begin{center}
		\begin{tabular}{|l||c|c|c|}
			\hline
			\bf{Regime} &  $\tau$ [s] &  $C(\tau)$ [1] & $a(\tau)$  [s$^{-1}$]  \\
			\hline
			\hline
			\text{(i) exponential decay} & 0 &  3.52 $\pm$ 0.08 & 1.16 $\pm$ 0.09 \\ 
			\text{(i) exponential decay} & 0.1 & 3.45 $\pm$ 0.01 & 1.10 $\pm$ 0.01 \\ 
			\text{(i) exponential decay} & 0.3 & 3.65 $\pm$ 0.07 & 1.69 $\pm$ 0.10 \\ 
			\text{(ii) damped oscillations} & 0.5 & 3.41 $\pm$ 0.07 & 2.07 $\pm$ 0.12 \\ 
			\text{(iii) exponential divergence} & 1.0 & 1.48 $\pm$ 0.05 & 1.23 $\pm$ 0.05 \\
			\hline
		\end{tabular}
	\end{center}
	\label{tab:fit} 
	\caption{The phenomenological parameters $C(\tau)$ and $a(\tau)$ for five values of the time delay $\tau$ corresponding to the three different dynamical regimes of the trimer isomerization. The presented values were obtained by fitting the formula \eqref{eq: trimer tst} to the BD data shown in Fig.~\ref{fig:transition rate trimer} b).}
\end{table}

In order to test the formula \eqref{eq: trimer tst}, we have fitted it to the transition rate $\kappa$ obtained from BD simulations as a function of the diffusion coefficient $D$. The results for different values of the time delay are shown in Fig.~\ref{fig:transition rate trimer} b) and the corresponding values of the coefficients $C(\tau)$ and $a(\tau)$, obtained from the fits, are given in Tab.~\ref{tab:fit}. The presented results prove that the transition rate exhibits exponential increase with the diffusion coefficient for a relatively broad range of values of the time delay, and thus the $D$ -dependence of $\kappa$ can be relatively well described by the Kramers-type ansatz \eqref{eq:kramers}. 

Our attempts to include the effects of the delay in the effective diffusion coefficient and energy barrier, analogously to the approach described in Sec. \ref{sec:dimer_rate}, where we made the replacements $\omega\rightarrow\omega_{\tau,\mathrm{ss}}$ and $D\rightarrow D_{\tau,\mathrm{ss}}$, did not lead to a significant change of the value for $C(\tau)$. We thus conclude that the deviations of $C(\tau)$ from unity are mainly caused by the assumption that the particle will almost in all cases cross the axis at the minimum of the potential energy. Further improvement would thus require a multidimensional transition rate theory, which is clearly beyond the scope of the present paper. 

%% file: Content/Extension.tex
\section{Extensions to other memory kernels} \label{Sec.: Extension}

So far, we have considered only the interactions involving a given positive delay time. But how robust are our analytical findings? Do they critically hinge on (possibly artificial) model details and break down upon some minor variation of the model definition? It turns out that most of the presented results can be directly applied also to other models with delay or memory. To see this, note that the central equation \eqref{lin SDDE} is equivalent to the generalized Langevin equation (GLE)
\begin{equation}
	\dot{x} = -\omega \int_{0}^{t} \mathrm{d}t' \phi(t-t') x(t') + \sigma\eta(t)   \label{GLE}
\end{equation}
with positive frequency $\omega > 0$ and the memory kernel given by $\phi(t)=\delta(t-\tau)$, $\tau>0$. Deriving a time-local Langevin equation from the GLE \eqref{GLE} with arbitrary $\phi(t)$ along the lines of \ref{app:noise_sol}, we obtain Eq.~\eqref{time local Langevin} with $\lambda(t)$ being the Green's function for Eq.~\eqref{GLE}, i.e. solving \eqref{GLE} with the initial condition $\lambda(0) = 1$ and $\lambda(t) = 0$, $t<0$. Therefore, all our results that do not depend on the specific form of the Green's function can be readily generalized to arbitrary $\phi(t)$ after substituting the Green's function \eqref{eq:lambda_matrix} for $\phi(t)=\delta(t-\tau)$ by the Green's function corresponding to the chosen $\phi(t)$. In the rest of this section, we review some paradigmatic examples of memory kernels $\phi(t)$, to provide readers with a set of examples of the potential generalizations we have in mind.

The simplest generalization of systems with the memory kernel $\phi(t)=\delta(t-\tau)$ are systems with multiple different time delays with the memory kernel
\begin{equation}
	\phi(t) = \sum_{i=1}^{N} \omega_i \delta(t-\tau_i).
\end{equation}
Properties of these systems are studied in Ref.~\cite{giuggioli2016fokker}. 

A slightly unusual but interesting variation that makes sense in the context of active matter employs a negative delay. The individual active agents may react to a future state of their neighborhood which they predict in the basis of its present state. For idealized systems capable of a perfect prediction, the GLE (or, equivalently, the linear SDDE) contains the memory kernel $\phi(t)=\delta(t+\tau)$, $\tau>0$ and it can be solved using the strategy described in \ref{App:A1}. The resulting Green's function
\begin{equation}
\lambda(t) = \sum_{k=0}^{\infty} \frac{(-\omega)^k}{k!}(t+k\tau)^k.
\label{eq:Bruwier}
\end{equation}
is the so called Bruwier series \cite{bruwier1930vequation} that is convergent for $|\omega|<|e\tau|$. Alternatively, the series \eqref{eq:Bruwier} can be written in the form \cite{perron1939bruwiersche} 
\begin{equation}
\lambda(t) = \frac{e^{s t}}{1-\tau s},
\end{equation}
where $s$ is the absolute value of the smallest root of the equation $\rho=-\omega e^{\tau \rho}$.

More realistic predictive systems might instead only have an imperfect knowledge of the future position and anticipate a position $x_{\rm pre}(t + \tau) \neq x(t + \tau)$. Therefore, we reformulate the deterministic part of the GLE \eqref{GLE} as
\begin{equation}
\dot{x}(t) = -\omega x_\mathrm{pre}(t+\tau).
\label{eq:lin_int}
\end{equation}
One of the reasonable strategies for predicting $x_\mathrm{pre}$ is to use the linear extrapolation
\begin{equation}
x_\mathrm{pre}(t+\tau) = x(t) + \tau \dot{x}(t),
\label{Xpre}
\end{equation}
which is identical to a small delay expansion of $x(t+\tau)$. The equation of motion \eqref{eq:lin_int} then assumes the time local form
\begin{equation}
(1+\omega\tau)\dot{x}(t) = -\omega x(t)
\end{equation}
which is solved by $x(t) = x_0 \lambda(t)$ with the exponentially decaying Green's function
\begin{equation}
	\lambda(t) = \exp \bigg [ - \frac{\omega}{1+\omega\tau} (t-t_0) \bigg ].
\end{equation}
The rescaled frequency $\omega_{\rm r} = \omega/(1+\omega\tau)$ decreases with increasing delay and thus the resulting dynamics in general exhibits slower relaxation and larger fluctuations (variance) than a system with vanishing time delay. Note that for conventional time delays [now corresponding to $\tau < 0$ in Eq.~\eqref{eq:lin_int}
], the presented first order approximation predicts dynamics with rescaled frequency $\omega_{\rm r}$ increasing with increasing time delay (for $\omega\tau < - 1$). One might conclude that such dynamics would lead to a smaller variance $\nu_{ss}$ for non-zero delays than for a vanishing delay, which contradicts our exact result \eqref{eq:SS_variance}. However, performing the first-order expansion in the equation $\dot{x}(t) = -kx(t-\tau) + \sigma \eta(t)$ including the noise term, one finds that both the rescaled frequency and diffusion coefficients are smaller than in the original equation and the resulting variance equals the first-order expansion of the exact result given in Eq.~\eqref{eq:SS_variance}. For details and further discussion of Taylor expansions in SDDE, see Ref.~\cite{guillouzic1999small}.

The most frequently used generic form of the memory kernel is the exponential 
\begin{equation}
	\phi(t) = b \exp( -bt )
\end{equation}
which is obtained, for example, after integrating out the momentum in the Langevin equation for position of an underdamped harmonic oscillator. This memory kernel leads to the corresponding Green's function
\begin{equation}
	\lambda(t) = \exp\left( -\frac{bt}{2} \right) \left[\cos(\Omega t) + \frac{b}{2\Omega} \sin(\Omega t) \right],
\label{eq:lambda_exp}
\end{equation}
where $\Omega = \sqrt{\omega^2 - b^2/4}$, which is reminiscent of the Green's function \eqref{eq:xexp} for the system with conventional time delay [$\phi(t)=\delta(t-\tau)$]. The difference is that, while the Green's function \eqref{eq:lambda_exp} always decays exponentially with time, the Green's function \eqref{eq:xexp} allows also for negative relaxation times and thus an exponential increase with time. Properties of the Green's function \eqref{eq:lambda_exp}, as well as those corresponding to a power-law memory, are discussed in more detail in Ref.~\cite{chase2016langevin}. 

%% file: Content/Conclusion.tex
\section{Conclusion and outlook} \label{Sec.:Conclusion}

Inspired by the surging interest in self-organized active matter and, more specifically, the experiments of Khadka et al. \cite{khadka2018active}, we considered $N$ Brownian particles interacting via time-delayed harmonic interactions and confined to a plane, as depicted in Fig.~\ref{fig:modelsketch} in Sec.~\ref{Sec.:Dynamics}. The system is described by the set \eqref{eq:Langevin_n} of $2N$ non-linear delayed Langevin equations and hence its dynamics is non-Markovian. At long times, the particles form highly symmetric dynamical molecular-like structures, depicted in Fig.~\ref{fig:CompareBondNumber} a) in Sec.~\ref{sec:structure}, which become increasingly compact for large $N$.

We have analyzed small systems of $N=2$ (dimer) and $N=3$ (trimer) particles analytically finding molecules with nearest-neighbor distance given by the equilibrium spring length $R$. To this end, we linearized the corresponding Langevin equations around the zero-temperature steady-state configurations, or, equivalently, around the minimum of the potential energy \eqref{eq:total_energy}. The linearized Langevin equations could be solved analytically, leading to Gaussian stationary probability densities with delay-dependent effective parameters. In the appendices, we provide analytical expressions for mean values, covariance matrix and time-correlation matrix for a multidimensional system of linear delayed Langevin equations. For the dimer and trimer, we have compared our analytical predictions with Brownian dynamics simulations of the complete model \eqref{eq:Langevin_n}. We have found good quantitative agreement in the parameter regimes where the system evolves relatively close to its minimum energy configuration, and good qualitative agreement otherwise (see Sec.~\ref{Sec.:Dynamics}).

Our analytical results for the dimer and trimer imply that these structures are stable only for small enough values of the product $k \tau$, where $k$ denotes the stifness of the potential. More precisely, these systems converge either exponentially or by exponentially-damped oscillations to corresponding steady states, or they exhibit exponentially diverging oscillations. Our analysis of systems with $N\geq 2$ by BD simulations, described in Sec.~\ref{sec:structure}, reveals that these dynamical regimes are stable beyond the linearization approximation and for an arbitrary number of particles. Specifically, we have found that the stability actually extends to larger values of $k\tau$ than predicted from the linearized equations, the critical value of the product $k\tau$ decaying approximately as $1/N$. Therefore, larger systems are more unstable than smaller ones, and the dependence of the stability on the particle number almost vanishes after rescaling the potential stiffness as $k \to k/N$. We conjecture that these instabilities are induced by the chosen form of the interaction which has infinite range and diverges with increasing inter-particle distance. In contrast, the model with constant forces, considered in Ref.~\cite{khadka2018active}, did not lead to unstable behavior.

Interpreting the inter-particle interactions as an action of a feedback control mechanism, we have, in Sec.~\ref{sec:entropy_fluxes}, used our analytical results for the dimer and trimer to evaluate the amount of entropy extracted by the feedback from (or information injected to) the system  in order to maintain the non-equilibrium structures. Interestingly enough, the entropy fluxes do not depend on the noise amplitude $D$ and hence they are discontinuous at $D = 0$, where the steady-state structures are stable without feedback and thus the entropy fluxes vanish.

Assuming the particles to be distinguishable, the steady-state structures (molecules) can form different isomers.
Their transition dynamics can provide rich additional insight into the energy landscape underlying the non-equilirbium structure formation. For the dimer and trimer, we have investigated how and when the transitions between the individual isomers can be described by transition state theory. For the dimer, we have applied our analytical results, based on the time-convolutionless transform leading to the time-local Fokker-Planck equation (FPE) \eqref{Eq:FPE_W1_text}, to construct several analytical approximations for the transition rate using Kramers' theory \cite{kramers1940original, hanggi1990review} and Bullerjahn's theory \cite{bullerjahn201spectroscopy}. We have also calculated the transition rate from the FPE numerically. Finally, we have compared the obtained predictions to results of BD simulations of the full problem. While the FPE gives the exact value of the transition rate for vanishing delay, our results show that the obtained rates agree with the true ones for small and moderate values of the delay only. We conjecture that this is caused by the fact that the classical absorbing boundary used in our numerical and analytical evaluation of the transition rate can not be used for larger values of $\tau$. Concerning the analytical results, the best agreement with the true rates was obtained by the Bullerjahn's formula \eqref{eq:rateBtinfty} with effective barrier height and diffusion coefficient taken from the time-local FPE \eqref{Eq:FPE_W1_text} and the prefactor rescaled according to Eq.~\eqref{eq:scaled_rates}. In the case of the trimer, we have confirmed by BD simulations that the transition rate increases exponentially with the noise strength $D$ even for longer delays and thus Kramers' or Bullerjahn's type predictions can be used also in this case. We plan to further investigate suitable absorbing boundary conditions for delayed systems to predict (at least numerically) transition rates also for large delays.

Finally in Sec.~\ref{Sec.: Extension}, we have
considered the robustness of our analytical results with respect to details of the realization of the delay. We demonstrated that most of the presented equations can be used also for systems with memory kernels different from that for discrete time delays, i.e. $\phi(t) = \delta(t-\tau)$. It is enough to substitute the Green's function $\lambda(t)$ \eqref{eq:lambda_matrix} corresponding the delayed Langevin equation by the Green's function corresponding to the memory kernel of interest. We reviewed some paradigmatic memory kernels and provided an outlook on the differences and similarities of the corresponding Green's functions. A more detailed study is left for future work.

As a further extension of our work, it would be interesting to consider physically more realistic interactions that vanish at large distances. Furthermore, we plan to investigate the reaction of the studied system to an external perturbation. Of particular interest could be the propagation and decay behavior of a local perturbation through the system, especially in case of large numbers of particles. Last but not least, we aim to investigate the behavior of the studied system under the action of an additional deterministic time-dependent driving and study the corresponding stochastic dynamics and thermodynamics.

\section*{Acknowledgments}
We acknowledge funding by Deutsche Forschungsgemeinschaft ({DFG}) via SPP 1726/1. VH gratefully acknowledges support by the Humboldt foundation and by the Czech Science Foundation (project No. 17-06716S). DG acknowledges funding by International Max Planck Research Schools (IMPRS). Furthermore, we thank Thomas Ihle and Sarah Loos for discussion.

%% file: Content/Appendix.tex
\appendix

\section{Solution of the Noiseless Problem}
\label{App:A1}

In this appendix we solve the multi-dimensional linear delay differential equation (LDDE)
\begin{equation}
	\dot{\mathbf{x}}(t) = - \omega\mathbf{x}(t-\tau ),
	\label{eq:LDSDE}
\end{equation}
where $\omega$ is a positive semi-definite matrix with real entries and $\mathbf{x}(t)$ is a column vector. Laplace transformation 
of this equation leads to the formula
\begin{equation}
s \tilde{\mathbf{x}}(s) - \mathbf{x}_0 = -\omega e^{-s \tau} \left[ \tilde{\mathbf{x}}(s) + \int_{-\tau}^0 \mathrm{d}t\ e^{-s t} \mathbf{x}(t) \right]
\end{equation}
with $\mathbf{x}_0\equiv \mathbf{x}(0)$. The solution of this equation for the Laplace transformed variable reads
\begin{multline}
\tilde{\mathbf{x}}(s) = \int_0^\infty \mathrm{d}t\ e^{-s t} \mathbf{x}(t) = \bigg(s \mathcal{I} + \omega e^{-s\tau}\bigg)^{-1} \bigg[ \mathbf{x}_0 - \omega e^{-s \tau} \int_{-\tau}^0 \mathrm{d}t\ e^{-s t} \mathbf{x}(t) \bigg]\\
=  \sum_{k=0}^{\infty} \frac{(-\omega)^k}{s^{k+1}} e^{-k s \tau} \bigg[ \mathbf{x}_0 - \omega e^{-s \tau} \int_{-\tau}^0 \mathrm{d}t\ e^{-s t} \mathbf{x}(t) \bigg],
\label{eq:xLT}
\end{multline}
where $\mathcal{I}$ denotes the identity matrix. In the last step, we have expanded the inverse matrix using the Neumann series. The inverse Laplace transform of the ratio 
$e^{-s \tau}/s^{k+1}$ is given by $(t-\tau)^k \theta(t-\tau)/k!$ \cite{Olver2010} and thus the formula~\eqref{eq:xLT} can be inverted. Finally, the solution of Eq.~\eqref{eq:LDSDE} is given by
\begin{equation}
	\mathbf{x}(t) = \lambda(t)\mathbf{x}_0 - \omega\int_{-\tau}^{0} \mathrm{d}s\ \lambda(t-s-\tau) \mathbf{x}(s),
	\label{eq:solution}
\end{equation}
where
\begin{equation}
	\lambda(t) = \sum_{k=0}^{\infty} \frac{(- \omega)^k}{k!} \left(t - k\tau\right)^k \theta (t-k\tau)
	\label{eq:lambda_matrix}
\end{equation}
is a matrix-valued function which solves Eq.~\eqref{eq:LDSDE} with the initial condition $\mathbf{x}(t)=0$ for all $t<0$ and $\mathbf{x}(0)=\mathcal{I}$. In the present paper, we always assume that the system is initialized at time $t$ at position $\mathbf{x}_0$ witha special history, namely $\mathbf{x}(t) = 0$ for all $t<0$, allowing us to simplify Eq.~\eqref{eq:solution} to $\mathbf{x}(t) = \lambda(t)\mathbf{x}_0$. 

The only fixed point of the delay differential equation~\eqref{eq:LDSDE} is $\mathbf{x}(t)=0$. In order to investigate its stability, it is useful to present an alternative 
solution of the LDDE~\eqref{eq:LDSDE} using the exponential ansatz $\mathbf{x}(t) \propto \exp\left(-\alpha t\right)$. Inserting this ansatz into Eq.~\eqref{eq:LDSDE} leads to the equation $\alpha = \omega\exp\left(\alpha \tau\right)$ for the matrix $\alpha$. Except for some notable exceptions \cite{Corless2007}, the solution of this equation is given by
\begin{equation}
\alpha = - \frac{1}{\tau} W\left(- \tau \omega\right),
\label{eq:lambert}
\end{equation}
where $W$ denotes the matrix valued Lambert $W$ function \cite{Yi2010}. The Lambert $W$ function is a multivalued complex function. The long-time behavior of solutions to Eq.~\eqref{eq:LDSDE} and thus also the stability of its fixed point are determined by the branch of $W$ yielding the largest real parts of the eigenvalues of the matrix $\alpha$. The corresponding values of the Lambert $W$ function 
strongly depend on the reduced delay $\tau \omega$.

\begin{figure}[ht!]
\centering
	\begin{tikzpicture}
	\node (img1)  {\includegraphics[width=0.4\columnwidth]{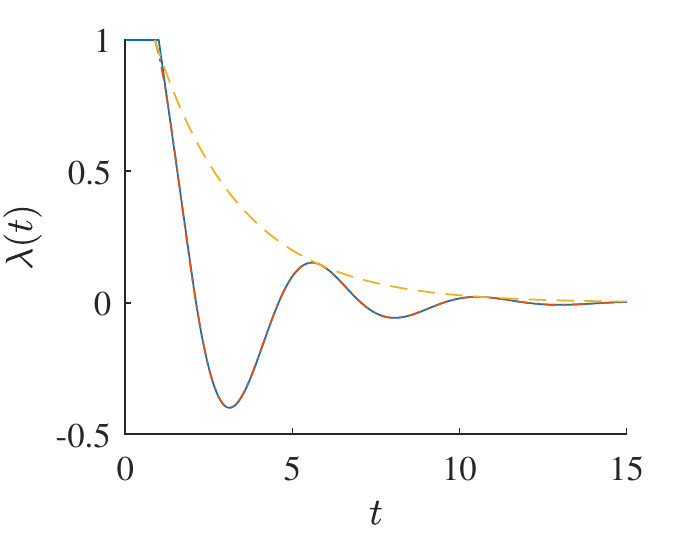}};
	\node[right=of img1, node distance=0.0cm, yshift=0cm,xshift=-1.0cm] (img2)
	{\includegraphics[width=0.4\columnwidth]{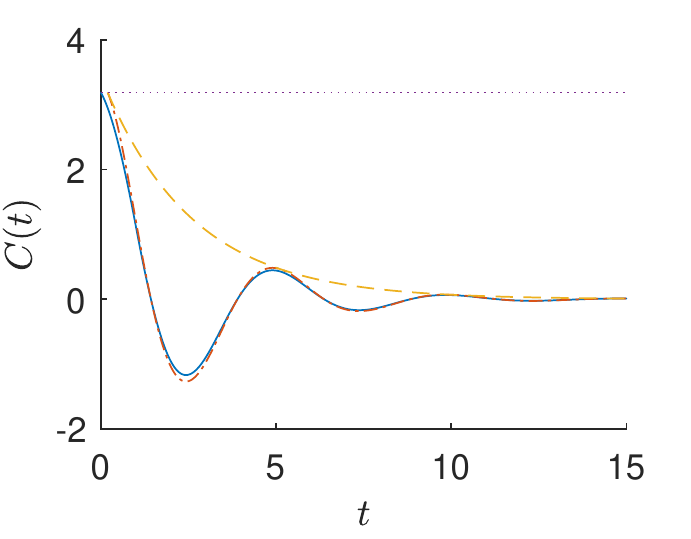}};
	\node[below=of img1, node distance=0.0cm, yshift=1cm,xshift=0.0cm] (img3)
	{\includegraphics[width=0.4\columnwidth]{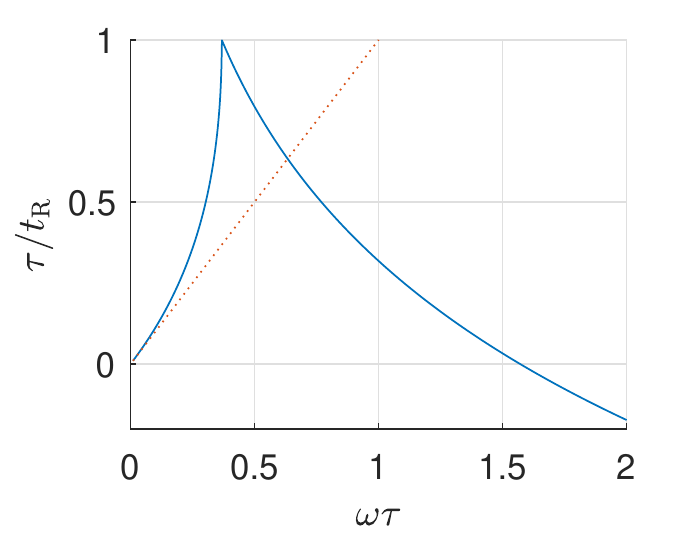}};
	\node[below=of img2, node distance=0.0cm, yshift=1cm,xshift=0.0cm] (img4)
	{\includegraphics[width=0.4\columnwidth]{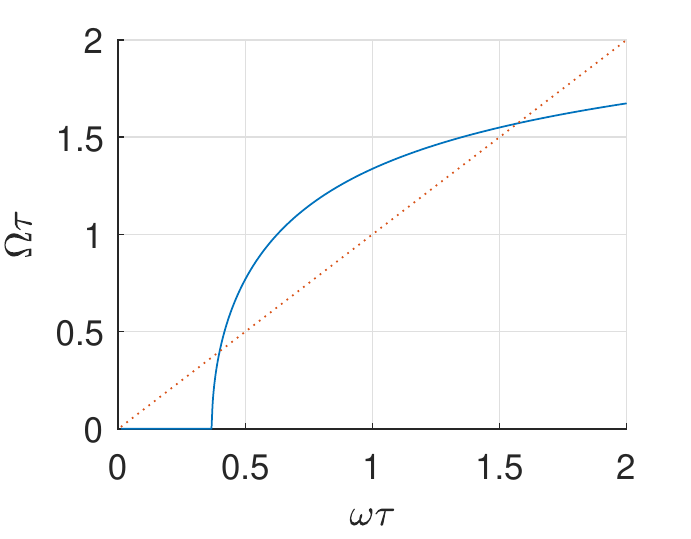}};
	\node[above=of img1, node distance=0.0cm, yshift=-1.8cm,xshift=-2.9cm] {a)};
	\node[above=of img2, node distance=0.0cm, yshift=-1.8cm,xshift=-2.9cm] {b)};	
	\node[above=of img3, node distance=0.0cm, yshift=-1.8cm,xshift=-2.9cm] {c)};
	\node[above=of img4, node distance=0.0cm, yshift=-1.8cm,xshift=-2.9cm] {d)};	
	\end{tikzpicture}	
\caption{Solid lines in the panels a) and b) show the Green's function $\lambda(t)$ \eqref{eq:lambda_matrix} for the one-dimensional case of the LDDE~\eqref{eq:LDSDE} and the time-correlation function $C(t)$ \eqref{eq:correlation_multiple_ode} for the one-dimensional linear SDDE~\eqref{lin SDDE} as functions of time, respectively. The dot-dashed lines describing the long-time behavior of these variables were calculated using the exponential solution~\eqref{eq:xexp} of Eq.~\eqref{eq:LDSDE} in the form $A \exp\left(-t/{t_{\rm R}}\right)\cos\left(\Omega t + \phi_0\right)$. The dashed lines delineate the overall exponential decay $\exp\left(-t/{t_{\rm R}}\right)$ of $\lambda(t)$ and $C(t)$. The dotted line in panel b) depicts the stationary value $C(0)$ of the variance given by Eq.~\eqref{eq:C_0_K_inf}. The lifetime $t_{\rm R}^{-1} = \Re \left({\alpha}\right)$ and the frequency $\Omega = \left|\Im \left({\alpha}\right)\right|$ of the oscillations in $\lambda(t)$ and $C(t)$ are shown in panels c) and d), respectively. The dotted lines in these panels serve just as an eye-guide depicting the linear behavior $\tau/t_{\rm R} = \Omega \tau = \omega\tau$. For the purposes of the appendix, we assume that the individual model parameters are dimensionless. Parameters used: $\omega = 0.9$ and $\tau = \sigma^2/2 = 1$.}	
	\label{fig:properties_of_lambda}	
\end{figure}

For example, in one dimension, where $\omega$ is a positive real number, the branch of the Lambert $W$ function with the largest real part
exhibits three qualitatively different regimes as a function of $\tau \omega$ leading to three different dynamical regimes of solutions
\begin{equation}
x(t) \propto \Re \left[ \exp(-\alpha t) \right] = \exp(-t/t_{\rm R})\cos\left(\Omega t\right),
\label{eq:xexp}
\end{equation}
$1/t_{\rm R} = \Re(\alpha)$, $\Omega = \left|\Im(\alpha)\right|$, to Eq.~\eqref{eq:LDSDE}. The boundaries between these regimes can be determined analytically \cite{bellman1963delay,michiels2007stability, guillouzic1999small}:
(i) $0<\tau \omega\le 1/e$, where $\alpha$ is real and positive and $x(t)$ decays exponentially to $0$ with lifetime $t_{\rm R}$; (ii) $1/e < \tau \omega< \pi/2$, where $\alpha$ is complex with a positive real part, producing exponentially damped oscillations of $x(t)$ with frequency $\Omega$ and lifetime $t_{\rm R}$; and (iii) $\pi/2< \tau \omega$, where $\alpha$ is complex with a negative real part, corresponding to exponentially diverging oscillations of $x(t)$ with frequency $\Omega$. For $\tau \omega = \pi/2$, $1/t_{\rm R} = 0$ and $\lambda(t)$ oscillates with frequency $\Omega$ without any decay.

In the panel a) of Fig.~\ref{fig:properties_of_lambda}, we show that for long times the Green's function $\lambda(t)$ for Eq.~\eqref{eq:LDSDE} is well approximated by the exponential solution~\eqref{eq:xexp}. The above described dynamical regimes are reflected in the behavior of the decay rate $1/t_{\rm R}$ and the frequency $\Omega$ of oscillations, plotted in the panels c) and d), respectively. The panel b) of the figure shows the steady auto correlation of $x(t)$ calculated in \ref{app:time_corr_matr}, which is also well described by the formula \eqref{eq:xexp}.

\section{Solution with Noise}
\label{App:B}
\subsection{One Dimension}
\label{app:noise_sol}

Let us now solve Eq.~\eqref{eq:LDSDE} with an additional noise term. For simplicity, we present the detailed derivation first in one dimension. We thus want to solve the equation
\begin{equation}
	\dot{x}(t) = - \omega x(t-\tau) + \sigma \eta(t),
	\label{lin SDDE}
\end{equation}
where $\eta(t)$ is a white noise fulfilling $\left<\eta(t)\right> = 0$ and  $\left<\eta(t)\eta(t')\right> = \delta(t-t')$. Due to the time delay, this is a time-nonlocal (and consequently non-Markovian) Langevin equation, which, however, can be transformed into a time-local Langevin equation with a colored noise via the so-called time-convolutionless transform \cite{giuggioli2016fokker}. The time-local equation can then be used to derive the Fokker-Planck equation for the PDFs for $x(t)$. In order to do that, we write the formal solution of Eq.~\eqref{lin SDDE}  as
\begin{equation}
	x(t) = x_0\lambda(t) + \sigma \int_0^t \mathrm{d}s\ \lambda(t-s) \eta(s) + \psi(t),  \label{solution SDDE}
\end{equation}
where $\lambda(t)$ is given by Eq.~\eqref{eq:lambda_matrix} and $\psi(t) = -\omega\int_{-\tau}^0 \mathrm{d}s\ \lambda (t-\tau-s)x(s)$ is determined by the initial condition $x(t)$ for $t<0$. As we show in the next section, the formula \eqref{solution SDDE} can already be used for the calculation of the time-correlation function for $x(t)$. Here, we differentiate the solution \eqref{solution SDDE} which yields the time-local Langevin equation
\begin{equation}
	\dot{x}(t) = -\omega_\tau(t) x(t) + b(t) + \sigma \xi(t).
\label{time local Langevin}
\end{equation}
Due to the time-nonlocal nature of Eq.~\eqref{lin SDDE}, the potential in the time-local equation \eqref{time local Langevin} possess the time-dependent stiffness
\begin{equation}
	\omega_\tau(t) \equiv -\frac{\dot{\lambda}(t)}{\lambda(t)}, \label{A(t)}
\end{equation} 
and the time-varying position of the minimum $- b(t)/\omega_\tau(t)$, with $b(t) \equiv \omega_\tau(t) \psi(t) + \dot{\psi}(t)$ vanishing for the special initial condition $x(t) = 0$ for all $t<0$. Furthermore, Eq.~\eqref{time local Langevin} includes the Gaussian colored noise $\xi(t) = \lambda(t) \frac{\mathrm{d}}{\mathrm{d}t} \int_0^t \mathrm{d}s \frac{\lambda(t-s)}{\lambda(t)} \eta(s)$ which satisfies $\langle \xi(t)\rangle =  0$ and
\begin{equation}
	\langle \xi(t)\xi(t')\rangle = \lambda(t) \lambda(t') \frac{\mathrm{d}}{\mathrm{d}t'} \int_0^{t'} \mathrm{d}s\  \frac{\mathrm{d}}{\mathrm{d}t} \frac{\lambda(t-s)\lambda(t'-s)}{\lambda(t) \lambda(t')},\quad t'<t.
\end{equation}
While Markov processes are completely determined by the the transition probability density $P_1(x,t|x_0,t_0)$ for going from the initial state $x_0$ at time $t_0$ to the final state $x$ at time $t$, non-Markov processes in general require a full hierarchy of joint probability densities. Nevertheless, similarly to the Markovian case, the Gaussian non-Markov process~\eqref{time local Langevin} is  completely determined by the joint probability distribution $P_2(x,t;x',t'|x_0,0)$ \cite{giuggioli2016fokker}.

The FPEs for the one- and two-time probability distributions $P_1(x,t|x_0,0)=\langle\delta[x-x(t)]\rangle$ and $P_2(x,t;x',t'|x_0,0)=\langle\delta[x-x(t)]\delta[x'-x(t')]\rangle$, where the averages are taken over all realizations of the process $x(t)$ departing from state $x_0$ at time $0$, are found to be
\begin{align}
	\frac{\partial}{\partial t}P_1 &= \frac{\partial}{\partial x} \bigg[ \omega_\tau(t)x - b(t) + 2D_\tau(t)\frac{\partial}{\partial x}\bigg] P_1,  \label{FPE W1}\\
	\frac{\partial}{\partial t}P_2 &= \frac{\partial}{\partial x} \bigg[ \omega_\tau(t)x - b(t) + c(t,t')\frac{\partial}{\partial x} + 2D_\tau(t)\frac{\partial}{\partial x}\bigg] P_2.   \label{FPE W2}
\end{align}
Similarly as the trap stiffness $\omega_\tau(t)$, also the effective diffusion coefficient, corresponding to the time-local description,  is time dependent if $\tau>0$. It reads
\begin{align}
	2D_\tau(t) \equiv& \frac{\sigma^2\lambda(t)^2}{2} \frac{\mathrm{d}}{\mathrm{d}t} \int_0^t \mathrm{d}s \frac{\lambda(s)^2}{\lambda(t)^2} 
	= \sigma^2 \left( \frac{\lambda(t)^2}{2} + \omega_\tau(t) \int_{0}^{t} \mathrm{d}s \lambda(s)^2 \right)  \nonumber\\
	=& \sigma^2\lambda(t)^2/2 + \omega_\tau(t) \nu(t)
\label{D(t)}
\end{align}
with the variance $\nu(t) = \sigma^2 \int_{0}^{t} \mathrm{d}s \lambda(s)^2$ and
$c(t,t') \equiv \sigma^2 \lambda(t) \frac{\mathrm{d}}{\mathrm{d}t} \int_0^{t'} \mathrm{d}s \lambda(t-s)\lambda(t'-s)/\lambda(t)$.

Because of the oscillatory nature of $\lambda(t)$ in the dynamical regimes (ii) and (iii), the coefficients $\omega_\tau$, $b$, $c$ and $D_\tau$ in the FPEs \eqref{FPE W1} and \eqref{FPE W2} change their signs and they can even diverge. These divergences, however, always mutually balance each other such that the solutions of the FPEs, as given by the Eqs.~\eqref{eq:W_1_prob_distr} and \eqref{eq:W_2_prob_distr}, are always reasonable \cite{giuggioli2016fokker}.

\subsection{Higher Dimensions}
\label{app:noise_sol3}

Let us now consider the problem
\begin{equation}
	\dot{\mathbf{x}}(t) = - \omega\mathbf{x}(t-\tau) + \sigma \bm{\eta}(t),
	\label{eq:lin_SDDE_multiple}
\end{equation}
with general matrices $\omega$ and $\sigma$ and the vector $\bm{\eta}(t)$ of white noises fulfilling $\left<\bm{\eta}(t)\right> = 0$ and $\left<\eta_i(t)\eta_j(t')\right> = \delta_{ij}\delta(t-t')$. Since this system of Langevin equations is linear, the one- and two-time probability distributions $P_1(\mathbf{x},t|\mathbf{x}_0 ,0 )$ and $P_2(\mathbf{x},t;\mathbf{x}',t'|\mathbf{x}_0 ,0 )$ for $\mathbf{x}(t)$ defined in the preceding section must be Gaussian \cite{Risken1996} as in the one-dimensional case and also the corresponding Fokker-Planck equations can be derived along similar lines as in one dimension. Instead of deriving these equations, we now provide a simpler alternative derivation of the properties of the Gaussian distribution
\begin{equation}
	P_1(\mathbf{x},t|\mathbf{x}_0,0) = \frac{1}{\sqrt{ (2\pi)^3 \det \mathcal{K}(t)}} \exp\bigg\{ -\frac{1}{2} \big(\mathbf{x} - \bm{\mu}(t) \big)^{\intercal}\cdot \mathcal{K}(t)^{-1} \cdot \big(\mathbf{x} - \bm{\mu}(t) \big) \bigg\}
	\label{eq:Gaussian_multiple}
\end{equation}
based solely on the formal solution of the Langevin system~\eqref{eq:lin_SDDE_multiple} 
\begin{equation}
\mathbf{x}(t) = \lambda(t) \mathbf{x}_0 + \int_0^t \mathrm{d}s\ \lambda(t-s) \sigma \bm{\eta}(s)
\label{eq:formal_sol_multiple}
\end{equation}
with the initial condition $\mathbf{x}(t) = 0$ for all $t<0$ and $\mathbf{x}(0) = \mathbf{x}_0$ and with the Green's function $\lambda(t)$ of the Langevin system given by Eq.~\eqref{eq:lambda_matrix}. 

The mean value $\bm{\mu}(t) = \langle \mathbf{x}(t) \rangle$ and the elements $\mathcal{K}_{ij}(t)$ of the covariance matrix $\mathcal{K}(t) = \left<\mathbf{x}(t) \mathbf{x}^{\intercal}(t) \right> - \left<\mathbf{x}(t)\right> \left< \mathbf{x}^{\intercal}(t) \right>$ defining the PDF~\eqref{eq:Gaussian_multiple} can be obtained by inserting $\mathbf{x}_i(t)$ from Eq.~\eqref{eq:formal_sol_multiple} into the definitions and averaging over the noise $\eta(t)$. The results are
\begin{equation}
\bm{\mu}(t) = \lambda(t)\mathbf{x}_0
\label{eq:mean_multiple}
\end{equation}
and
\begin{equation}
	\mathcal{K}(t) = \int_{0}^{t} \mathrm{d}s\ \lambda(t-s) \sigma \sigma^{\intercal}\lambda^{\intercal}(t-s). \label{eq:covariance_multiple}
\end{equation}
These formulas can be generalized straightforwardly to arbitrary initial conditions, where $\mathbf{x}(t)$ for $t\le 0$ is drawn from some probability distribution $P[\mathbf{x}(t),t\le 0]$. Then the formal solution of the system~\eqref{lin SDDE} reads 
\begin{eqnarray}
\mathbf{x}(t) &=& \mathbf{y}(t) + \int_0^t \mathrm{d}s\ \lambda(t-s) \sigma \bm{\eta}(s),\\
	\mathbf{y}(t) &=& \lambda(t) \mathbf{x}(0) - \omega\int_{-\tau}^{0} \mathrm{d}s\ \lambda(t-s-\tau) \mathbf{x}(s),
\end{eqnarray}
The mean value $\bm{\mu}(t)$ is given by $\bm{\mu}(t) = \left<\mathbf{y}(t)\right>_{\mathbf{x}(t\le 0)}$,
and $\mathcal{K}(t) =  \left<\mathbf{y}(t)\left[\mathbf{y}(t)\right]^{\intercal} \right>_{\mathbf{x}(t\le 0)} + \int_{0}^{t} \mathrm{d}s\ \lambda(t-s) \sigma \sigma^{\intercal}\lambda^{\intercal}(t-s)$. The averages $\left<\bullet\right>_{\mathbf{x}(t\le 0)}$ above are taken with respect to the PDF $P[\mathbf{x}(t),t\le 0]$. The long-time behavior of the covariance matrix \eqref{eq:covariance_multiple} is studied in the next section of this appendix.

\section{Time-Correlation Matrix and Stationary Covariance Matrix}
\label{app:time_corr_matr}

The coefficients in the Gaussian two-time PDF $P_2(\mathbf{x},t;\mathbf{x}',t'|\mathbf{x}_0 ,0)$ can be obtained in a similar manner as in \ref{app:noise_sol3}. Here, we calculate only the stationary space-time correlation matrix $\mathcal{C}(t) = \lim_{s\to \infty}\left[\left<\mathbf{x}(s+t) \mathbf{x}^{\intercal}(s) \right>-\left<\mathbf{x}(s+t)\right> \left<\mathbf{x}^{\intercal}(s) \right>\right] = \lim_{s\to \infty}\left<\mathbf{x}(s+t) \mathbf{x}^{\intercal}(s) \right>$ which exists only if $\lim_{s\to \infty} \left<\mathbf{y}(s)\right> = \lim_{s\to \infty}\left<\mathbf{x}(s)\right> = 0$. Its matrix elements can be calculated in an analogous fashion as the elements of $\mathcal{K}(t)$. The result
\begin{equation}
\mathcal{C}(t) = \lim_{s\to\infty} \int_{0}^{s} \mathrm{d}s'\ \lambda(s+t-s') \sigma \sigma^{\intercal}\lambda^{\intercal}(s-s')
\label{eq:correlation_multiple}
\end{equation}
can be evaluated numerically. It is possible to rewrite it in a simpler form. Taking the derivative of $\mathcal{C}(t)$ with respect to $t$ reveals that for $t>0$ the correlation matrix obeys the same delay differential equation as the Green's function $\lambda(t)$, i.e.
\begin{equation}
\dot{\mathcal{C}}(t) = -\omega\mathcal{C}(t-\tau),\quad t>0.
\label{eq:correlation_multiple_ode}
\end{equation}
The restriction $t>0$ for validity of this equation comes from discontinuity (and thus non-differentiability) of $\lambda(t)$ at $t=0$. The solution to Eq.~\eqref{eq:correlation_multiple_ode} is given by Eq.~\eqref{eq:solution} and hence the stationary space-time correlation matrix can be written as
\begin{equation}
	\mathcal{C}(t) = \lambda(t)\mathcal{C}_0 - \omega\int_{-\tau}^{0} \mathrm{d}s\ \lambda(t-s-\tau) \mathcal{C}(s),
	\label{eq:correlation_multiple_best}
\end{equation}
where $\mathcal{C}_0 = \mathcal{C}(0) \lim_{t\to \infty}\mathcal{K}(t)$ is given by the long time limit of the covariance matrix. The stationary correlation matrix is thus solely determined by the unknown initial condition $\mathcal{C}(t)$, $t\in [-\tau,0]$. Fortunately, this initial condition can be calculated using the approach of Frank et al. \cite{frank2003fokker} who calculated the time-correlation function $\mathcal{C}(t)$ for $t\in[0,\tau]$ in one-dimension (see also Ref.~\cite{loos2017force}). They utilized the symmetry $\mathcal{C}(t) = \mathcal{C}(-t)$ following from Eq.~\eqref{eq:correlation_multiple} to rewrite the delay differential equation~\eqref{eq:correlation_multiple_ode} as $\dot{\mathcal{C}}(t) = -\omega\mathcal{C}(\tau-t)$ for $t \in (0,\tau)$. Taking the derivative of this equation and using Eq.~\eqref{eq:correlation_multiple_ode} yields the second order \emph{ordinary} differential equation 
\begin{equation}
\ddot{\mathcal{C}}(t) = - \omega^2 \mathcal{C}(t),\quad t \in (0,\tau)
\label{eq:correlation_multiple_best_initial_eq}
\end{equation}
for the initial condition $\mathcal{C}(t)=\mathcal{C}(-t)$, $t\in [-\tau,0]$. The solution of this equation reads
$\mathcal{C}(t) = \mathcal{C}_0 \cos(\omega t) + \dot{\mathcal{C}}_0 \omega^{-1} \sin(\omega t)$, where we still need to determine the unknown coefficients $\mathcal{C}_0 = \mathcal{C}(0)$ and $\dot{\mathcal{C}}_0 = \lim_{t\to 0+}\dot{\mathcal{C}}(t)$. To this end, we need to evaluate independently $\mathcal{C}(t)$ and/or $\dot{\mathcal{C}}(t)$ for two times $t\in(0,\tau)$. Specifically, we show at the end of this appendix that $\mathcal{C}(\tau) = 0.5 \omega^{-1}\sigma \sigma^{\intercal}$ and $\dot{\mathcal{C}}_0 = - 0.5 \sigma \sigma^{\intercal}$. Using these results, the final expression for the correlation matrix for $t\in [-\tau,\tau]$ reads
\begin{equation}
\mathcal{C}(t) = \mathcal{C}_0 \cos(\omega t) - 0.5 \sigma \sigma^{\intercal} \omega^{-1} \sin(\omega |t|)
\label{C_tau}
\end{equation}
with the initial value
\begin{equation}
\mathcal{C}(0) = \mathcal{C}_0 = \lim_{s\to\infty} \mathcal{K}(s)
= \frac{1}{2} \left[
\omega^{-1} \sigma \sigma^{\intercal} + 
\sigma \sigma^{\intercal} \omega^{-1} \sin(\omega\tau)
\right] \cos^{-1}(\omega\tau)
\label{eq:C_0_K_inf}
\end{equation} 
given by the stationary value of the covariance matrix. The whole time-dependence of $\mathcal{C}(t)$ for $t \ge 0$ is thus described by the formulas \eqref{eq:correlation_multiple_best}, \eqref{C_tau}, and \eqref{eq:C_0_K_inf}. The correlation matrix for negative times then follows from the symmetry $\mathcal{C}(t)=\mathcal{C}(-t)$. Finally, let us note than in the one-dimensional case, where $\omega$ and $\sigma$ stand for real numbers and, more generally, if the matrices $\omega^{-1}$ and $\sigma \sigma^{\intercal}$ commute, we can rewrite Eq.~\eqref{eq:C_0_K_inf} as
\begin{equation}
\mathcal{C}_0 = \frac{\sigma \sigma^{\intercal}}{2\omega} \left[
\mathcal{I}  + 
\sin(\omega\tau)
\right] \cos^{-1}(\omega\tau),
\label{eq:C_0_K_inf_scalar}
\end{equation} 
where $\mathcal{I}$ denotes the identity matrix. An example of the time-correlation function for a one dimensional system is depicted in Fig.~\ref{fig:properties_of_lambda} b) in \ref{App:A1}.

The expression for $C(\tau)$ can be obtained by multiplying Eq.~\eqref{eq:lin_SDDE_multiple} by $\mathbf{x}^{\intercal}(s)$, averaging the result over the noise, using the assumed stationarity of the process
implying the formula $\dot{\mathcal{C}}(0) = \mathrm{d} \left[\lim_{s\to\infty} \left<\mathbf{x}(s)\mathbf{x}^{\intercal}(s) \right>\right]/\mathrm{d}s = 0$, and applying the symmetry $\mathcal{C}(t)=\mathcal{C}(-t)$. The result is $\mathcal{C}(\tau) = \omega^{-1}\sigma\left<\bm{\eta}(s)\mathbf{x}^{\intercal}(s)\right> = 0.5 \omega^{-1}\sigma \sigma^{\intercal}$, where the last equality comes after inserting the formal solution \eqref{eq:formal_sol_multiple} for $\mathbf{x}(t)$ into the average, using the covariance of the noise, and noticing that in the resulting integral we integrate over half of the emerging $\delta$-function only. The expression for $\dot{\mathcal{C}}_0$ then comes simply from Eq.~\eqref{eq:correlation_multiple_ode}, which is invalid for $t = 0$, but can be used for $t$ arbitrarily close to 0 from the right, and using the result for $\mathcal{C}(\tau)$.

%% file: main.bbl
\providecommand{\newblock}{}
\begin{thebibliography}{10}
\expandafter\ifx\csname url\endcsname\relax
  \def\url#1{{\tt #1}}\fi
\expandafter\ifx\csname urlprefix\endcsname\relax\def\urlprefix{URL }\fi
\providecommand{\eprint}[2][]{\url{#2}}

\bibitem{cavagna2010starling}
Cavagna A, Cimarelli A, Giardina I, Parisi G, Santagati R, Stefanini F and
  Viale M 2010 {\em Proceedings of the National Academy of Sciences\/} {\bf
  107} 11865--11870

\bibitem{ben1994bacteria}
Ben-Jacob E, Schochet O, Tenenbaum A, Cohen I, Czirok A and Vicsek T 1994 {\em
  Nature\/} {\bf 368} 46

\bibitem{zhang2010bacteria}
Zhang H~P, Be'er A, Florin E~L and Swinney H~L 2010 {\em Proceedings of the
  National Academy of Sciences\/} {\bf 107} 13626--13630

\bibitem{elgeti2015swimmers}
Elgeti J, Winkler R~G and Gompper G 2015 {\em Reports on Progress in Physics\/}
  {\bf 78} 056601

\bibitem{ramaswamy2010mechanics}
Ramaswamy S 2010 {\em Annual Review of Condensed Matter Physics\/} {\bf 1}
  323--345

\bibitem{vicsek2012collective}
Vicsek T and Zafeiris A 2012 {\em Physics Reports\/} {\bf 517} 71--140

\bibitem{vicsek1995model}
Vicsek T, Czir{\'o}k A, Ben-Jacob E, Cohen I and Shochet O 1995 {\em Physical
  Review Letters\/} {\bf 75} 1226

\bibitem{attanasi2014information}
Attanasi A, Cavagna A, Del~Castello L, Giardina I, Grigera T~S, Jeli{\'c} A,
  Melillo S, Parisi L, Pohl O, Shen E {\em et~al.\/} 2014 {\em Nature
  Physics\/} {\bf 10} 691

\bibitem{piwowarczyk2018collective}
Piwowarczyk R, Selin M, Ihle T and Volpe G 2018 {\em arXiv preprint
  arXiv:1803.06026\/}

\bibitem{mijalkov2016engineering}
Mijalkov M, McDaniel A, Wehr J and Volpe G 2016 {\em Physical Review X\/} {\bf
  6} 011008

\bibitem{khadka2018active}
Khadka U, Holubec V, Yang H and Cichos F 2018 {\em Nature Communications\/}
  {\bf 9} 3864

\bibitem{Brambilla2013}
Brambilla M, Ferrante E, Birattari M and Dorigo M 2013 {\em Swarm
  Intelligence\/} {\bf 7} 1--41

\bibitem{bauerle2018quorum}
B{\"a}uerle T, Fischer A, Speck T and Bechinger C 2018 {\em Nature
  Communications\/} {\bf 9}

\bibitem{gibbs1981optics}
Gibbs H~M, Hopf F~A, Kaplan D~L and Shoemaker R~L 1981 {\em Physical Review
  Letters\/} {\bf 46} 474

\bibitem{arecchi1992optics}
Arecchi F~T, Giacomelli G, Lapucci A and Meucci R 1992 {\em Physical Review
  A\/} {\bf 45} R4225

\bibitem{masoller2002feedback}
Masoller C 2002 {\em Physical Review Letters\/} {\bf 88} 034102

\bibitem{kanter2010optics}
Kanter I, Aviad Y, Reidler I, Cohen E and Rosenbluh M 2010 {\em Nature
  Photonics\/} {\bf 4} 58

\bibitem{Baraban2013}
Baraban L, Streubel R, Makarov D, Han L, Karnaushenko D, Schmidt O~G and
  Cuniberti G 2013 {\em ACS Nano\/} {\bf 7} 1360--1367

\bibitem{Qian2013}
Qian B, Montiel D, Bregulla A, Cichos F and Yang H 2013 {\em Chemical
  Science\/} {\bf 4}(4) 1420--1429

\bibitem{walther2008janus}
Walther A and M{\"u}ller A~H~E 2008 {\em Soft Matter\/} {\bf 4} 663--668

\bibitem{jiang2010janus}
Jiang H~R, Yoshinaga N and Sano M 2010 {\em Physical Review Letters\/} {\bf
  105} 268302

\bibitem{kramers1940original}
Kramers H~A 1940 {\em Physica\/} {\bf 7} 284--304

\bibitem{hanggi1990review}
H{\"a}nggi P, Talkner P and Borkovec M 1990 {\em Reviews of Modern Physics\/}
  {\bf 62} 251

\bibitem{bullerjahn201spectroscopy}
Bullerjahn J~T, Sturm S and Kroy K 2014 {\em Nature Communications\/} {\bf 5}
  4463

\bibitem{bellman1963delay}
Bellman R and Cooke K~L 1963 {\em Delay differential equations\/} vol~6
  (Academic Press)

\bibitem{atay2010complex}
Atay F~M 2010 {\em Complex time-delay systems: theory and applications\/}
  (Berlin: Springer)

\bibitem{michiels2007stability}
Michiels W and Niculescu S~I 2007 {\em Stability and stabilization of
  time-delay systems: an eigenvalue-based approach\/} (Society for Industrial
  and Applied Mathematics)

\bibitem{foss1996multistability}
Foss J, Longtin A, Mensour B and Milton J 1996 {\em Physical Review Letters\/}
  {\bf 76} 708

\bibitem{longtin2009stochastic}
Longtin A 2009 Stochastic delay-differential equations {\em Complex time-delay
  systems\/} (Springer) pp 177--195

\bibitem{guillouzic1999small}
Guillouzic S, L'Heureux I and Longtin A 1999 {\em Physical Review E\/} {\bf 59}
  3970

\bibitem{frank2003fokker}
Frank T~D, Beek P~J and Friedrich R 2003 {\em Physical Review E\/} {\bf 68}
  021912

\bibitem{loos2017force}
Loos S and Klapp S~H~L 2017 {\em Physical Review E\/} {\bf 96} 012106

\bibitem{frank2005delay}
Frank T~D 2005 {\em Physical Review E\/} {\bf 71} 031106

\bibitem{kuchler1992delay}
K{\"u}chler U 1992 {\em Stochastics: An International Journal of Probability
  and Stochastic Processes\/} {\bf 40} 23

\bibitem{giuggioli2016fokker}
Giuggioli L, McKetterick T~J, Kenkre V and Chase M 2016 {\em Journal of Physics
  A: Mathematical and Theoretical\/} {\bf 49} 384002

\bibitem{adelman1976fokker}
Adelman S~A 1976 {\em The Journal of Chemical Physics\/} {\bf 64} 124--130

\bibitem{fox1977generalized}
Fox R~F 1977 {\em Journal of Mathematical Physics\/} {\bf 18} 2331--2335

\bibitem{hanggi1978correlation}
H{\"a}nggi P 1978 {\em Zeitschrift f{\"u}r Physik B Condensed Matter\/} {\bf
  31} 407--416

\bibitem{sancho1982analytical}
Sancho J~M, San~Miguel M, Katz S~L and Gunton J~D 1982 {\em Physical Review
  A\/} {\bf 26} 1589

\bibitem{hernandez1983joint}
Hern{\'a}ndez-Machado A, Sancho J~M, San~Miguel M and Pesquera L 1983 {\em
  Zeitschrift f{\"u}r Physik B Condensed Matter\/} {\bf 52} 335--343

\bibitem{gopalsamy2013population}
Gopalsamy K 2013 {\em Stability and Oscillations in Delay Differential
  Equations of Population Dynamics\/} vol~74 (Berlin/Heidelberg: Springer
  Science \& Business Media)

\bibitem{mao2005population}
Mao X, Yuan C and Zou J 2005 {\em Journal of Mathematical Analysis and
  Applications\/} {\bf 304} 296--320

\bibitem{voss200economic}
Voss H~U and Kurths J 2002 Analysis of economic delayed-feedback dynamics {\em
  Modelling and Forecasting Financial Data\/} (Springer) pp 327--349

\bibitem{stoica2005finance}
Stoica G 2005 {\em Proceedings of the American Mathematical Society\/} {\bf
  133} 1837--1841

\bibitem{mackey1989price}
Mackey M~C 1989 {\em Journal of Economic Theory\/} {\bf 48} 497--509

\bibitem{gao2009finance}
Gao Q and Ma J 2009 {\em Nonlinear Dynamics\/} {\bf 58} 209

\bibitem{kyrychko2010engineering}
Kyrychko Y~N and Hogan S~J 2010 {\em Journal of Vibration and Control\/} {\bf
  16} 943--960

\bibitem{beuter1993feedback}
Beuter A, B{\'e}lair J and Labrie C 1993 {\em Bulletin of mathematical
  biology\/} {\bf 55} 525--541

\bibitem{chen1997coordination}
Chen Y, Ding M and Kelso J~A~S 1997 {\em Physical Review Letters\/} {\bf 79}
  4501

\bibitem{Novak2008}
Nov{\' a}k B and Tyson J~J 2008 {\em Nature Reviews Molecular Cell Biology\/}
  {\bf 9} 981

\bibitem{haken2007brain}
Haken H 2007 {\em Brain Dynamics: an Introduction to Models and Simulations\/}
  (Berlin/Heidelberg: Springer Science \& Business Media)

\bibitem{marcus1989stability}
Marcus C~M and Westervelt R~M 1989 {\em Physical Review A\/} {\bf 39} 347

\bibitem{sompolinsky1991visual}
Sompolinsky H, Golomb D and Kleinfeld D 1991 {\em Physical Review A\/} {\bf 43}
  6990

\bibitem{rosinberg2015thermodyn}
Rosinberg M~L, Munakata T and Tarjus G 2015 {\em Physical Review E\/} {\bf 91}
  042114

\bibitem{van2018thermodynamic}
Van~Vu T and Hasegawa Y 2018 {\em arXiv preprint arXiv:1809.06610\/}

\bibitem{loos2019heat}
Loos S~A and Klapp S~H 2019 {\em Scientific Reports\/} {\bf 9} 2491

\bibitem{seifert2012thermodyn}
Seifert U 2012 {\em Reports on Progress in Physics\/} {\bf 75} 126001

\bibitem{sekimoto2010stochastic}
Sekimoto K 2010 {\em Stochastic Energetics\/} vol 799 (Berlin: Springer)

\bibitem{Kubo1966}
Kubo R 1966 {\em Reports on Progress in Physics\/} {\bf 29} 255--284

\bibitem{Stephens1963random}
Stephens M 1963 {\em Biometrika\/} {\bf 50} 385--390

\bibitem{Peruani2007}
Peruani F and Morelli L~G 2007 {\em Physical Review Letters\/} {\bf 99}(1)
  010602

\bibitem{Selmke2018theory}
Selmke M, Khadka U, Bregulla A~P, Cichos F and Yang H 2018 {\em Physical
  Chemistry Chemical Physics\/} {\bf 20} 10502--10520

\bibitem{Risken1996}
Risken H and Frank T 1996 {\em Fokker-Planck Equation: Methods of Solution and
  Applications\/} 2nd ed Springer Series in Synergetics (Berlin Heidelberg:
  Springer-Verlag)

\bibitem{bouchet2010thermodynamics}
Bouchet F, Gupta S and Mukamel D 2010 {\em Physica A: Statistical Mechanics and
  its Applications\/} {\bf 389} 4389--4405

\bibitem{fuchs2016thermodynamics}
Fuchs J, Goldt S and Seifert U 2016 {\em Europhysics Letters\/} {\bf 113} 60009

\bibitem{Frank2016}
Frank T 2016 {\em Physics Letters A\/} {\bf 380} 1341 -- 1351

\bibitem{Seifert2005}
Seifert U 2005 {\em Phys. Rev. Lett.\/} {\bf 95}(4) 040602

\bibitem{grote1980ght}
Grote R~F and Hynes J~T 1980 {\em The Journal of Chemical Physics\/} {\bf 73}
  2715--2732

\bibitem{pollak1990vtst}
Pollak E 1990 {\em The Journal of Chemical Physics\/} {\bf 93} 1116--1124

\bibitem{grabert1988delay}
Grabert H and Linkwitz S 1988 {\em Physical Review A\/} {\bf 37} 963

\bibitem{guillouzic2000rate}
Guillouzic S, L'Heureux I and Longtin A 2000 {\em Physical Review E\/} {\bf 61}
  4906

\bibitem{curtin2004delay}
Curtin D, Hegarty S~P, Goulding D, Houlihan J, Busch T, Masoller C and Huyet G
  2004 {\em Physical Review E\/} {\bf 70} 031103

\bibitem{Ornigotti2018}
Ornigotti L, Ryabov A, Holubec V and Filip R 2018 {\em Physical Review E\/}
  {\bf 97}(3) 032127

\bibitem{Holubec2018}
Holubec V, Kroy K and Steffenoni S 2019 {\em Physical Review E\/} {\bf 99}(3)
  032117

\bibitem{Siiler2018}
\ifmmode~\check{S}\else \v{S}\fi{}iler M, Ornigotti L, Brzobohat\'y O, J\'akl
  P, Ryabov A, Holubec V, Zem\'anek P and Filip R 2018 {\em Phys. Rev. Lett.\/}
  {\bf 121}(23) 230601

\bibitem{bruwier1930vequation}
Bruwier L 1930 Sur l'\'{e}quation fonctionelle $
  y^{(n)}(x)+a_{1}y^{(n-1)}(x+c)+...+a_{n-1}y'(x+\overline{n-1}c)+a_{n}y(x+nc)=0$
  {\em Comptes Rendus du Congres National des Sciences\/} pp 91--97

\bibitem{perron1939bruwiersche}
Perron O 1939 {\em Mathematische Zeitschrift\/} {\bf 45} 127--141

\bibitem{chase2016langevin}
Chase M, McKetterick T~J, Giuggioli L and Kenkre V 2016 {\em The European
  Physical Journal B\/} {\bf 89} 87

\bibitem{Olver2010}
Olver F~W~J, Lozier D~W, Boisvert R~F and Clark C~W 2010 {\em NIST Handbook of
  Mathematical Functions Hardback and CD-ROM\/} (Cambridge University Press)

\bibitem{Corless2007}
Corless R~M, Ding H, Higham N~J and Jeffrey D~J 2007 The solution of $s exp(s)
  = a$ is not always the {L}ambert $w$ function of $a$ {\em Proceedings of the
  2007 International Symposium on Symbolic and Algebraic Computation\/} ISSAC
  '07 (New York, NY, USA: ACM) pp 116--121

\bibitem{Yi2010}
Yi S, Nelson P~W and Ulsoy A~G 2010 {\em Time-Delay Systems: Analysis and
  Control Using the Lambert W Function\/} (Singapur: World Scientific)

\end{thebibliography}
